\documentclass[aps, prd, reprint, nofootinbib, groupedaddress, showpacs]{revtex4-1}
\usepackage{amsmath,amssymb, amsthm,amstext}
\usepackage{natbib}
\usepackage{graphicx}
\usepackage{color}
\usepackage{array}
\usepackage{bm}
\usepackage{multirow}
\usepackage[breaklinks,colorlinks, citecolor=blue]{hyperref}
\usepackage{pifont, ragged2e}
\usepackage{enumitem}

\def\be{\begin{equation}}
\def\ee{\end{equation}}
\def\mE{\mathcal E}
\def\Planck{\emph{Planck} }

\newcommand*\Bell{\ensuremath{\boldsymbol\ell}}
\newcommand{\xmark}{\text{\ding{55}}}

\begin{document}
\title{Approximate Likelihood Approaches for Detecting the Influence of Primordial
Gravitational Waves in Cosmic Microwave Background Polarization}
\author{Zhen Pan}
\email{zhpan@ucdavis.edu}
\affiliation{Department of Physics, University of California,
One Shields Avenue, Davis, CA, 95616}
\author{Ethan Anderes}
\email{anderes@ucdavis.edu}
\affiliation{Department of Statistics, University of California,
One Shields Avenue, Davis, CA, 95616, USA}
\author{Lloyd Knox}
\email{lknox@ucdavis.edu}
\affiliation{Department of Physics, University of California,
One Shields Avenue, Davis, CA, 95616, USA}

\date{\today}

\begin{abstract}
One of the major targets for next-generation cosmic microwave background (CMB)
experiments is the detection of the primordial B-mode signal. Planning is under
way for Stage-IV experiments that are projected to have instrumental noise small
enough to make lensing and foregrounds the dominant source of uncertainty for estimating
the tensor-to-scalar ratio $r$ from polarization maps.
This makes delensing a crucial part of future CMB polarization science.
In this paper we  present a likelihood method for estimating the
tensor-to-scalar ratio $r$ from CMB polarization observations, which
combines the benefits of a full-scale likelihood approach with the tractability
of the quadratic delensing technique.
This method is a pixel space, all order likelihood analysis of the quadratic delensed B modes,
and it essentially builds upon the quadratic
delenser by taking into account all order lensing and pixel space anomalies.
Its tractability relies on a crucial factorization of the pixel space covariance matrix of the
polarization observations which allows one to compute the full Gaussian approximate likelihood profile,
as a function of $r$, at the same computational cost of a single likelihood evaluation.
\end{abstract}


\maketitle

\section{\label{sec:intro}Introduction}
The inflation paradigm has successfully explained the origin of primordial density
perturbations that grew into the Cosmic Microwave Background (CMB)
anisotropies and large scale structure we
observe \citep[e.g.][]{Mukhanov1981, Guth1981, Linde1982, Albrecht1982, Lidsey1997, Lyth1999}.
A key prediction of inflation is the background of primordial
gravitational waves (GWs) or tensor-mode perturbations
\citep[e.g.][]{Starobinskii1979, Rubakov1982, Fabbri1983,%
Abbott1984, Starobinskii1985},
which imprints a unique polarization pattern,
called a primordial B mode, on the CMB anisotropies
\cite{Stebbins1996, Kamionkowski1997a, Kamionkowski1997,Seljak1997,%
Seljak1997a,Zaldarriaga1997a}.
Further, detection of a nearly scale-invariant background of
GWs would severely challenge non-inflationary models
\cite[e.g.][]{Khoury2001, Khoury2001a,Khoury2003a,Steinhardt2002, Boyle2004a}.
The strength of primordial gravitational waves or tensor-mode power is commonly
quantified by the tensor-to-scalar ratio $r$.
Joint analysis of BICEP2/Keck Array and \Planck data yields an upper bound
$r < 0.12$ at $95\%$ confidence level \citep{BICEP2/Keck2015},
the bound is slightly tightened  when the \Planck high-$\ell$
polarization data are included \citep{PlanckCollaborationXX2015},
and BICEP2/Keck Collaboration gives the latest upper bound $r < 0.09$ at
$95\%$ confidence level \citep{BICEP2/Keck2016}.
Fourth generation experiments,
including COrE, LiteBird, and CMB Stage-IV,  are expected to
constrain $r$ with uncertainty $\sigma(r)\simeq 0.001$
\citep{COrECollaboration2011, LiteBirdCollaboration2014,%
LiteBird2016, s42016,Cabass2016, Kamionkowski2016, Delabrouille2017}.

The primordial B modes are contaminated by several sources:
emission from galactic dust and other foregrounds
\cite{Hildebrand1999, Draine2004, Benoit2004,Mortonson2014, Niemack2015,%
PlanckCollaborationXXX2016, PlanckCollaborationL2016, Krachmalnicoff2016},
instrumental noise, and gravitational lensing of scalar CMB perturbations.
The B modes generated by gravitational lensing of the CMB have  been detected
\citep{Hanson2013, Ade2014, VanEngelen2014, Story2014, BICEP2/Keck2015, PlanckCollaboration2015a}.
The lensed B-mode power spectrum is nearly a constant at small multipoles ($\ell \lesssim 1000$) and
therefore manifests as an effective white noise with amplitude $\sim 5$ $\mu$K-arcmin
\citep{Lewis2006, Sherwin2015}.
For CMB Stage-IV, we expect to decrease the instrumental noise to $\sim 1\ \mu$K-arcmin \citep{s42016}.
Then, the lensing B noise (and foregrounds) would become the dominant noise source
limiting the primordial B-mode survey.

Fortunately,  the lensing B noise is well understood.
Up to leading order, one can effectively delense observed B modes by utilizing
a quadratic combination of observed E modes and an estimate of the lensing potential field $\phi^{\rm est}$
\citep{Knox2002, Kesden2002, Seljak2003a, Simard2015a, Sherwin2015}.
We find that the validity of the quadratic delenser crucially depends on a partial cancellation of
higher order lensing terms (see Section \ref{sec:app4} for details).
However, in the regime of low instrumental noise and small lensing potential field estimate uncertainty,
higher-order lensing terms, ignored by the quadratic delensing technique,
can have an appreciable effect. These higher order terms not only induce
a delensing bias but also contain information on primordial B modes.
In addition, experimental complexities such as non-stationary noise and sky cuts
become non-trivial for spectral based methods such as the quadratic delenser.

As an alternative, a full-scale likelihood analysis of the tensor-to-scalar ratio $r$ can,
in principle, optimally account for all the Gaussian and non-Gaussian information in the CMB observations.
Unfortunately, a full likelihood analysis requires computation resources beyond
what is available in the near future.
In this paper, we introduce a likelihood approximation which is modified from the full-scale
likelihood, so as to be computationally tractable.

We start with introducing a Gaussian likelihood incorporating all the 2-point information.
A key element of our likelihood analysis is the covariance matrix of the polarization maps.
For each data pair ($d_i, d_j$) ($d$ can be $Q$ or $U$),  its covariance
depends on the primordial polarization power spectra $C_\ell^{EE}$ and $C_\ell^{BB,r}$,
lensing potential field $\phi(x)$ (and instrumental noise $N^{QQ}$ and $N^{UU}$),
where the E-mode power $C_\ell^{EE}$ has been well constrained
\citep[e.g.][]{PlanckCollaborationXI2015, PlanckCollaborationXIII2015},
while the primordial B-mode signal has not been detected.
Following the prediction of the inflation paradigm,
we assume the tensor perturbations to be scale-invariant and Gaussian.
Hence all the primordial B-mode  information is encoded in the single parameter $r$, the
tensor-to-scalar ratio at $k=0.05$ Mpc$^{-1}$.
Then the covariance matrix $\tilde \Sigma_{r,\phi}$
depends on the unknown parameter $r$ and the underlying lensing potential field $\phi(x)$,
where $\phi(x)$ can be estimated from exterior tracers, e.g. cosmic infrared background (CIB)
\citep{Song2003, Dole2006, PlanckCollaborationXVII2014,%
PlanckCollaboration2015a, Larsen2016, Manzotti2017} or
from intrinsic CMB via, e.g. quadratic estimators \citep{Hu2001, Hu2002b}
or Bayesian approach \citep{Hirata2003a, Hirata2003,Anderes2011, Anderes2015, Millea2017}.
We obtain a covariance matrix $\Sigma_r$ depending only on $r$ by marginalizing $\tilde \Sigma_{r,\phi}$
over uncertainties in the lensing potential field estimate $\phi^{\rm est}$.
With this full covariance matrix $\Sigma_r$, it is then straightforward to compute the likelihood of $r$
for given data vector $d$ by approximating $d$ as a Gaussian vector.

In principle, this Gaussian likelihood method can exploit
all the 2-point $r$ information from the polarization maps,
but usually the computation resource demands are still excessive. For example,
to constrain $r$ from some polarization maps with $p$ pixels,
we need to compute the full likelihood profile $L(r|d)$ as a function of $r$,
which in practice requires computing the likelihood on a range of $r$ values,
say $50$  values evenly distributed in the interval $[0, 0.2]$.
For each different $r$, we need to compute the quadratic form $d^\intercal\Sigma_r^{-1} d$ and
the determinant $\det(\Sigma_r)$, due to the $r$ dependence of the covariance matrix.
In any realistic experiments with $p \gtrsim 10^4$,
it is a huge amount of work to compute and invert the covariance matrix of dimension $2p\times 2p$
for that many $r$ values, where the factor $2$ comes from two observables $Q$ and $U$ on each pixel.

In this paper, we present a modified Gaussian likelihood tailoring
the full-scale likelihood analysis so as to be computationally tractable.
The method consists of two parts.
In the first part, we decompose the covariance matrix $\Sigma_r$ as $\Sigma^{\rm en}+ r\Sigma^{\rm b}$,
where $\Sigma^{\rm en}$ is the contribution from E modes and instrumental noise,
and $r\Sigma^{\rm b}$ is the contribution from B modes. This decomposition allows us to compute
the covariance $\Sigma_r$, as a function of $r$,
at the same computational cost of a single covariance matrix computation.
In the second part, we suppress data size by tracking only $s$
high signal-to-noise modes, say the large-scale quadratic delensed B modes.
We project out the lensing-generated B modes and obtain the delensed modes $B^{\rm del}$
from the polarization data $d$ via a projection matrix $v$,
$(^0B^{\rm del}_{\Bell})_s = (v^\intercal)_{s\times 2p} d_{2p}$,
with $s\sim 500$ and the upper left index $^0$ denoting the projected data vector
limited to the $s$ lowest frequency  modes available.
Then, the covariance matrix of the projected data vector  $^0B^{\rm del}_{\Bell}$
is given by $v^\intercal \Sigma_r v$,
with which the computation of the $r$ likelihood  $L(r|^0B^{\rm del}_{\Bell} )$
given the projected data vector turns out to be tractable.
This method can be naturally extended to incorporate higher frequency modes,
as we describe in Section \ref{sec:like}.

The paper is organized as follows. We introduce the quadratic delenser
and the likelihood based delenser in Section \ref{sec:delen} and Section \ref{sec:like}, respectively.
In Section \ref{sec:sims}, we apply the two $r$ constraining techniques
on simulations mimicking Stage III and IV CMB surveys,
and compare their $r$ constraints.
We conclude with Section \ref{sec:summary}. For reference, we derive the analytic expression
for the covariance matrices of the polarization maps, and the eigenvalue method for inverting large matrices
in Appendices \ref{sec:app1}, \ref{sec:app2} and \ref{sec:app3}, respectively.

\section{Quadratic Delenser}
\label{sec:delen}
For simplicity, we assume no contamination of foregrounds throughout this paper.
Then the observed B modes can generally be expressed as
\be
B^{\rm obs} = B^{r} + B^{\rm len} + N^B,
\ee
where $B^r, B^{\rm len}, N^B$ are primordial B signal, lensing B noise
and instrumental B noise, respectively. To constrain the primordial B signal,
delensing is essential, where we obtain an estimate of the lensing B noise and subtract it off from the
observed B modes. Here we introduce a quadratic delenser.

Accurate to the leading order of the lensing potential $\phi_{\Bell}$,
$B^{\rm len}$ is the convolution of the lensing potential and primordial E modes, i.e.,
\be
\label{eq:convol}
B^{\rm len}_{\Bell} = \int \frac{d^2\Bell'}{2\pi} \Bell'\cdot(\Bell-\Bell') \sin(2\varphi_{\Bell,\Bell'}) E_{\Bell'}\phi_{\Bell-\Bell'}.
\ee
Usually, the underlying lensing potential is not  known {\it a priori}, but can be estimated from either intrinsic CMB
or from external tracers. From an estimated lensing potential $\phi^{\rm est}_{\Bell}$
and observed modes $E_{\Bell}^{\rm obs}$, we construct a quadratic estimate of the lensing B noise
\be
\label{eq:wconvol}
B^{\rm len, est}_{\Bell} = \int \frac{d^2\Bell'}{2\pi} f_{\Bell,\Bell'} \Bell' \cdot(\Bell-\Bell')\sin(2\varphi_{\Bell,\Bell'})  E^{\rm obs}_{\Bell'} \phi^{\rm est}_{\Bell-\Bell'},
\ee
where $\varphi_{\Bell,\Bell'} = \varphi_{\Bell} - \varphi_{\Bell'}$,
$E^{\rm obs}$ is the observed E modes (to the lowest order, the difference between
lensed E and primordial E can be neglected),
and the weighting function $f_{\Bell,\Bell'}$ is to be determined by minimizing the residual,
$B_{\Bell}^{\rm res} = B^{\rm len}_{\Bell} - B^{\rm len, est}_{\Bell}$.
If we define the correlation coefficient of $\phi_{\Bell}$ and $\phi^{\rm est}_{\Bell}$
\be
\label{eq:corr}
\rho_\ell =  \frac{C_\ell^{\phi, \phi^{\rm est}}}{\sqrt{C_\ell^{\phi\phi} C_\ell^{\phi^{\rm est}\phi^{\rm est}}}},
\ee
the optimal weight at leading order was proved to be \citep{Sherwin2015}
\be
f_{\Bell,\Bell'} = \frac{C_{l'}^{EE}}{C_{l'}^{EE} + N_{l'}^{EE}} \rho_{|\Bell-\Bell'|}^2,
\ee
which enables a minimal residual power spectrum
\be
\label{eq:respower}
\begin{aligned}
C_{\Bell}^{BB, {\rm res}}
& =\int \frac{d^2\Bell'}{(2\pi)^2} \left[\Bell'\cdot(\Bell-\Bell') \sin(2\varphi_{\Bell,\Bell'})  \right]^2\\
&\times C_{\Bell'}^{EE} C^{\phi\phi}_{\Bell-\Bell'} (1-f_{\Bell,\Bell'}).
\end{aligned}
\ee

After subtracting off the template $B_{\Bell}^{\rm len, est}$, we obtain a quadratic delensed B-mode map
\be
\label{eq:bdel}
B_{\Bell}^{\rm del} = B_{\Bell}^{\rm obs} - B_{\Bell}^{\rm len, est} = B_{\Bell}^r + B_{\Bell}^{\rm res} + N_{\Bell}^B,
\ee
and its power spectrum
\be
\label{eq:delpower}
C_\ell^{BB, {\rm del}} = C_\ell^{BB,r} + C_\ell^{BB, {\rm res}} + N^{BB}_\ell.
\ee
From the delensed B modes, one can better constrain $r$ due to the suppressed lensing B noise.
Note that in the evaluation of
the residual lensing B power
$C_\ell^{BB, {\rm res}} $ of Eq. (\ref{eq:respower}) we have made two approximations:
1) we keep only the linear order lensing in $B^{\rm len}$;
2) we completely ignore the lensing in $E^{\rm obs}$.

\section{Gaussian Likelihood Delenser}
\label{sec:like}

In contrast to the quadratic delenser, the likelihood analysis works on observables
$Q^{\rm obs}$ and $U^{\rm obs}$ in pixel space.
Concatenating the polarization data on all pixels yields a length-$2p$ data vector
\be
d = [Q^{\rm obs}(x_1) \cdots Q^{\rm obs}(x_p), U^{\rm obs}(x_1) \cdots  U^{\rm obs}(x_p)]^\intercal,
\ee
with $p$ being the number of pixels.
We first evaluate the covariance matrix of the data vector,
which depends on the primordial polarization power spectra $C_\ell^{EE}$ and $C_\ell^{BB,r}$,
lensing potential field $\phi(x)$ (and instrumental noise). With $C_\ell^{EE}$ being well-determined,
and $C_\ell^{BB,r}$ being parametrized by the tensor-to-scalar ratio $r$,
we marginalize the covariance matrix $\tilde \Sigma_{r,\phi}$ over uncertainties in $\phi(x)$ estimate,
and obtain a covariance matrix $\Sigma_r$ depending only on $r$. Then it is straightforward to compute the approximate likelihood of $r$ for given data $d$
by approximating $d$ as a Gaussian vector, i.e.,
\be
\label{eq:likeli}
    -2\log L(r|d) = d^\intercal\Sigma^{-1}_r d + \log \det\Sigma_r,
\ee
up to a constant term.

\subsection{Comparison with the Quadratic Delenser}
Before delving into the details of the Gaussian likelihood delenser,
it would be useful to do a brief comparison with the quadratic delenser (see Table \ref{table1}):
\begin{itemize}
    \item The Quadratic Delenser works on the delensed modes $B^{\rm del}_{\Bell}$ in Fourier space,
    and approximates these modes as stationary and Gaussian, i.e.,
    \[
    \begin{pmatrix}
        \vdots \\
        B^{\rm del}_{\Bell}\\
        \vdots
    \end{pmatrix} \sim
    N(0, \begin{bmatrix}
    \ddots & 0 & 0 \\
    0 & C^{BB,\rm del}_{\ell} & 0\\
    0 & 0 & \ddots
    \end{bmatrix}),
    \]
    where the power spectrum $C^{BB,\rm del}_{\ell}$, derived in Eq. (\ref{eq:delpower}), only
    takes into account the leading-order in $\phi$.
    Therefore, the quadratic delenser exploits the 2-point information in
    a biased way by ignoring the non-stationarity and higher-order lensing in the power spectrum.

    \item The Gaussian Likelihood  Delenser works on the observables $d$ in pixel space, and approximates the data
    vector $d$ as Gaussian after marginalizing over uncertainties in the $\phi$ estimate. In the computation of
    the covariance matrix $\Sigma_r$, all-order lensing is taken into account
    and no stationarity assumption is made. Therefore, the Gaussian likelihood
    delenser naturally incorporates all the 2-point information.
\end{itemize}

\begin{table}
\begin{tabular}{ m{2.4cm} |c| c}
      Delenser & Quadratic Delenser  & Gaussian Likelihood  \\ \hline
     working space  & Fourier & pixel\\ \hline
      power spectrum / cov. matrix & leading order & all order\\ \hline
     non-stationarity& $\xmark $ & $\checkmark$ \\
     non-Gaussianity &  $\xmark $ & $\xmark $ \\ \hline
\end{tabular}
\caption{A brief comparison of the two delensers.}
\label{table1}
\end{table}

The Gaussian likelihood is potentially favored in several aspects,
but usually is computationally excessive.
As explained in the Introduction, the bottleneck of the likelihood analysis is
the covariance matrix $\Sigma_r$ related computation,
which is of large size $2p\times 2p$, and is a function of $r$.
Here we introduce a modified Gaussian likelihood method.
The method consists of two parts, covariance decomposition and data compression,
where the former allows us to compute the covariance matrix $\Sigma_r$,
as a function of $r$, at the computation cost of a single covariance matrix computation;
and the latter allows us to compress the covariance matrix
by tracking a small number of high $S/N$ modes.

\subsection{Covariance Decomposition}
To avoid repeating the computation of the covariance matrix $\Sigma_r$ for each different $r$,
we find it is possible to single out the $r$ dependence by decomposing the  covariance matrix as
\be
\label{eq:dec}
\Sigma_r = \Sigma^{\rm en} + r \Sigma^{\rm b},
\ee
where $\Sigma^{\rm en}$ is the contribution from E modes and instrumental noise,
and $r \Sigma^{\rm b}$ is the contribution from primordial B modes.
With this decomposition, we can obtain the covariance matrix $\Sigma_r$ as a function
of $r$, as long as the $r$-independent components $\Sigma^{\rm en}$ and  $\Sigma^{\rm b}$ are obtained.

For the covariance decomposition of Eq. (\ref{eq:dec}),
we first decompose observables $Q^{\rm obs}$ and $U^{\rm obs}$ as linear combinations of E modes and B modes.
Stokes parameters $Q$ and $U$ are related to coordinate independent quantities $E$ and $B$ via
\citep{Stebbins1996, Kamionkowski1997a, Kamionkowski1997, Seljak1997a}
\be
\label{eq:queb}
\begin{aligned}
Q_{\Bell} &= - \cos(2\varphi_{\Bell}) E_{\Bell} + \sin(2\varphi_{\Bell}) B_{\Bell}, \\
U_{\Bell} &= - \sin(2\varphi_{\Bell}) E_{\Bell} - \cos(2\varphi_{\Bell}) B_{\Bell}.
\end{aligned}
\ee
We define the following modulated E/B modes
\be
\begin{aligned}
SE_{\Bell} & \equiv - \sin(2\varphi_{\Bell}) E_{\Bell}, \quad  & SB_{\Bell}  \equiv +\sin(2\varphi_{\Bell}) B_{\Bell},\\
CE_{\Bell} & \equiv - \cos(2\varphi_{\Bell}) E_{\Bell}, \quad  & CB_{\Bell} \equiv -\cos(2\varphi_{\Bell}) B_{\Bell},
\end{aligned}
\ee
then the observables $Q^{\rm obs}$ and $U^{\rm obs}$ are consequently expressed as
\be
\begin{aligned}
Q^{\rm obs}(x) &= \widetilde{CE}(x) + \sqrt{r} \widetilde{SB^0}(x) + N^Q(x),\\
U^{\rm obs}(x) &= \widetilde{SE}(x) + \sqrt{r} \widetilde{CB^0}(x) + N^U(x),
\end{aligned}
\ee
where $B^0$ denotes fiducial B modes with unity power spectrum $C_\ell^{BB,r=1}$,
tildes denote lensed fields $\widetilde X(x) = X(x+\nabla\phi(x))$ ($X = CE, SE, CB, SB$),
and $N^{Q,U}$ is the $Q/U$ noise.

With above decomposition, we find the data vector $d$ is Gaussian with covariance $\tilde\Sigma_{r, \phi}$
for given $r$ and $\phi(x)$, i.e., $d \sim  N(0, \tilde\Sigma_{r, \phi})$, where
\be
\tilde\Sigma_{r, \phi} \equiv
\left[
\begin{tabular}{cc}
  $\tilde \Sigma^{Q^{\rm obs},Q^{\rm obs}}$ & $\tilde\Sigma^{Q^{\rm obs},U^{\rm obs}}$  \\
  $\tilde \Sigma^{Q^{\rm obs},U^{\rm obs}}$ & $\tilde\Sigma^{U^{\rm obs},U^{\rm obs}}$
  \end{tabular}
\right]_{r,\phi}
\ee
and the covariance matrix is naturally expressible in the form of Eq. (\ref{eq:dec}), i.e.,
\be
\label{eq:elements}
\begin{aligned}
    \tilde \Sigma^{Q^{\rm obs},Q^{\rm obs}} &= \tilde\Sigma^{CE,CE} + r \tilde\Sigma^{SB^0, SB^0} + \Sigma^{NQ,NQ},\\
    \tilde \Sigma^{Q^{\rm obs},U^{\rm obs}} &= \tilde\Sigma^{CE,SE} + r \tilde\Sigma^{SB^0, CB^0} ,\\
    \tilde \Sigma^{U^{\rm obs},U^{\rm obs}} &= \tilde\Sigma^{SE,SE} + r \tilde\Sigma^{CB^0, CB^0} + \Sigma^{NU,NU}.
\end{aligned}
\ee

In a more practical case, we only have an estimate of lensing potential $\phi_{\Bell}^{\rm est}$
which is a noisy version of the true $\phi_{\Bell}$,
i.e., $\phi_{\Bell}^{\rm est} = \phi_{\Bell} + n^\phi_{\Bell}$,
where $n^\phi_{\Bell}$ is the uncertainty of the $\phi$ estimate
and its power spectrum $N_\ell^{\phi\phi}$ is usually an output of the lensing estimator used.
For an unbiased estimator with Gaussian uncertainty, one can write $n^\phi_{\Bell}\sim N(0, N_\ell^{\phi\phi})$.
Then the correlation coefficient of $\phi$ and $\phi^{\rm est}$ defined in Eq. (\ref{eq:corr}) now
is explicitly known as
\be
\rho_\ell =  \sqrt{\frac{C_\ell^{\phi\phi}}{C_\ell^{\phi\phi} + N_\ell^{\phi\phi}}}.
\ee
In this context, one can treat $\phi^{\rm est}$ as data and compute the
posterior on $\phi$ given $\phi^{\rm est}$, i.e.,
\be
P(\phi_{\Bell} | \phi^{\rm est}_{\Bell})
\sim N\left(\mu\phi_{\Bell},
C_{\ell}^{n\phi,n\phi}\right).
\ee
with
\be
\label{eq:wf}
\begin{aligned}
\mu \phi_{\Bell} &= \frac{C_\ell^{\phi\phi}}{C_\ell^{\phi\phi} + N_\ell^{\phi\phi}} \phi_{\Bell}^{\rm est}
=\rho_\ell^2 \phi_{\Bell}^{\rm est}, \\
C_{\ell}^{n\phi,n\phi}& = \frac{C_\ell^{\phi\phi}}{C_\ell^{\phi\phi} + N_\ell^{\phi\phi}}N_\ell^{\phi\phi}
=\rho_\ell^2  N_\ell^{\phi\phi},
\end{aligned}
\ee
Therefore a sample $\phi_{\Bell}  \sim P(\phi_{\Bell} | \phi^{\rm est}_{\Bell})$ can be writen as
\be
\phi_{\Bell} = \mu\phi_{\Bell} + n\phi_{\Bell},
\ee
with $n\phi_{\Bell}\sim N(0, C_\ell^{n\phi,n\phi})$.

Marginalizing $\tilde \Sigma_{r,\phi}$ over $n\phi$, we obtain a covariance matrix only depending on $r$, i.e.,
$\Sigma_r \equiv \langle \tilde \Sigma_{r,\phi} \rangle_{n\phi}$,
where its analytic form is presented in Appendices \ref{sec:app1} and \ref{sec:app2}.
The computation of its inverse matrix $\Sigma_r^{-1}$ is presented in Appendix \ref{sec:app3}.

\subsection{Data Compression}
\subsubsection{idea}
To compress the data, we project the original length-$2p$ data vector $d$ to
a length-$s$ data ($s\sim 500$) via a projection matrix $v$, and apply the likelihood analysis on the
projected data $\hat d_s = (v^\intercal)_{s\times 2p} d_{2p}$. Let $\Sigma_r$ be the covariance matrix of data vector $d$,
then $v^\intercal\Sigma_r v$ is the covariance matrix of projected data vector $\hat d$, i.e.,
$d\sim N\left(0, \Sigma_r\right)$,
and
$\hat d \sim N\left(0, v^\intercal \Sigma_r v \right)$.
Then the likelihood of $r$ given projected data $\hat d$ is simply
\be
\label{eq:projlikeli}
    -2\log L(r|\hat d)
    = \hat d^\intercal (v^\intercal\Sigma_r v)^{-1} \hat d + \log \det(v^\intercal\Sigma_r v),
\ee
up to a constant term.

\begin{figure}
\centering
\includegraphics[height=1.8in]{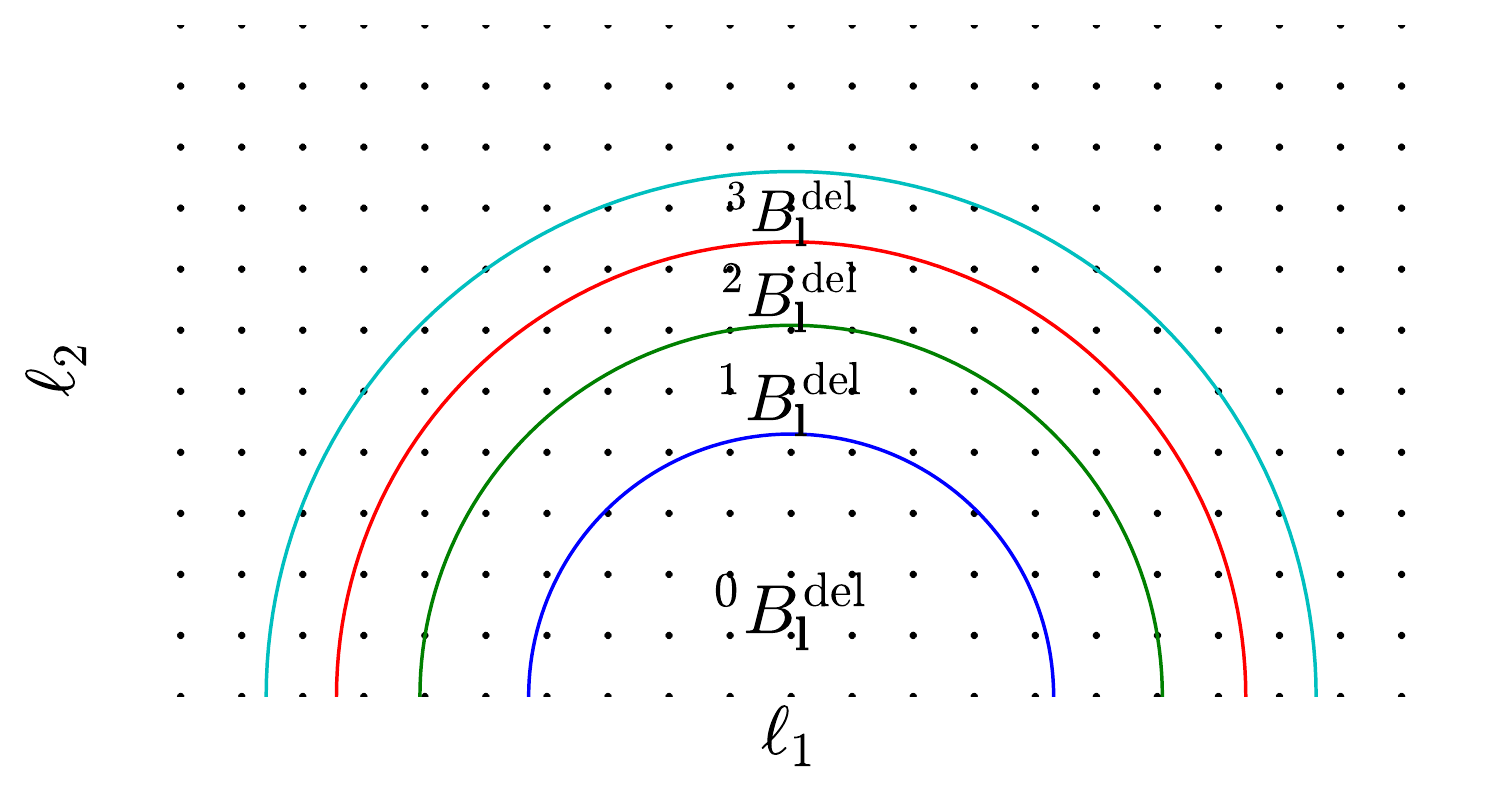}
\caption{ \label{fig:Bdel}
The modes covered by each different projected vector $^i B_{\Bell}^{\rm del} (i = 0, 1, 2, \cdots)$.
Here we only show the modes with $\ell_2 \geq 0$, since our observables $Q^{\rm obs}$
and $U^{\rm obs}$ are real numbers.}
\end{figure}

The goal is to find a projection matrix $v$ such that $\hat d$ is highly informative for $r$.
Since the primordial B modes at large scales are less contaminated by the lensing B noise,
a natural choice is to project the polarization data to the large-scale quadratic delensed modes
defined in Equation (\ref{eq:bdel}),
i.e., $\hat d^0 =$ $^0B_{\Bell}^{\rm del}$, where the upper index $^0$ denotes the projected data vector
limited to the $s$ lowest frequency  modes available (Figure \ref{fig:Bdel}).
The method can be naturally extended to higher frequency modes,
$\hat d^i = \ ^iB^{\rm del}_{\Bell}$ $(i = 1,2,\dots)$
with $\Bell$ running over the $s$ next/next-next/$\dots$ lowest frequency modes.
With these projected data vectors $\hat d^i = \ ^iB^{\rm del}_{\Bell}= (v^\intercal)^i d$,
the total likelihood is given by
\be
\label{eq:totlikeli}
    \log L(r|d) \approx \sum_{i=0}^{i_{\rm max}}\log L(r|\hat d^i),
\ee
assuming negligible correlation for different projected vectors.\footnote{
The large-scale delensed B modes are no longer the highest $S/N$ modes,
when foregrounds, contaminating the primordial B modes more at large scales, are considered.
Our methodology is flexible. In principle, modes could be selected that
minimizes noise and residual foreground contamination.}
We will confirm the validity of ignoring the cross correlation via simulations in Section \ref{sec:sims}.

We find that the modified Gaussian likelihood
method works better if we incorporate the same number of E modes
and delensed B modes in each projected vector,
i.e.,
\be
\label{eq:ebproj}
[\hat d^i]_{2s} =
\begin{bmatrix}
    (^iB^{\rm del}_{\Bell})_s\\
    (^iE^{\rm obs}_{\Bell})_s
\end{bmatrix}
= \begin{bmatrix}
    (v_b^\intercal)^i_{s\times 2p}\\
    (v_e^\intercal)^i_{s\times 2p}
\end{bmatrix}  d_{2p} = (v^\intercal)^i_{2s\times 2p} d_{2p}.
\ee

\subsubsection{projection matrix}
In this subsection, we focus on the computation of the projection matrix.
As described in Section \ref{sec:delen}, the quadratic delenser is actually a linear operator,
i.e.,
\be
\label{eq:qdrdel}
\begin{aligned}
d=(Q^{\rm obs}, U^{\rm obs})^\intercal &\xrightarrow{{\rm Eq.}(\ref{eq:queb})}  (E^{\rm obs}, B^{\rm obs}) ,  \\
 & \xrightarrow{{\rm Eq.}(\ref{eq:wconvol})} (B^{\rm len, est}, B^{\rm obs}) , \\
 & \xrightarrow {{\rm Eq.}(\ref{eq:bdel}) } B^{\rm obs}- B^{\rm len, est} = B^{\rm del}.
\end{aligned}
\ee
Therefore we can formally write the quadratic delensing as
$(B^{\rm del}_{\Bell})_{2p} = \mathcal P_{2p\times 2p} d_{2p}$,
where $\mathcal P$ is a concatenation of the three linear operations above,
and its matrix elements can be found by
recording the impulse response of the delensed modes
to each element in the data vector.
For example, we first do the quadratic delensing to a ``data vector"
$\delta_1 = (1,0, \dots,0)^\intercal_{2p}$
and denote the corresponding delensed modes as $(B^{\rm del}_{\Bell}|_{\delta_1})$,
i.e.,
\be
(B^{\rm del}_{\Bell}|_{\delta_1})_{2p} = \mathcal P_{2p\times2p} (\delta_1)_{2p} = \textrm{1st col. of}\ \mathcal P.
\ee
where $(B^{\rm del}_{\Bell}|_{\delta_1})$ is obtained via delensing of Eq. (\ref{eq:qdrdel}).
In this way, we obtain the matrix $\mathcal P$.

It is clear that the projection matrices of vectors $ ^iB^{\rm del}_{\Bell}$
correspond to row blocks of $\mathcal P$.
Explicitly, we write $(B^{\rm del}_{\Bell})_{2p} = \mathcal P_{2p\times 2p} d_{2p}$ as
\be
\begin{pmatrix}
    (^0B_{\Bell}^{\rm del})_s\\
    (^1B_{\Bell}^{\rm del})_s\\
    (^2B_{\Bell}^{\rm del})_s\\
    \dots
\end{pmatrix}
= \mathcal P_{2p\times 2p} d_{2p}
= \begin{pmatrix}
    (^0\mathcal P)_{s\times 2p}\\
    (^1\mathcal P)_{s\times 2p}\\
    (^2\mathcal P)_{s\times 2p}\\
    \dots
\end{pmatrix} d_{2p}.
\ee
and therefore we obtain $(v_b^\intercal)^i_{s\times 2p} = (^i\mathcal P)_{s\times 2p}$.
The projection matrices $(v_e^\intercal)^i$ can be obtained in a similar way.

\section{Simulations}
\label{sec:sims}
In this section, we apply the quadratic delenser and the modified Gaussian likelihood  method
on CMB polarization simulations, and compare the resulting $r$ constraints.
The fiducial cosmology we use is a flat $\Lambda$CDM cosmology with
a baryon density $\omega_{\rm b} = 0.02246$, a cold dark matter density $\omega_{\rm c} = 0.1185$,
a reionization optical depth $\tau = 0.1283$,
an angular size of sound horizon at recombination $100\theta_{\star} = 1.0410$,
an amplitude and a spectral index of the primordial scalar the perturbation power spectrum
$10^9A_{\rm s} = 2.1333, n_{\rm s} = 0.9686$,
and a tensor-to-scalar ratio $r$ in the range of $[0.001, 0.1]$.
For each different $r$, we simulate $500$ realizations of primordial polarization fields $Q(x)$ and $U(x)$,
then lense these fields via the same lensing potential field $\phi(x)$.
All the power spectra used in simulations are computed from the Boltzmann code  {\tt CLASS} \citep{Lesg2011}.

\subsection{Two Surveys}

\begin{table}
\scalebox{1.3} {
\begin{tabular}{ l r|c| c |c}
     &  & $\Delta_{\rm T} (\mu$K-arcmin) & $\theta_{\rm FWHM}$ & $f_{\rm sky}$ \\ \hline
    \multirow{2}{*}{Lb}  &  $\phi$ & 0.5  & $2'$ & \multirow{2}{*}{$2.7\%$}\\ \cline{2-4}
                         &  $B$    & 0.5 & $10'$ & \\ \hline
    \multirow{2}{*}{La}  &  $\phi$ & 1  & $2'$ & \multirow{2}{*}{$2.7\%$}\\ \cline{2-4}
                        &  $B$    & 1 & $10'$ & \\ \hline
    \multirow{2}{*}{N}  &  $\phi$ & 10  & $2'$  & \multirow{2}{*}{$2.7\%$} \\ \cline{2-4}
                        &  $B$ & 10 & $10'$ & \\ \hline
\end{tabular}
}
\caption{The three scenarios we simulated.}
\label{table}
\end{table}

\begin{figure}
\centering
\includegraphics[height=3in]{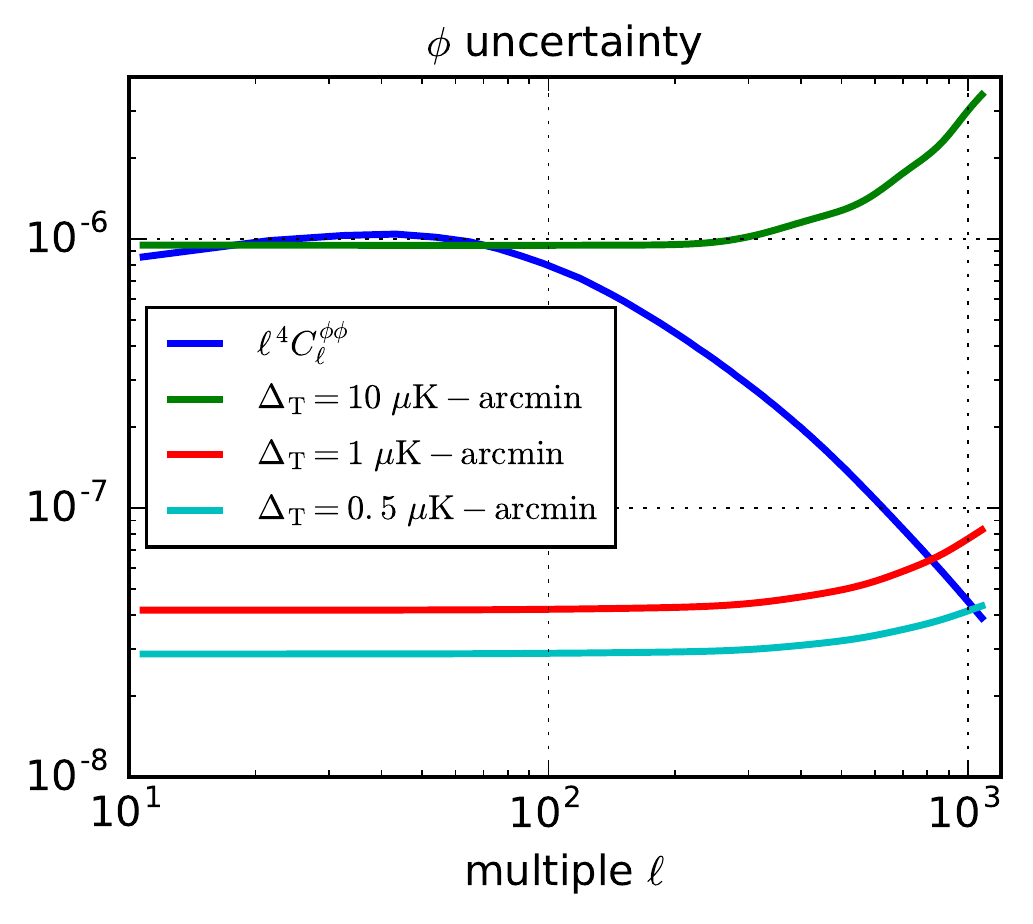}
\caption{ \label{fig:phi_noise}
The $\phi(x)$ reconstruction noises for Scenario N ($\Delta_{\rm T} = 10$ $\mu$K-arcmin), La
($\Delta_{\rm T} = 1$ $\mu$K-arcmin) and Lb ($\Delta_{\rm T} = 0.5$ $\mu$K-arcmin) surveys.}
\end{figure}

We consider a survey strategy consisting of two different surveys of the same area of sky, differing
in angular resolution. The main goal of the higher-resolution survey is to allow for a reconstruction of
the lensing potential. Such reconstructions benefit from reaching an angular scale comparable to the typical
lensing deflection angle of $\sim 2$ arcmin. In contrast, the primordial B-mode signal is on fairly large angular scales of
greater than a degree. In principal, one high-resolution survey could be used both for the lensing reconstruction
and for sensitivity to the primordial B-mode signal. However, there are advantages to using a survey dedicated
to the large-scale signals. These advantages do not appear in the idealized analyses that we perform here,
as they are related to systematic error control and foreground cleaning, as we now explain briefly.
The large-scale survey can be achieved with a smaller telescope with a simplified optics chain.
Having a smaller telescope facilitates boresight rotation, which BICEP2/Keck have used for null
tests to bound certain systematic errors. Foreground cleaning is also likely to be more of a challenge
at larger angular scales than it is for the smaller-angular scales with the bulk of the lensing information,
and serves as a further driver of differences in optimal design for the two surveys.
For these reasons a two-survey approach is likely to be a part of the strawman concepts for
the CMB Stage-IV instrument soon to emerge from the CMB Stage-IV Concept Definition Taskforce.

In this paper, we simulate three scenarios.
We consider a scenario `N' in which the B-mode instrument noise power
is larger than the B-mode lensing power, and two scenarios
`La' and `Lb' with the opposite situation.
The more sensitive scenarios La and Lb are motivated by potential CMB-S4 scenarios.
Each scenario consists of two surveys, a high-resolution survey for $\phi$ reconstruction
and a low-resolution survey capturing the B-mode signal, covering the same patch of the sky
(see Table \ref{table} for the survey configurations in detail).
For the high-resolution surveys, the lensing potential reconstruction noise
expected from the EB quadratic estimator \citep{Hu2001, Hu2002b,Anderes2013}
is shown in Figure \ref{fig:phi_noise}.

\begin{figure*}
\centering
\includegraphics[height=4in]{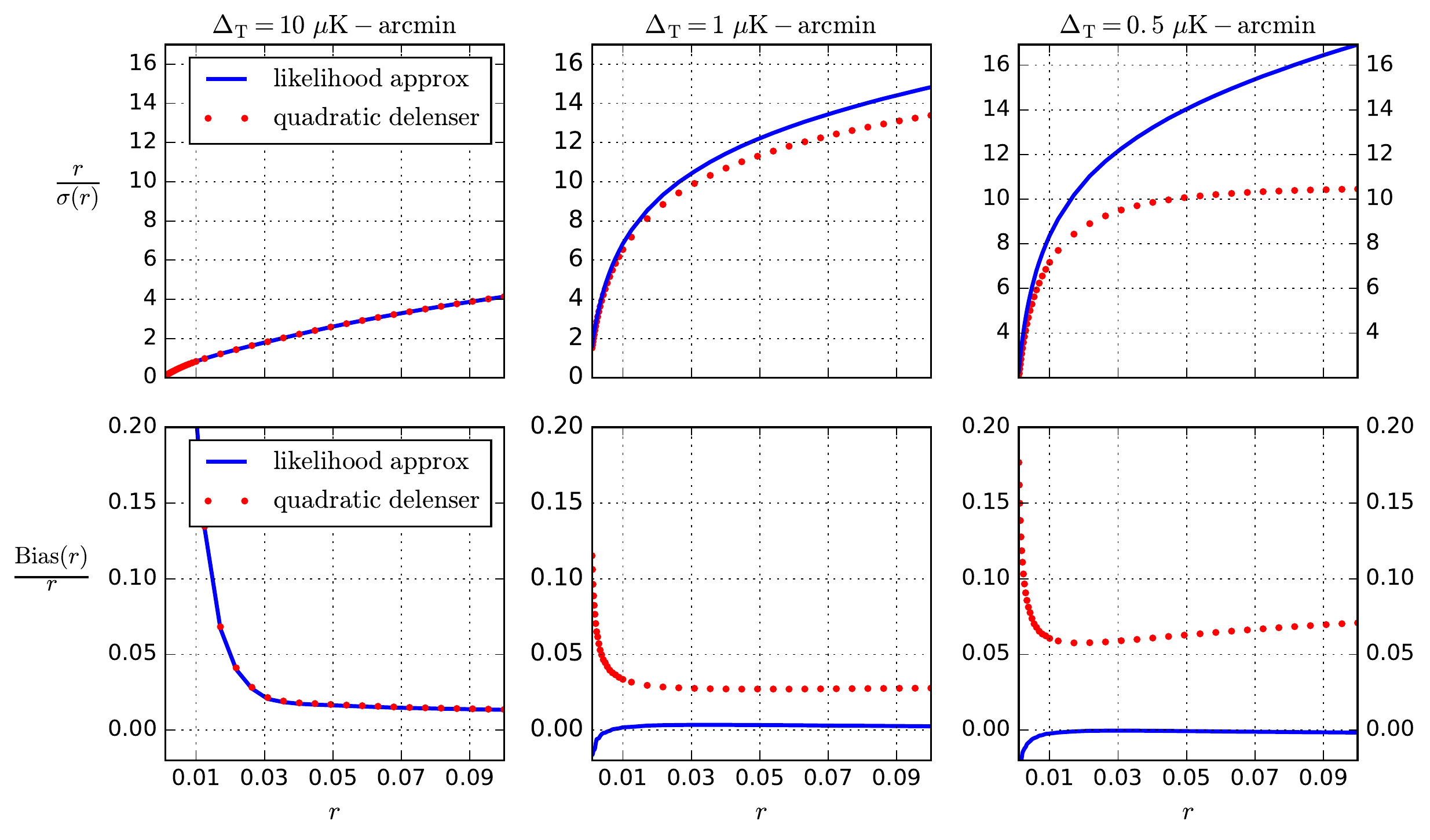}
\caption{\label{fig:f3}
Upper three panels show the detection levels $r/\sigma(r)$ expected from surveys of Scenario N, La, and Lb,
and lower three panels show the corresponding bias levels ${\rm Bias(r)}/r$.}
\end{figure*}

\subsection{$r$ Constraints}
For each simulated CMB realization, we first reconstruct the lensing potential field from
the $\phi$ survey (\emph{high} resolution survey)
using the EB quadratic estimator,
then use the reconstructed lensing field $\phi^{\rm est}(x)$ to delense the
\emph{low} resolution polarization maps using the quadratic delenser (Section \ref{sec:delen})
and the modified Gaussian likelihood method (Section \ref{sec:like}),
and finally compare their $r$ constraints from the two delensers.\footnote{For intrinsic estimators,
the reconstructed lensing potential field and its reconstruction noise are
correlated with the fields being delensed, and the correlation is expected to bias
the delensing. Fortunately, corresponding debias techniques have been extensively investigated
and used \citep[see e.g.][]{Teng2011,Namikawa2015,Sehgal2016,Carron2017}.
To avoid unnecessary complexity, we choose not to directly use the reconstructed field
$\phi^{\rm est}(x)$ for delensing,
instead use a simulated one $\phi^{\rm est}(x) = \phi(x) + n^\phi(x)$,
with $\phi(x)$ being the true lensing potential field and $n^\phi(x)$ being Gaussian noise
with power expected from the EB quadratic estimator (Figure \ref{fig:phi_noise}).}

In Figure \ref{fig:f3}, we show the detection level $r/\sigma(r)$
and bias level ${\rm Bias}(r)/r$ obtained from the quadratic delenser
and from the modified Gaussian likelihood method,
where $\sigma(r)$ and ${\rm Bias}(r)$ are the standard error and
the average bias of the $500$ best-fit $r$ values (from $500$ CMB realizations), respectively.
For Scenario N, both methods  obtain  similar $r$ detection levels,
while the modified Gaussian likelihood method shows its advantages in the
Scenario La and Lb.
We find that in the regime of low map noise  ($\lesssim 1$ $\mu$K-arcmin),
the bias of the modified Gaussian likelihood method is appreciably smaller than
that of the quadratic delenser (see next subsection for the detailed bias analysis
for the quadratic delenser).

For Scenario La with map noise $\Delta_{\rm T} = 1$ $\mu$K-arcmin and
sky coverage $f_{\rm sky} = 2.7\%$, we expect to detect the primordial B-mode signal at
$\sim 1\ \sigma$ level for $r = 0.001$ and at $\sim 15 \ \sigma$ level for $r = 0.1$.
The lower noise Scenario Lb with map noise $\Delta_{\rm T} = 0.5$ $\mu$K-arcmin
and the same sky coverage, only marginally increases the detection level,
due to the saturation of cosmic variance.

\subsection{Bias Analysis for the Quadratic Delenser}
\label{sec:app4}
In this subsection, we aim to quantify the bias of the quadratic delenser
introduced by ignoring the lensing in E modes and higher order lensing in
B modes.\footnote{ In principle, ignoring the non-stationarity of the delensed B modes also induces
some bias to the $r$ constraint. But we will see this bias is negligible. }
For clarity, we use the following notation to denote
the connection between lensed and primordial variables
\be
\begin{aligned}
\widetilde E = E + \delta E_{\text{\rm from E}} + \delta E_{\text{\rm from B}} , \\
\widetilde B = B + \delta B_{\text{\rm from E}} + \delta B_{\text{\rm from B}} ,
\end{aligned}
\ee
where $\delta X_{\text{\rm from Y}}$ is the lensing in (lensed) $X$ from  (primordial) $Y$.
In addition, $\delta E_{\text{\rm from B}}$ and $\delta B_{\text{\rm from B}}$ are much smaller
than their counterparts $\delta E_{\text{\rm from E}} $ and $\delta B_{\text{\rm from E}}$, so we simply ignore them
in this subsection.

\begin{figure*}
\centering
\includegraphics[height=2.8in]{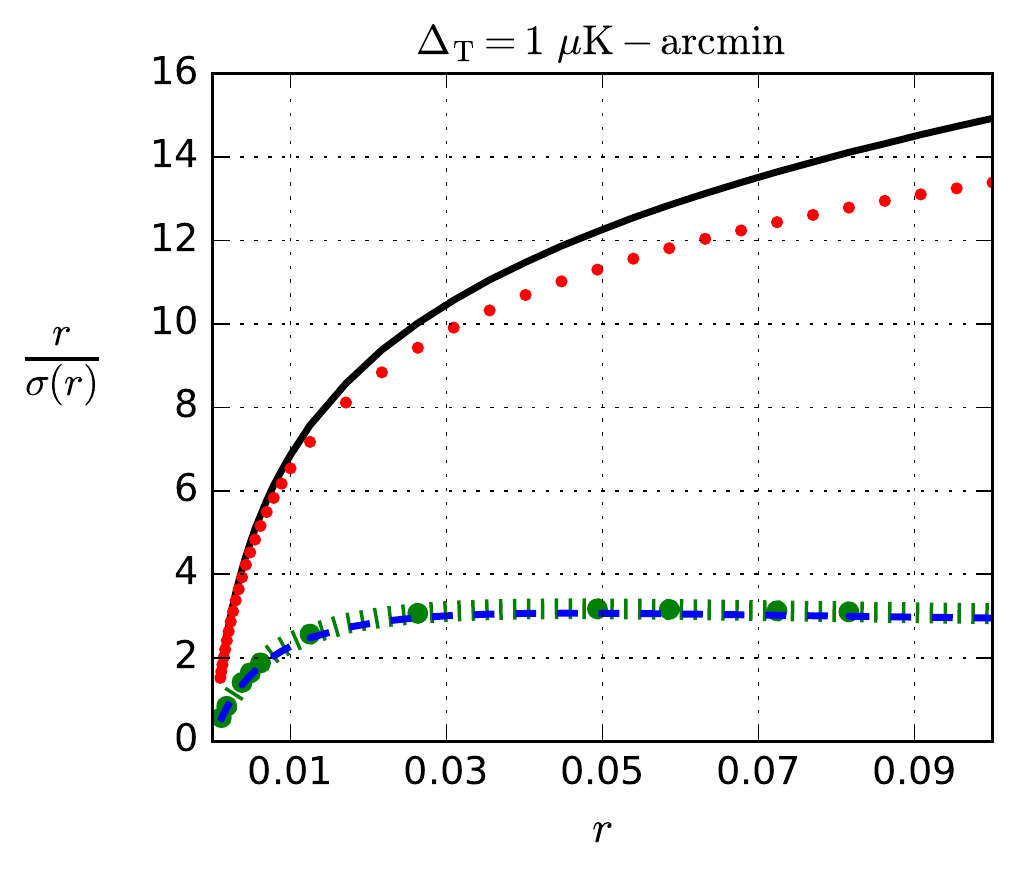}
\includegraphics[height=2.8in]{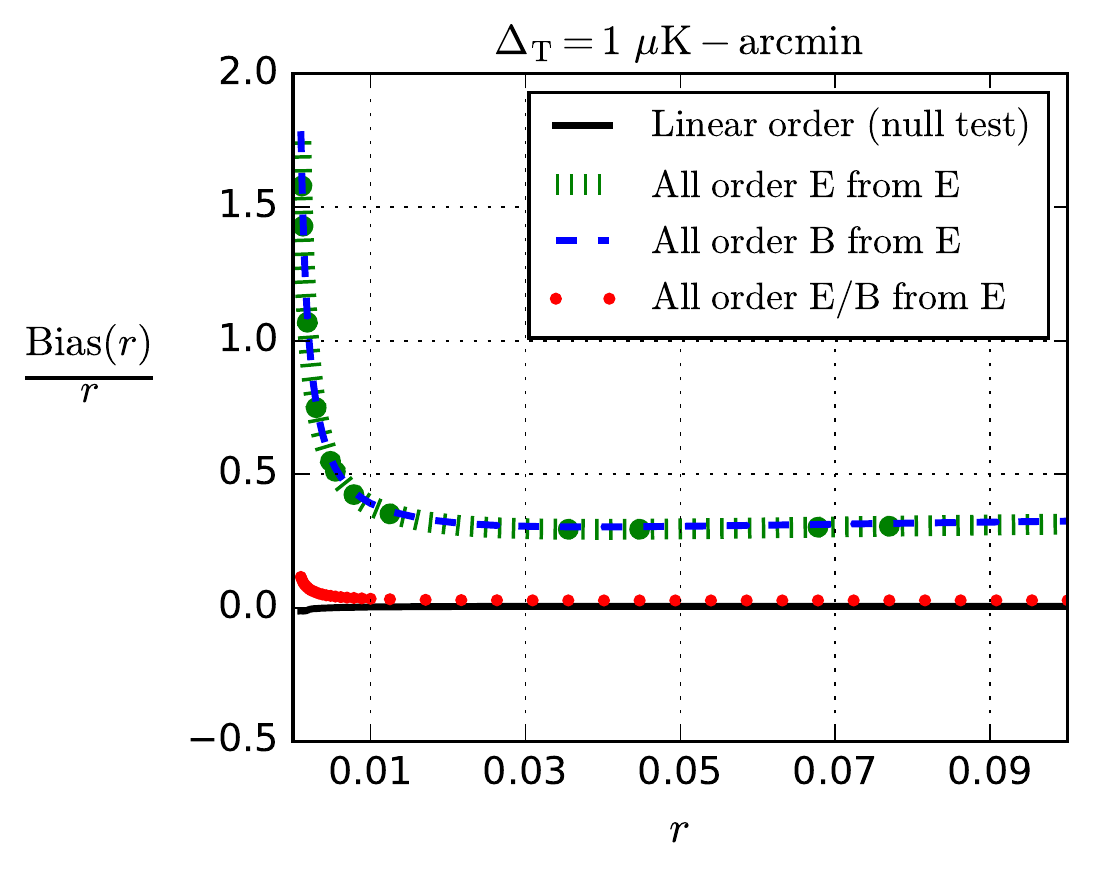}
\caption{\label{fig:f4}
Bias analysis of the quadratic delenser via simulations under different assumptions:
(black/solid lines) null test assuming $E^{\rm obs} = E + N^E$ and
$B^{\rm obs} = B + \delta^1B_{\text{\rm from E}} + N^B$;
(green/bar lines) all order E from E test assuming
$E^{\rm obs} = E + \delta E_{\text{\rm from E}} + N^E$,
$B^{\rm obs} = B + \delta^1B_{\text{\rm from E}} + N^B$;
(blue/dashed lines) all order B from E test assuming
$E^{\rm obs} = E + N^E$ and
$B^{\rm obs} = B +  \delta B_{\text{\rm from E}} + N^B$;
(red/dots) all order E/B from E test assuming
$E^{\rm obs} = E + \delta E_{\text{\rm from E}} + N^E$ and
$B^{\rm obs} = B +  \delta B_{\text{\rm from E}} + N^B$.
}
\end{figure*}

\begin{enumerate}[label=(\roman*)]
    \item First we do a null test.
In accordance with the two approximations made in the quadratic delenser (Section \ref{sec:delen}),
we completely drop lensing in E modes and only keep linear order lensing in B modes, i.e.,
we simulate polarization maps assuming $E^{\rm obs} = E + N^E$
and $B^{\rm obs} = B + \delta^1B_{\text{\rm from E}} + N^B$, where
$N^{E/B}$ is the E/B map noise, and  $\delta^1B_{\text{\rm from E}}$ is the linear order lensing in B from E.
As expected, we find the quadratic delenser is not biased in this context
(Figure \ref{fig:f4}, black/solid lines).\footnote{From the null test,
where we ignore the non-stationarity of the delensed B modes,
we conclude that the bias induced by ignoring  the non-stationarity
is negligible.}\footnote{Comparing the detection level of the null test (solid line in the left panel of Figure \ref{fig:f4}),
and the detection level of the modified Gaussian likelihood (solid line in the second panel of Figure \ref{fig:f3}),
we find the two matches exactly. Therefore, we confirm the validity of
the two major approximations used in the modified Gaussian likelihood:
only keeping a few projected data vectors,
and igoring the cross relation between different projected data vectors.}

    \item To scrutinize the bias introduced by ignoring lensing in E modes,
we keep all order lensing in E modes and linear order lensing in B modes,
i.e., we simulate polarization maps assuming
$E^{\rm obs} = E +  \delta E_{\text{\rm from E}} + N^E$
and $B^{\rm obs} = B^r + \delta^1B_{\text{\rm from E}} + N^B$.
In this context, the quadratic delenser is highly biased (Figure \ref{fig:f4}, green/bar lines).

    \item In the same way, to test the bias introduced by ignoring high order lensing terms in B modes,
we ignore lensing in E modes and keep all order lensing in B modes, i.e.,
we do simulations assuming $E^{\rm obs} = E  + N^E$
and $B^{\rm obs} = B + \delta B_{\text{\rm from E}} + N^B$.
In this context, we also find the quadratic delenser is highly biased.
More interestingly, we find that the bias level almost exactly matches that of ignoring lensing in E modes
(Figure \ref{fig:f4}, blue/dashed lines).

    \item The final step is to check the interaction between the two bias terms from (iii) and (iv).
For this purpose, we keep all order lensing in E modes and all order lensing in B modes,
i.e., we simulate polarization maps assuming $E^{\rm obs} = E + \delta E_{\text{\rm from E}} + N^E$,
and $B^{\rm obs} = B +  \delta B_{\text{\rm from E}} + N^B$.
We find that the two bias contributions cancel to a high precision
and therefore the net bias is strongly suppressed (Figure \ref{fig:f4}, red/dots).

To make sense of the bias cancellation, we do a simple magnitude analysis.
In the quadratic delenser, we delense the B modes via a quadratic template subtraction
$B^{\rm res} = B_{\text{\rm from E}}- E^{\rm obs}*\phi^{\rm est}$,
and assume a residual power spectrum
$C^{BB,{\rm res}}_\ell = \langle |\delta^1 B_{\text{\rm from E}}- E*\phi^{\rm est}|_\ell^2 \rangle$,
where $*$ denotes the convolution defined in Equation (\ref{eq:convol},\ref{eq:wconvol}),
in the $\delta^1 B_{\text{\rm from E}}$ term we ignore
the second (and higher) order lensing in B modes $\delta^2 B_{\text{\rm from E}}$,
and in the $E*\phi$ term we ignore the difference of $E$ and $\widetilde E$.
Therefore the template subtraction used has an error
$\delta^2 B_{\text{\rm from E}}- \delta^1 E_{\text{\rm from E}} * \phi$,
where both error terms are of the same order $O(E\phi^2)$ considering that
\be
\begin{aligned}
&(\delta^2 B_{\text{\rm from E}})_{\Bell} \\
=& -\frac{1}{2}\int \frac{d^2 {\Bell_1}d^2 {\Bell_2}}{(2\pi)^2}
[({\Bell_1}+ {\Bell})\cdot({\Bell_1}+ {\Bell_2})]  \\
& [({\Bell_1}+ {\Bell})\cdot {\Bell_2}] E_{{\Bell_1}+ {\Bell}} \sin(2\varphi_{{\Bell_1}+ {\Bell}, {\Bell}})
\phi^*_{{\Bell_1}+ {\Bell_2}} \phi_{\Bell_2},
\end{aligned}
\ee
and
\be
\begin{aligned}
&(\delta^1 E_{\text{\rm from E}})_{\Bell} \\
= & \int \frac{d^2 {\Bell'}}{2\pi} {\Bell'} \cdot ({\Bell'}+ \Bell) E_{{\Bell'}+\Bell}
\cos(2\varphi_{{\Bell'}+\Bell,\Bell}) \phi^*_{\Bell'}.
\end{aligned}
\ee

\end{enumerate}

To summarize, in the quadratic delenser,
\be
B^{\rm del}_{\Bell} = B^r_{\Bell} + B^{\rm res}_{\Bell} + N^B_{\Bell},
\ee
we have ignored lensing in E modes and high order lensing in B modes when estimating
the residual power spectrum $\langle |B^{\rm res}_{\Bell}|^2\rangle$.
We find that each of the two approximation introduces a strong bias in the $r$ estimate,
while the two bias contributions cancel to a high precision,
and the validity of the quadratic delenser sensitively depends on the cancellation.

According to the above analysis, the bias in the residual power estimate
in principle is independent of primordial B-mode signal $B^r$,
therefore we naively expect a $r$-independent bias ${\rm Bias}(r)$
and therefore a bias level ${\rm Bias}(r)/r$ decaying with growing $r$,
which is indeed the behavior we observe for
$r\lesssim 0.01$ (Figure \ref{fig:f3} and \ref{fig:f4}).
But the bias level does not dies down for even greater $r$,
since the $r$ constraints become more sensitive to higher frequency
regime where the bias is stronger. Here we give an informal analysis
of the  bias level behavior.
From a single delensed B modes $B^{\rm del}_{\Bell}$,
we can estimate the primordial B-mode power spectrum
with root variance $\Delta C_{\Bell} = C^{BB,r}_{\Bell} + C^{BB,{\rm res}}_{\Bell} + N^{BB}_{\Bell}$,
and  consequently estimate $r$ with mean value
\be
r^{\rm est}_{\Bell} = \frac{|B^{\rm del}_{\Bell}|^2 - C^{BB,{\rm res}}_{\Bell}- N^{BB}_{\Bell}}{C^{BB,r=1}_{\Bell}},
\ee
and
with root variance $\sigma_{\Bell}(r) = \Delta C_{\Bell}/C^{BB,r=1}_{\Bell}$.
These estimators from different modes can be added with inverse-variance weighting
$r^{\rm est} = \sum_{\Bell} W_{\Bell} r^{\rm est}_{\Bell}$,
where
\be
W_{\Bell}(r) = \frac{\frac{1}{\sigma^2_{\Bell}(r)}}{\sum_{\Bell} \frac{1}{\sigma^2_{\Bell}(r)}}.
\ee
It is straightforward to understand that  $W_{\Bell}(r)$ increases with $r$ for large $|{\Bell}|$
where $C^{BB,{\rm res}}_{\Bell}+ N^{BB}_{\Bell}$ dominates $\Delta C_{\Bell}$,
and decreases with $r$ for small $|{\Bell}|$ where $C^{BB,r}_{\Bell} $ dominates $\Delta C_{\Bell}$.
In addition, we know that the quadratic delenser is a biased estimator, i.e.,
\be
\langle |B^{\rm del}_{\Bell}|^2 \rangle = C^{BB,{\rm res}}_{\Bell} + N^{BB}_{\Bell} + C^{BB,r}_{\Bell} + C^{BB,{\rm bias}}_{\Bell},
\ee
and
\be
\langle r^{\rm est}_{\Bell} \rangle = \frac{C^{BB,r}_{\Bell} + C^{BB,{\rm bias}}_{\Bell} }{C^{BB,r=1}_{\Bell}}
= r + r^{\rm bias}_{\Bell},
\ee
where $r^{\rm bias}_{\Bell}$ increases with $|{\Bell}|$.
Therefore, we have $\langle r^{\rm est}\rangle
= \sum_{\Bell} W_{\Bell} \langle r^{\rm est}_{\Bell}\rangle
= r   + \sum_{\Bell} W_{\Bell}(r) r^{\rm bias}_{\Bell}
= r   + {\rm Bias}(r)$,
with ${\rm Bias}(r)$ increasing with $r$.
It also explains the increasing bias level ${\rm Bias}(r)$ with
decreasing map noise $N^{BB}_{\Bell}$ (see Figure \ref{fig:f3}).
Note that we do not expect the quadratic delenser
to exactly match the inverse-variance weighted estimator described above, but the latter
should a good proxy for interpreting the bias behavior.

\subsection{Non-stationary Noise }
\label{sec:nonsta}

\begin{figure*}
\centering
\includegraphics[height=2in]{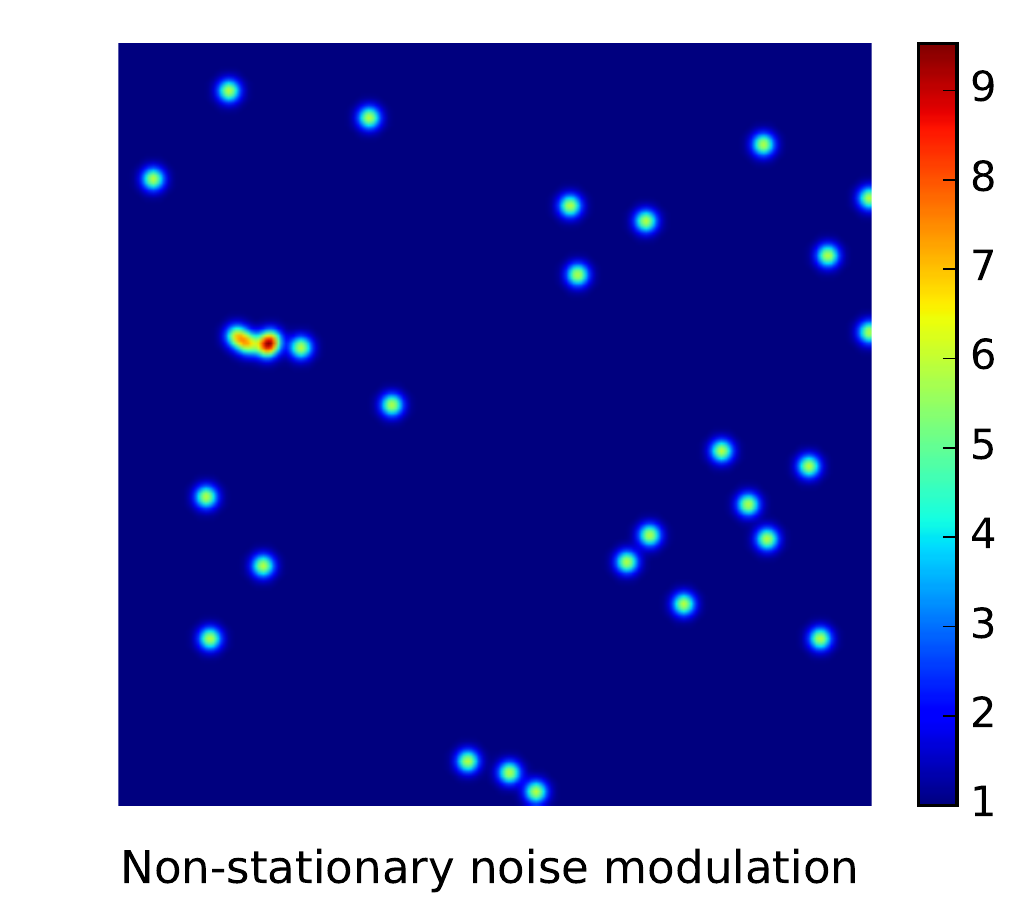}%
\includegraphics[height=2in]{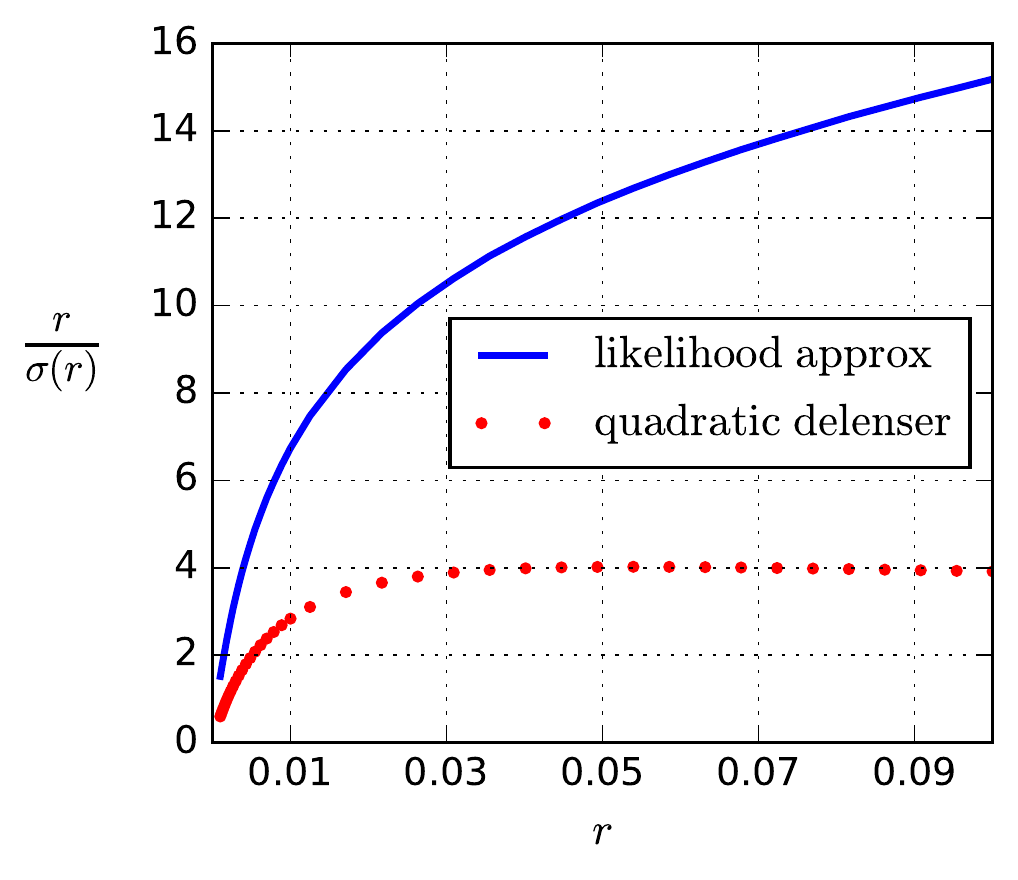}%
\includegraphics[height=2in]{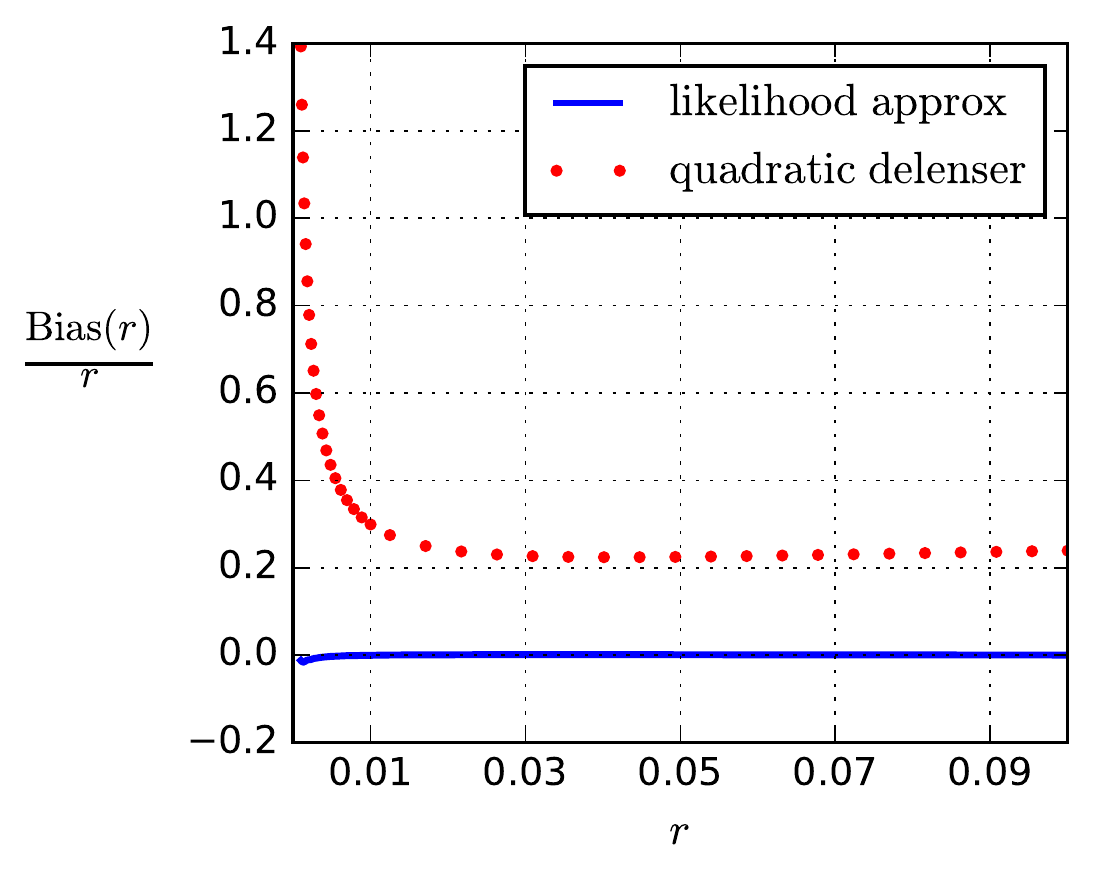}
\caption{\label{fig:nonsta}
The impact of non-stationary map noise on the $r$ constraints for the Scenario La experiments
($\Delta_{\rm T} = 1\ \mu$K-arcmin).
Left panel: the non-stationary noise modulation $\sigma(\mathbf x)$.
Middle/Right panel: the detection/bias level of $r$ constraints inferred from the two delensers.}
\end{figure*}

The modified Gaussian likelihood  works not only as a correction to
the quadratic template subtraction estimator,
but also shows its advantage in dealing with realistic experiment complexities,
e.g., non-stationary noise and sky cuts.
Here we explore an example of non-stationary noise with pixel
dependent noise, i.e., $\left< n(\mathbf x) n(\mathbf y)\right>
= \sigma^2(\mathbf x)\Delta_{\rm P}^2 \delta_D(\mathbf x-\mathbf y)$,
with $\Delta_{\rm P}=\sqrt{2} \ \Delta_{\rm T} = \sqrt{2}\ \mu$K-arcmin, and
$\sigma(\mathbf x)$  a pixel-dependent modulation (Figure \ref{fig:nonsta}).
We expect the likelihood based estimator to work robustly in the presence of non-stationary noise,
as long as we take the pixel dependent noise into account
when calculating the covariance matrix of noise (see Appendix \ref{sec:app2}).
But the non-stationary noise becomes troublesome for
the quadratic delenser in  Fourier space.
\footnote{In the case of non-stationary noise,
the noise power spectrum loses the protection of symmetry,
i.e, $\langle n_{\Bell}n_{\Bell'}\rangle = N_{\Bell,\Bell'}$ now depends on
both multipoles instead of their linear combination $\Bell,\Bell'$.
If we were to correctly use the quadratic delenser,
then the residual power evaluation in Equation (\ref{eq:respower})
becomes difficult, and is out of the scope of this paper.
Here we simply (but incorrectly) assume the stationary noise power spectrum in Equation (\ref{eq:respower}),
and test how the non-stationary noise biases the $r$ constraint from the quadratic delenser.}

Applying the two estimators on simulations with non-stationary noise,
we find that the modified Gaussian likelihood method works as well as in the case of stationary noise,
while the quadratic delenser is significantly biased (Figure \ref{fig:nonsta}).

\section{Summary and Conclusions}
\label{sec:summary}

Delensing is a crucial part for future CMB experiments aiming to detect a primordial B-mode signal.
Up to linear order, one can effectively delense observed B-modes by utilizing
a quadratic combination of observed E-modes and an estimate of the lensing potential.
This is the underlying idea of the quadratic delenser.
However, in the regime of small map noise,
the lensing in E modes, and higher order lensing in B modes
ignored by the quadratic delenser, significantly bias the $r$ constraint.
We investigated the bias induced by each of the two approximations via simulations,
finding that each of two approximations induce a large bias, while the two bias terms
partly cancel and therefore the net bias is moderately suppressed.
The validity of the quadratic delenser sensitively depends on the cancellation.

Alternatively,  a full-scale likelihood analysis of the tensor-to-scalar ratio $r$ can,
in principle, optimally account for all the $r$ information in the CMB observations and
remedy possible bias problems.
Unfortunately, a full likelihood analysis requires computation resources beyond
what is available in the near future.
In this paper, we presented a modified Gaussian likelihood method.
This method consists of two parts, covariance decomposition and data compression.
In the first part, we decomposed the covariance matrix
in the form of $\Sigma_r = \Sigma^{\rm en} + r \Sigma^{\rm b}$,
which allows us to compute the covariance matrix $\Sigma_r$, as a function of $r$, at the computational cost
of a single covariance matrix evaluation.
In the second part, we compressed the data size by keeping only $s\sim 500$
high signal-to-noise modes, say the large-scale quadratic delensed B modes.
We obtained these B modes from polarization data $d$ via a projection matrix $v$,
$(^0B^{\rm del}_{\Bell})_s = (v^\intercal)_{s\times 2p} d_{2p}$,
and applied the likelihood analysis on the projected data vector.
This method can be naturally extended to incorporate higher frequency modes.

Finally, we applied the quadratic delenser and
the modified Gaussian likelihood method on simulated CMB observations mimicking experiments of
Scenario N, La, and Lb, and compare the resulting $r$ constraints.
We found that, the two methods have similar performance in constraining $r$
for Scenario N, while the quadratic delenser does not perform
as well for the lower-noise Scenario La and Lb due to a strong $r$ constraint bias
in the regime of low map noise. For Scenario La, we expected to detect the
primordial B-mode signal at $\sim 1 \sigma$ level for $r=0.001$, and at $\sim 15 \sigma$ level for $r=0.1$,
from the modified Gaussian likelihood method.
For Scenario Lb with even lower map noise and the same sky coverage,
the detection level only marginally increases due to the saturation of cosmic variance.
Therefore it would be valuable to optimize the survey configurations (${\Delta_{\rm T}, f_{\rm sky}}$)
for the coming CMB experiments given a fixed amount of survey time \citep{s42016}.

We also explored the impact of realistic experiment complexities:
in the presence of non-stationary noise, the modified Gaussian likelihood method also works robustly
as long as we slightly modify the noise covariance matrix to take into account the pixel dependent noise.

\begin{acknowledgements}
ZP is supported by UC Davis Dissertation Year Fellowship.
EA acknowledges support from NSF CAREER grant DMS-1252795.
This work made extensive use of the NASA Astrophysics Data System and
of the {\tt astro-ph} preprint archive at {\tt arXiv.org}.
\end{acknowledgements}

\bibliography{ms}

\begin{thebibliography}{71}%
\makeatletter
\providecommand \@ifxundefined [1]{%
 \@ifx{#1\undefined}
}%
\providecommand \@ifnum [1]{%
 \ifnum #1\expandafter \@firstoftwo
 \else \expandafter \@secondoftwo
 \fi
}%
\providecommand \@ifx [1]{%
 \ifx #1\expandafter \@firstoftwo
 \else \expandafter \@secondoftwo
 \fi
}%
\providecommand \natexlab [1]{#1}%
\providecommand \enquote  [1]{``#1''}%
\providecommand \bibnamefont  [1]{#1}%
\providecommand \bibfnamefont [1]{#1}%
\providecommand \citenamefont [1]{#1}%
\providecommand \href@noop [0]{\@secondoftwo}%
\providecommand \href [0]{\begingroup \@sanitize@url \@href}%
\providecommand \@href[1]{\@@startlink{#1}\@@href}%
\providecommand \@@href[1]{\endgroup#1\@@endlink}%
\providecommand \@sanitize@url [0]{\catcode `\\12\catcode `\$12\catcode
  `\&12\catcode `\#12\catcode `\^12\catcode `\_12\catcode `\%12\relax}%
\providecommand \@@startlink[1]{}%
\providecommand \@@endlink[0]{}%
\providecommand \url  [0]{\begingroup\@sanitize@url \@url }%
\providecommand \@url [1]{\endgroup\@href {#1}{\urlprefix }}%
\providecommand \urlprefix  [0]{URL }%
\providecommand \Eprint [0]{\href }%
\providecommand \doibase [0]{http://dx.doi.org/}%
\providecommand \selectlanguage [0]{\@gobble}%
\providecommand \bibinfo  [0]{\@secondoftwo}%
\providecommand \bibfield  [0]{\@secondoftwo}%
\providecommand \translation [1]{[#1]}%
\providecommand \BibitemOpen [0]{}%
\providecommand \bibitemStop [0]{}%
\providecommand \bibitemNoStop [0]{.\EOS\space}%
\providecommand \EOS [0]{\spacefactor3000\relax}%
\providecommand \BibitemShut  [1]{\csname bibitem#1\endcsname}%
\let\auto@bib@innerbib\@empty
\bibitem [{\citenamefont {Mukhanov}\ and\ \citenamefont
  {Chibisov}(1981)}]{Mukhanov1981}%
  \BibitemOpen
  \bibfield  {author} {\bibinfo {author} {\bibfnamefont {V.}~\bibnamefont
  {Mukhanov}}\ and\ \bibinfo {author} {\bibfnamefont {G.}~\bibnamefont
  {Chibisov}},\ }\href {http://www.jetpletters.ac.ru/ps/1510/article_23079.pdf}
  {\enquote {\bibinfo {title} {{Quantum fluctuations and a nonsingular
  universe}},}\ } (\bibinfo {year} {1981})\BibitemShut {NoStop}%
\bibitem [{\citenamefont {Guth}(1981)}]{Guth1981}%
  \BibitemOpen
  \bibfield  {author} {\bibinfo {author} {\bibfnamefont {A.~H.}\ \bibnamefont
  {Guth}},\ }\href {\doibase 10.1103/PhysRevD.23.347} {\bibfield  {journal}
  {\bibinfo  {journal} {Phys. Rev. D}\ }\textbf {\bibinfo {volume} {23}},\
  \bibinfo {pages} {347} (\bibinfo {year} {1981})}\BibitemShut {NoStop}%
\bibitem [{\citenamefont {Linde}(1982)}]{Linde1982}%
  \BibitemOpen
  \bibfield  {author} {\bibinfo {author} {\bibfnamefont {A.~D.}\ \bibnamefont
  {Linde}},\ }\href {\doibase 10.1016/0370-2693(82)91219-9} {\bibfield
  {journal} {\bibinfo  {journal} {Phys. Lett. B}\ }\textbf {\bibinfo {volume}
  {108}},\ \bibinfo {pages} {389} (\bibinfo {year} {1982})},\ \Eprint
  {http://arxiv.org/abs/arXiv:1011.1669v3} {arXiv:arXiv:1011.1669v3}
  \BibitemShut {NoStop}%
\bibitem [{\citenamefont {Albrecht}\ and\ \citenamefont
  {Steinhardt}(1982)}]{Albrecht1982}%
  \BibitemOpen
  \bibfield  {author} {\bibinfo {author} {\bibfnamefont {A.}~\bibnamefont
  {Albrecht}}\ and\ \bibinfo {author} {\bibfnamefont {P.~J.}\ \bibnamefont
  {Steinhardt}},\ }\href {\doibase 10.1103/PhysRevLett.48.1220} {\bibfield
  {journal} {\bibinfo  {journal} {Phys. Rev. Lett.}\ }\textbf {\bibinfo
  {volume} {48}},\ \bibinfo {pages} {1220} (\bibinfo {year}
  {1982})}\BibitemShut {NoStop}%
\bibitem [{\citenamefont {Lidsey}\ \emph {et~al.}(1997)\citenamefont {Lidsey},
  \citenamefont {Liddle}, \citenamefont {Kolb}, \citenamefont {Copeland},
  \citenamefont {Barreiro},\ and\ \citenamefont {Abney}}]{Lidsey1997}%
  \BibitemOpen
  \bibfield  {author} {\bibinfo {author} {\bibfnamefont {J.~E.}\ \bibnamefont
  {Lidsey}}, \bibinfo {author} {\bibfnamefont {A.~R.}\ \bibnamefont {Liddle}},
  \bibinfo {author} {\bibfnamefont {E.~W.}\ \bibnamefont {Kolb}}, \bibinfo
  {author} {\bibfnamefont {E.~J.}\ \bibnamefont {Copeland}}, \bibinfo {author}
  {\bibfnamefont {T.}~\bibnamefont {Barreiro}}, \ and\ \bibinfo {author}
  {\bibfnamefont {M.}~\bibnamefont {Abney}},\ }\href {\doibase
  10.1103/RevModPhys.69.373} {\bibfield  {journal} {\bibinfo  {journal} {Rev.
  Mod. Phys.}\ }\textbf {\bibinfo {volume} {69}},\ \bibinfo {pages} {373}
  (\bibinfo {year} {1997})}\BibitemShut {NoStop}%
\bibitem [{\citenamefont {Lyth}\ and\ \citenamefont {Riotto}(1999)}]{Lyth1999}%
  \BibitemOpen
  \bibfield  {author} {\bibinfo {author} {\bibfnamefont {D.~H.}\ \bibnamefont
  {Lyth}}\ and\ \bibinfo {author} {\bibfnamefont {A.}~\bibnamefont {Riotto}},\
  }\href {\doibase 10.1016/S0370-1573(98)00128-8} {\bibfield  {journal}
  {\bibinfo  {journal} {Phys. Rep.}\ }\textbf {\bibinfo {volume} {314}},\
  \bibinfo {pages} {1} (\bibinfo {year} {1999})}\BibitemShut {NoStop}%
\bibitem [{\citenamefont {Starobinskii}(1979)}]{Starobinskii1979}%
  \BibitemOpen
  \bibfield  {author} {\bibinfo {author} {\bibfnamefont {A.~A.}\ \bibnamefont
  {Starobinskii}},\ }\href {http://adsabs.harvard.edu/abs/1979ZhPmR..30..719S}
  {\bibfield  {journal} {\bibinfo  {journal} {J. Exp. Theor. Phys. Lett.}\
  }\textbf {\bibinfo {volume} {30}},\ \bibinfo {pages} {719} (\bibinfo {year}
  {1979})}\BibitemShut {NoStop}%
\bibitem [{\citenamefont {Rubakov}\ \emph {et~al.}(1982)\citenamefont
  {Rubakov}, \citenamefont {Sazhin},\ and\ \citenamefont
  {Veryaskin}}]{Rubakov1982}%
  \BibitemOpen
  \bibfield  {author} {\bibinfo {author} {\bibfnamefont {V.~A.}\ \bibnamefont
  {Rubakov}}, \bibinfo {author} {\bibfnamefont {M.~V.}\ \bibnamefont {Sazhin}},
  \ and\ \bibinfo {author} {\bibfnamefont {A.~V.}\ \bibnamefont {Veryaskin}},\
  }\href {\doibase 10.1016/0370-2693(82)90641-4} {\bibfield  {journal}
  {\bibinfo  {journal} {Phys. Lett. B}\ }\textbf {\bibinfo {volume} {115}},\
  \bibinfo {pages} {189} (\bibinfo {year} {1982})}\BibitemShut {NoStop}%
\bibitem [{\citenamefont {Fabbri}\ and\ \citenamefont
  {Pollock}(1983)}]{Fabbri1983}%
  \BibitemOpen
  \bibfield  {author} {\bibinfo {author} {\bibfnamefont {R.}~\bibnamefont
  {Fabbri}}\ and\ \bibinfo {author} {\bibfnamefont {M.~D.}\ \bibnamefont
  {Pollock}},\ }\href {\doibase 10.1016/0370-2693(83)91322-9} {\bibfield
  {journal} {\bibinfo  {journal} {Phys. Lett. B}\ }\textbf {\bibinfo {volume}
  {125}},\ \bibinfo {pages} {445} (\bibinfo {year} {1983})}\BibitemShut
  {NoStop}%
\bibitem [{\citenamefont {Abbott}\ and\ \citenamefont
  {Wise}(1984)}]{Abbott1984}%
  \BibitemOpen
  \bibfield  {author} {\bibinfo {author} {\bibfnamefont {L.~F.}\ \bibnamefont
  {Abbott}}\ and\ \bibinfo {author} {\bibfnamefont {M.~B.}\ \bibnamefont
  {Wise}},\ }\href {\doibase 10.1016/0550-3213(84)90329-8} {\bibfield
  {journal} {\bibinfo  {journal} {Nucl. Physics, Sect. B}\ }\textbf {\bibinfo
  {volume} {244}},\ \bibinfo {pages} {541} (\bibinfo {year}
  {1984})}\BibitemShut {NoStop}%
\bibitem [{\citenamefont {Starobinskii}(1985)}]{Starobinskii1985}%
  \BibitemOpen
  \bibfield  {author} {\bibinfo {author} {\bibfnamefont {A.}~\bibnamefont
  {Starobinskii}},\ }\href@noop {} {\bibfield  {journal} {\bibinfo  {journal}
  {Sov. Astron. Lett.}\ }\textbf {\bibinfo {volume} {11}},\ \bibinfo {pages}
  {133} (\bibinfo {year} {1985})}\BibitemShut {NoStop}%
\bibitem [{\citenamefont {Stebbins}(1996)}]{Stebbins1996}%
  \BibitemOpen
  \bibfield  {author} {\bibinfo {author} {\bibfnamefont {A.}~\bibnamefont
  {Stebbins}},\ }\href {http://arxiv.org/abs/astro-ph/9609149} {\bibfield
  {journal} {\bibinfo  {journal} {eprint}\ ,\ \bibinfo {pages} {27}} (\bibinfo
  {year} {1996})},\ \Eprint {http://arxiv.org/abs/9609149v1} {arXiv:9609149v1
  [arXiv:astro-ph]} \BibitemShut {NoStop}%
\bibitem [{\citenamefont {Kamionkowski}\ \emph
  {et~al.}(1997{\natexlab{a}})\citenamefont {Kamionkowski}, \citenamefont
  {Kosowsky},\ and\ \citenamefont {Stebbins}}]{Kamionkowski1997a}%
  \BibitemOpen
  \bibfield  {author} {\bibinfo {author} {\bibfnamefont {M.}~\bibnamefont
  {Kamionkowski}}, \bibinfo {author} {\bibfnamefont {A.}~\bibnamefont
  {Kosowsky}}, \ and\ \bibinfo {author} {\bibfnamefont {A.}~\bibnamefont
  {Stebbins}},\ }\href {\doibase 10.1103/PhysRevLett.78.2058} {\bibfield
  {journal} {\bibinfo  {journal} {Phys. Rev. Lett.}\ }\textbf {\bibinfo
  {volume} {78}},\ \bibinfo {pages} {2058} (\bibinfo {year}
  {1997}{\natexlab{a}})},\ \Eprint {http://arxiv.org/abs/9609132}
  {arXiv:9609132 [astro-ph]} \BibitemShut {NoStop}%
\bibitem [{\citenamefont {Kamionkowski}\ \emph
  {et~al.}(1997{\natexlab{b}})\citenamefont {Kamionkowski}, \citenamefont
  {Kosowsky},\ and\ \citenamefont {Stebbins}}]{Kamionkowski1997}%
  \BibitemOpen
  \bibfield  {author} {\bibinfo {author} {\bibfnamefont {M.}~\bibnamefont
  {Kamionkowski}}, \bibinfo {author} {\bibfnamefont {A.}~\bibnamefont
  {Kosowsky}}, \ and\ \bibinfo {author} {\bibfnamefont {A.}~\bibnamefont
  {Stebbins}},\ }\href {\doibase 10.1103/PhysRevD.55.7368} {\bibfield
  {journal} {\bibinfo  {journal} {Phys. Rev. D}\ }\textbf {\bibinfo {volume}
  {55}},\ \bibinfo {pages} {7368} (\bibinfo {year} {1997}{\natexlab{b}})},\
  \Eprint {http://arxiv.org/abs/9611125} {arXiv:9611125 [astro-ph]}
  \BibitemShut {NoStop}%
\bibitem [{\citenamefont {Seljak}\ and\ \citenamefont
  {Zaldarriaga}(1997)}]{Seljak1997}%
  \BibitemOpen
  \bibfield  {author} {\bibinfo {author} {\bibfnamefont {U.}~\bibnamefont
  {Seljak}}\ and\ \bibinfo {author} {\bibfnamefont {M.}~\bibnamefont
  {Zaldarriaga}},\ }\href {\doibase 10.1103/PhysRevLett.78.2054} {\bibfield
  {journal} {\bibinfo  {journal} {Phys. Rev. Lett.}\ }\textbf {\bibinfo
  {volume} {78}},\ \bibinfo {pages} {2054} (\bibinfo {year} {1997})},\ \Eprint
  {http://arxiv.org/abs/9609169} {arXiv:9609169 [astro-ph]} \BibitemShut
  {NoStop}%
\bibitem [{\citenamefont {Seljak}(1997)}]{Seljak1997a}%
  \BibitemOpen
  \bibfield  {author} {\bibinfo {author} {\bibfnamefont {U.}~\bibnamefont
  {Seljak}},\ }\href {\doibase 10.1086/304123} {\bibfield  {journal} {\bibinfo
  {journal} {Astrophys. J.}\ }\textbf {\bibinfo {volume} {482}},\ \bibinfo
  {pages} {6} (\bibinfo {year} {1997})},\ \Eprint
  {http://arxiv.org/abs/9608131} {arXiv:9608131 [astro-ph]} \BibitemShut
  {NoStop}%
\bibitem [{\citenamefont {Zaldarriaga}\ and\ \citenamefont
  {Seljak}(1997)}]{Zaldarriaga1997a}%
  \BibitemOpen
  \bibfield  {author} {\bibinfo {author} {\bibfnamefont {M.}~\bibnamefont
  {Zaldarriaga}}\ and\ \bibinfo {author} {\bibfnamefont {U.}~\bibnamefont
  {Seljak}},\ }\href {\doibase 10.1103/PhysRevD.55.1830} {\bibfield  {journal}
  {\bibinfo  {journal} {Phys. Rev. D}\ }\textbf {\bibinfo {volume} {55}},\
  \bibinfo {pages} {1830} (\bibinfo {year} {1997})}\BibitemShut {NoStop}%
\bibitem [{\citenamefont {Khoury}\ \emph
  {et~al.}(2001{\natexlab{a}})\citenamefont {Khoury}, \citenamefont {Ovrut},
  \citenamefont {Steinhardt},\ and\ \citenamefont {Turok}}]{Khoury2001}%
  \BibitemOpen
  \bibfield  {author} {\bibinfo {author} {\bibfnamefont {J.}~\bibnamefont
  {Khoury}}, \bibinfo {author} {\bibfnamefont {B.~A.}\ \bibnamefont {Ovrut}},
  \bibinfo {author} {\bibfnamefont {P.~J.}\ \bibnamefont {Steinhardt}}, \ and\
  \bibinfo {author} {\bibfnamefont {N.}~\bibnamefont {Turok}},\ }\href
  {\doibase 10.1103/PhysRevD.66.046005} {\bibfield  {journal} {\bibinfo
  {journal} {Phys. Rev. D}\ }\textbf {\bibinfo {volume} {66}},\ \bibinfo
  {pages} {046005} (\bibinfo {year} {2001}{\natexlab{a}})},\ \Eprint
  {http://arxiv.org/abs/0109050} {arXiv:0109050 [hep-th]} \BibitemShut
  {NoStop}%
\bibitem [{\citenamefont {Khoury}\ \emph
  {et~al.}(2001{\natexlab{b}})\citenamefont {Khoury}, \citenamefont {Ovrut},
  \citenamefont {Steinhardt},\ and\ \citenamefont {Turok}}]{Khoury2001a}%
  \BibitemOpen
  \bibfield  {author} {\bibinfo {author} {\bibfnamefont {J.}~\bibnamefont
  {Khoury}}, \bibinfo {author} {\bibfnamefont {B.~A.}\ \bibnamefont {Ovrut}},
  \bibinfo {author} {\bibfnamefont {P.~J.}\ \bibnamefont {Steinhardt}}, \ and\
  \bibinfo {author} {\bibfnamefont {N.}~\bibnamefont {Turok}},\ }\href
  {\doibase 10.1103/PhysRevD.64.123522} {\bibfield  {journal} {\bibinfo
  {journal} {Phys. Rev. D}\ }\textbf {\bibinfo {volume} {64}},\ \bibinfo
  {pages} {123522} (\bibinfo {year} {2001}{\natexlab{b}})},\ \Eprint
  {http://arxiv.org/abs/0103239} {arXiv:0103239 [hep-th]} \BibitemShut
  {NoStop}%
\bibitem [{\citenamefont {Khoury}\ \emph {et~al.}(2003)\citenamefont {Khoury},
  \citenamefont {Steinhardt},\ and\ \citenamefont {Turok}}]{Khoury2003a}%
  \BibitemOpen
  \bibfield  {author} {\bibinfo {author} {\bibfnamefont {J.}~\bibnamefont
  {Khoury}}, \bibinfo {author} {\bibfnamefont {P.~J.}\ \bibnamefont
  {Steinhardt}}, \ and\ \bibinfo {author} {\bibfnamefont {N.}~\bibnamefont
  {Turok}},\ }\href {\doibase 10.1103/PhysRevLett.91.161301} {\bibfield
  {journal} {\bibinfo  {journal} {Phys. Rev. Lett.}\ }\textbf {\bibinfo
  {volume} {91}},\ \bibinfo {pages} {161301} (\bibinfo {year} {2003})},\
  \Eprint {http://arxiv.org/abs/0302012} {arXiv:0302012 [astro-ph]}
  \BibitemShut {NoStop}%
\bibitem [{\citenamefont {Steinhardt}\ and\ \citenamefont
  {Turok}(2002)}]{Steinhardt2002}%
  \BibitemOpen
  \bibfield  {author} {\bibinfo {author} {\bibfnamefont {P.~J.}\ \bibnamefont
  {Steinhardt}}\ and\ \bibinfo {author} {\bibfnamefont {N.}~\bibnamefont
  {Turok}},\ }\href {\doibase 10.1103/PhysRevD.65.126003} {\bibfield  {journal}
  {\bibinfo  {journal} {Phys. Rev. D}\ }\textbf {\bibinfo {volume} {65}},\
  \bibinfo {pages} {126003} (\bibinfo {year} {2002})},\ \Eprint
  {http://arxiv.org/abs/0111098} {arXiv:0111098 [hep-th]} \BibitemShut
  {NoStop}%
\bibitem [{\citenamefont {Boyle}\ \emph {et~al.}(2004)\citenamefont {Boyle},
  \citenamefont {Steinhardt},\ and\ \citenamefont {Turok}}]{Boyle2004a}%
  \BibitemOpen
  \bibfield  {author} {\bibinfo {author} {\bibfnamefont {L.~A.}\ \bibnamefont
  {Boyle}}, \bibinfo {author} {\bibfnamefont {P.~J.}\ \bibnamefont
  {Steinhardt}}, \ and\ \bibinfo {author} {\bibfnamefont {N.}~\bibnamefont
  {Turok}},\ }\href {\doibase 10.1103/PhysRevD.69.127302} {\bibfield  {journal}
  {\bibinfo  {journal} {Phys. Rev. D}\ }\textbf {\bibinfo {volume} {69}},\
  \bibinfo {pages} {127302} (\bibinfo {year} {2004})},\ \Eprint
  {http://arxiv.org/abs/0307170v1} {arXiv:0307170v1 [arXiv:hep-th]}
  \BibitemShut {NoStop}%
\bibitem [{\citenamefont {{BICEP2/Keck Array and Planck
  Collaborations}}(2015)}]{BICEP2/Keck2015}%
  \BibitemOpen
  \bibfield  {author} {\bibinfo {author} {\bibnamefont {{BICEP2/Keck Array and
  Planck Collaborations}}},\ }\href {\doibase 10.1103/PhysRevLett.114.101301}
  {\bibfield  {journal} {\bibinfo  {journal} {Phys. Rev. Lett.}\ }\textbf
  {\bibinfo {volume} {114}},\ \bibinfo {pages} {101301} (\bibinfo {year}
  {2015})},\ \Eprint {http://arxiv.org/abs/1502.00612} {arXiv:1502.00612}
  \BibitemShut {NoStop}%
\bibitem [{\citenamefont {{Planck Collaboration
  XX}}(2016)}]{PlanckCollaborationXX2015}%
  \BibitemOpen
  \bibfield  {author} {\bibinfo {author} {\bibnamefont {{Planck Collaboration
  XX}}},\ }\href {\doibase 10.1051/0004-6361/201525898} {\bibfield  {journal}
  {\bibinfo  {journal} {Astron. Astrophys.}\ }\textbf {\bibinfo {volume}
  {594}},\ \bibinfo {pages} {A20} (\bibinfo {year} {2016})},\ \Eprint
  {http://arxiv.org/abs/1502.02114} {arXiv:1502.02114} \BibitemShut {NoStop}%
\bibitem [{\citenamefont {{BICEP2/Keck Array
  Collaborations}}(2016)}]{BICEP2/Keck2016}%
  \BibitemOpen
  \bibfield  {author} {\bibinfo {author} {\bibnamefont {{BICEP2/Keck Array
  Collaborations}}},\ }\href {\doibase 10.1103/PhysRevLett.116.031302}
  {\bibfield  {journal} {\bibinfo  {journal} {Phys. Rev. Lett.}\ }\textbf
  {\bibinfo {volume} {116}},\ \bibinfo {pages} {031302} (\bibinfo {year}
  {2016})}\BibitemShut {NoStop}%
\bibitem [{\citenamefont {{COrE Collaboration}}(2011)}]{COrECollaboration2011}%
  \BibitemOpen
  \bibfield  {author} {\bibinfo {author} {\bibnamefont {{COrE
  Collaboration}}},\ }\href {http://arxiv.org/abs/1102.2181
  http://adsabs.harvard.edu/cgi-bin/nph-data_query?bibcode=2011arXiv1102.2181T&link_type=ABSTRACT%5Cnpapers://ee00755c-a478-4d4e-a50b-ef01ee7b9957/Paper/p10382}
  {\bibfield  {journal} {\bibinfo  {journal} {eprint}\ } (\bibinfo {year}
  {2011})},\ \Eprint {http://arxiv.org/abs/1102.2181} {arXiv:1102.2181}
  \BibitemShut {NoStop}%
\bibitem [{\citenamefont {{LiteBird
  Collaboration}}(2014)}]{LiteBirdCollaboration2014}%
  \BibitemOpen
  \bibfield  {author} {\bibinfo {author} {\bibnamefont {{LiteBird
  Collaboration}}},\ }\href {\doibase 10.1007/s10909-013-0996-1} {\bibfield
  {journal} {\bibinfo  {journal} {J. Low Temp. Phys.}\ }\textbf {\bibinfo
  {volume} {176}},\ \bibinfo {pages} {733} (\bibinfo {year} {2014})},\ \Eprint
  {http://arxiv.org/abs/1311.2847} {arXiv:1311.2847} \BibitemShut {NoStop}%
\bibitem [{\citenamefont {Ishino}\ \emph {et~al.}(2016)\citenamefont {Ishino},
  \citenamefont {Akiba}, \citenamefont {Arnold}, \citenamefont {Barron},
  \citenamefont {Borrill}, \citenamefont {Chendra}, \citenamefont {Chinone},
  \citenamefont {Cho}, \citenamefont {Cukierman}, \citenamefont {de~Haan},
  \citenamefont {Dobbs}, \citenamefont {Dominjon}, \citenamefont {Dotani},
  \citenamefont {Elleflot}, \citenamefont {Errard}, \citenamefont {Fujino},
  \citenamefont {Fuke}, \citenamefont {Funaki}, \citenamefont {Goeckner-Wald},
  \citenamefont {Halverson}, \citenamefont {Harvey}, \citenamefont {Hasebe},
  \citenamefont {Hasegawa}, \citenamefont {Hattori}, \citenamefont {Hattori},
  \citenamefont {Hazumi}, \citenamefont {Hidehira}, \citenamefont {Hill},
  \citenamefont {Hilton}, \citenamefont {Holzapfel}, \citenamefont {Hori},
  \citenamefont {Hubmayr}, \citenamefont {Ichiki}, \citenamefont {Imada},
  \citenamefont {Inatani}, \citenamefont {Inoue}, \citenamefont {Inoue},
  \citenamefont {Irie}, \citenamefont {Irwin}, \citenamefont {Ishitsuka},
  \citenamefont {Jeong}, \citenamefont {Kanai}, \citenamefont {Karatsu},
  \citenamefont {Kashima}, \citenamefont {Katayama}, \citenamefont {Kawano},
  \citenamefont {Kawasaki}, \citenamefont {Keating}, \citenamefont
  {Kernasovskiy}, \citenamefont {Keskitalo}, \citenamefont {Kibayashi},
  \citenamefont {Kida}, \citenamefont {Kimura}, \citenamefont {Kimura},
  \citenamefont {Kisner}, \citenamefont {Kohri}, \citenamefont {Komatsu},
  \citenamefont {Komatsu}, \citenamefont {Kuo}, \citenamefont {Kuromiya},
  \citenamefont {Kusaka}, \citenamefont {Lee}, \citenamefont {Li},
  \citenamefont {Linder}, \citenamefont {Maki}, \citenamefont {Matsuhara},
  \citenamefont {Matsumura}, \citenamefont {Matsuoka}, \citenamefont
  {Matsuura}, \citenamefont {Mima}, \citenamefont {Minami}, \citenamefont
  {Mitsuda}, \citenamefont {Nagai}, \citenamefont {Nagasaki}, \citenamefont
  {Nagata}, \citenamefont {Nakajima}, \citenamefont {Nakamura}, \citenamefont
  {Namikawa}, \citenamefont {Naruse}, \citenamefont {Nishibori}, \citenamefont
  {Nishijo}, \citenamefont {Nishino}, \citenamefont {Noda}, \citenamefont
  {Noguchi}, \citenamefont {Ogawa}, \citenamefont {Ogburn}, \citenamefont
  {Oguri}, \citenamefont {Ohta}, \citenamefont {Okada}, \citenamefont
  {Okamoto}, \citenamefont {Okamura}, \citenamefont {Otani}, \citenamefont
  {Pisano}, \citenamefont {Rebeiz}, \citenamefont {Richards}, \citenamefont
  {Sakai}, \citenamefont {Sakurai}, \citenamefont {Sato}, \citenamefont {Sato},
  \citenamefont {Segawa}, \citenamefont {Sekiguchi}, \citenamefont {Sekimoto},
  \citenamefont {Sekine}, \citenamefont {Seljak}, \citenamefont {Sherwin},
  \citenamefont {Shimizu}, \citenamefont {Shinozaki}, \citenamefont {Shu},
  \citenamefont {Stompor}, \citenamefont {Sugai}, \citenamefont {Sugita},
  \citenamefont {Suzuki}, \citenamefont {Suzuki}, \citenamefont {Suzuki},
  \citenamefont {Tajima}, \citenamefont {Takada}, \citenamefont {Takakura},
  \citenamefont {Takano}, \citenamefont {Takatori}, \citenamefont {Takei},
  \citenamefont {Tanabe}, \citenamefont {Tomaru}, \citenamefont {Tomita},
  \citenamefont {Turin}, \citenamefont {Uozumi}, \citenamefont {Utsunomiya},
  \citenamefont {Uzawa}, \citenamefont {Wada}, \citenamefont {Watanabe},
  \citenamefont {Westbrook}, \citenamefont {Whitehorn}, \citenamefont {Yamada},
  \citenamefont {Yamamoto}, \citenamefont {Yamasaki}, \citenamefont
  {Yamashita}, \citenamefont {Yoshida}, \citenamefont {Yoshida},\ and\
  \citenamefont {Yotsumoto}}]{LiteBird2016}%
  \BibitemOpen
  \bibfield  {author} {\bibinfo {author} {\bibfnamefont {H.}~\bibnamefont
  {Ishino}}, \bibinfo {author} {\bibfnamefont {Y.}~\bibnamefont {Akiba}},
  \bibinfo {author} {\bibfnamefont {K.}~\bibnamefont {Arnold}}, \bibinfo
  {author} {\bibfnamefont {D.}~\bibnamefont {Barron}}, \bibinfo {author}
  {\bibfnamefont {J.}~\bibnamefont {Borrill}}, \bibinfo {author} {\bibfnamefont
  {R.}~\bibnamefont {Chendra}}, \bibinfo {author} {\bibfnamefont
  {Y.}~\bibnamefont {Chinone}}, \bibinfo {author} {\bibfnamefont
  {S.}~\bibnamefont {Cho}}, \bibinfo {author} {\bibfnamefont {A.}~\bibnamefont
  {Cukierman}}, \bibinfo {author} {\bibfnamefont {T.}~\bibnamefont {de~Haan}},
  \bibinfo {author} {\bibfnamefont {M.}~\bibnamefont {Dobbs}}, \bibinfo
  {author} {\bibfnamefont {A.}~\bibnamefont {Dominjon}}, \bibinfo {author}
  {\bibfnamefont {T.}~\bibnamefont {Dotani}}, \bibinfo {author} {\bibfnamefont
  {T.}~\bibnamefont {Elleflot}}, \bibinfo {author} {\bibfnamefont
  {J.}~\bibnamefont {Errard}}, \bibinfo {author} {\bibfnamefont
  {T.}~\bibnamefont {Fujino}}, \bibinfo {author} {\bibfnamefont
  {H.}~\bibnamefont {Fuke}}, \bibinfo {author} {\bibfnamefont {T.}~\bibnamefont
  {Funaki}}, \bibinfo {author} {\bibfnamefont {N.}~\bibnamefont
  {Goeckner-Wald}}, \bibinfo {author} {\bibfnamefont {N.}~\bibnamefont
  {Halverson}}, \bibinfo {author} {\bibfnamefont {P.}~\bibnamefont {Harvey}},
  \bibinfo {author} {\bibfnamefont {T.}~\bibnamefont {Hasebe}}, \bibinfo
  {author} {\bibfnamefont {M.}~\bibnamefont {Hasegawa}}, \bibinfo {author}
  {\bibfnamefont {K.}~\bibnamefont {Hattori}}, \bibinfo {author} {\bibfnamefont
  {M.}~\bibnamefont {Hattori}}, \bibinfo {author} {\bibfnamefont
  {M.}~\bibnamefont {Hazumi}}, \bibinfo {author} {\bibfnamefont
  {N.}~\bibnamefont {Hidehira}}, \bibinfo {author} {\bibfnamefont
  {C.}~\bibnamefont {Hill}}, \bibinfo {author} {\bibfnamefont {G.}~\bibnamefont
  {Hilton}}, \bibinfo {author} {\bibfnamefont {W.}~\bibnamefont {Holzapfel}},
  \bibinfo {author} {\bibfnamefont {Y.}~\bibnamefont {Hori}}, \bibinfo {author}
  {\bibfnamefont {J.}~\bibnamefont {Hubmayr}}, \bibinfo {author} {\bibfnamefont
  {K.}~\bibnamefont {Ichiki}}, \bibinfo {author} {\bibfnamefont
  {H.}~\bibnamefont {Imada}}, \bibinfo {author} {\bibfnamefont
  {J.}~\bibnamefont {Inatani}}, \bibinfo {author} {\bibfnamefont
  {M.}~\bibnamefont {Inoue}}, \bibinfo {author} {\bibfnamefont
  {Y.}~\bibnamefont {Inoue}}, \bibinfo {author} {\bibfnamefont
  {F.}~\bibnamefont {Irie}}, \bibinfo {author} {\bibfnamefont {K.}~\bibnamefont
  {Irwin}}, \bibinfo {author} {\bibfnamefont {H.}~\bibnamefont {Ishitsuka}},
  \bibinfo {author} {\bibfnamefont {O.}~\bibnamefont {Jeong}}, \bibinfo
  {author} {\bibfnamefont {H.}~\bibnamefont {Kanai}}, \bibinfo {author}
  {\bibfnamefont {K.}~\bibnamefont {Karatsu}}, \bibinfo {author} {\bibfnamefont
  {S.}~\bibnamefont {Kashima}}, \bibinfo {author} {\bibfnamefont
  {N.}~\bibnamefont {Katayama}}, \bibinfo {author} {\bibfnamefont
  {I.}~\bibnamefont {Kawano}}, \bibinfo {author} {\bibfnamefont
  {T.}~\bibnamefont {Kawasaki}}, \bibinfo {author} {\bibfnamefont
  {B.}~\bibnamefont {Keating}}, \bibinfo {author} {\bibfnamefont
  {S.}~\bibnamefont {Kernasovskiy}}, \bibinfo {author} {\bibfnamefont
  {R.}~\bibnamefont {Keskitalo}}, \bibinfo {author} {\bibfnamefont
  {A.}~\bibnamefont {Kibayashi}}, \bibinfo {author} {\bibfnamefont
  {Y.}~\bibnamefont {Kida}}, \bibinfo {author} {\bibfnamefont {N.}~\bibnamefont
  {Kimura}}, \bibinfo {author} {\bibfnamefont {K.}~\bibnamefont {Kimura}},
  \bibinfo {author} {\bibfnamefont {T.}~\bibnamefont {Kisner}}, \bibinfo
  {author} {\bibfnamefont {K.}~\bibnamefont {Kohri}}, \bibinfo {author}
  {\bibfnamefont {E.}~\bibnamefont {Komatsu}}, \bibinfo {author} {\bibfnamefont
  {K.}~\bibnamefont {Komatsu}}, \bibinfo {author} {\bibfnamefont {C.-L.}\
  \bibnamefont {Kuo}}, \bibinfo {author} {\bibfnamefont {S.}~\bibnamefont
  {Kuromiya}}, \bibinfo {author} {\bibfnamefont {A.}~\bibnamefont {Kusaka}},
  \bibinfo {author} {\bibfnamefont {A.}~\bibnamefont {Lee}}, \bibinfo {author}
  {\bibfnamefont {D.}~\bibnamefont {Li}}, \bibinfo {author} {\bibfnamefont
  {E.}~\bibnamefont {Linder}}, \bibinfo {author} {\bibfnamefont
  {M.}~\bibnamefont {Maki}}, \bibinfo {author} {\bibfnamefont {H.}~\bibnamefont
  {Matsuhara}}, \bibinfo {author} {\bibfnamefont {T.}~\bibnamefont
  {Matsumura}}, \bibinfo {author} {\bibfnamefont {S.}~\bibnamefont {Matsuoka}},
  \bibinfo {author} {\bibfnamefont {S.}~\bibnamefont {Matsuura}}, \bibinfo
  {author} {\bibfnamefont {S.}~\bibnamefont {Mima}}, \bibinfo {author}
  {\bibfnamefont {Y.}~\bibnamefont {Minami}}, \bibinfo {author} {\bibfnamefont
  {K.}~\bibnamefont {Mitsuda}}, \bibinfo {author} {\bibfnamefont
  {M.}~\bibnamefont {Nagai}}, \bibinfo {author} {\bibfnamefont
  {T.}~\bibnamefont {Nagasaki}}, \bibinfo {author} {\bibfnamefont
  {R.}~\bibnamefont {Nagata}}, \bibinfo {author} {\bibfnamefont
  {M.}~\bibnamefont {Nakajima}}, \bibinfo {author} {\bibfnamefont
  {S.}~\bibnamefont {Nakamura}}, \bibinfo {author} {\bibfnamefont
  {T.}~\bibnamefont {Namikawa}}, \bibinfo {author} {\bibfnamefont
  {M.}~\bibnamefont {Naruse}}, \bibinfo {author} {\bibfnamefont
  {T.}~\bibnamefont {Nishibori}}, \bibinfo {author} {\bibfnamefont
  {K.}~\bibnamefont {Nishijo}}, \bibinfo {author} {\bibfnamefont
  {H.}~\bibnamefont {Nishino}}, \bibinfo {author} {\bibfnamefont
  {A.}~\bibnamefont {Noda}}, \bibinfo {author} {\bibfnamefont {T.}~\bibnamefont
  {Noguchi}}, \bibinfo {author} {\bibfnamefont {H.}~\bibnamefont {Ogawa}},
  \bibinfo {author} {\bibfnamefont {W.}~\bibnamefont {Ogburn}}, \bibinfo
  {author} {\bibfnamefont {S.}~\bibnamefont {Oguri}}, \bibinfo {author}
  {\bibfnamefont {I.}~\bibnamefont {Ohta}}, \bibinfo {author} {\bibfnamefont
  {N.}~\bibnamefont {Okada}}, \bibinfo {author} {\bibfnamefont
  {A.}~\bibnamefont {Okamoto}}, \bibinfo {author} {\bibfnamefont
  {T.}~\bibnamefont {Okamura}}, \bibinfo {author} {\bibfnamefont
  {C.}~\bibnamefont {Otani}}, \bibinfo {author} {\bibfnamefont
  {G.}~\bibnamefont {Pisano}}, \bibinfo {author} {\bibfnamefont
  {G.}~\bibnamefont {Rebeiz}}, \bibinfo {author} {\bibfnamefont
  {P.}~\bibnamefont {Richards}}, \bibinfo {author} {\bibfnamefont
  {S.}~\bibnamefont {Sakai}}, \bibinfo {author} {\bibfnamefont
  {Y.}~\bibnamefont {Sakurai}}, \bibinfo {author} {\bibfnamefont
  {Y.}~\bibnamefont {Sato}}, \bibinfo {author} {\bibfnamefont {N.}~\bibnamefont
  {Sato}}, \bibinfo {author} {\bibfnamefont {Y.}~\bibnamefont {Segawa}},
  \bibinfo {author} {\bibfnamefont {S.}~\bibnamefont {Sekiguchi}}, \bibinfo
  {author} {\bibfnamefont {Y.}~\bibnamefont {Sekimoto}}, \bibinfo {author}
  {\bibfnamefont {M.}~\bibnamefont {Sekine}}, \bibinfo {author} {\bibfnamefont
  {U.}~\bibnamefont {Seljak}}, \bibinfo {author} {\bibfnamefont
  {B.}~\bibnamefont {Sherwin}}, \bibinfo {author} {\bibfnamefont
  {T.}~\bibnamefont {Shimizu}}, \bibinfo {author} {\bibfnamefont
  {K.}~\bibnamefont {Shinozaki}}, \bibinfo {author} {\bibfnamefont
  {S.}~\bibnamefont {Shu}}, \bibinfo {author} {\bibfnamefont {R.}~\bibnamefont
  {Stompor}}, \bibinfo {author} {\bibfnamefont {H.}~\bibnamefont {Sugai}},
  \bibinfo {author} {\bibfnamefont {H.}~\bibnamefont {Sugita}}, \bibinfo
  {author} {\bibfnamefont {J.}~\bibnamefont {Suzuki}}, \bibinfo {author}
  {\bibfnamefont {T.}~\bibnamefont {Suzuki}}, \bibinfo {author} {\bibfnamefont
  {A.}~\bibnamefont {Suzuki}}, \bibinfo {author} {\bibfnamefont
  {O.}~\bibnamefont {Tajima}}, \bibinfo {author} {\bibfnamefont
  {S.}~\bibnamefont {Takada}}, \bibinfo {author} {\bibfnamefont
  {S.}~\bibnamefont {Takakura}}, \bibinfo {author} {\bibfnamefont
  {K.}~\bibnamefont {Takano}}, \bibinfo {author} {\bibfnamefont
  {S.}~\bibnamefont {Takatori}}, \bibinfo {author} {\bibfnamefont
  {Y.}~\bibnamefont {Takei}}, \bibinfo {author} {\bibfnamefont
  {D.}~\bibnamefont {Tanabe}}, \bibinfo {author} {\bibfnamefont
  {T.}~\bibnamefont {Tomaru}}, \bibinfo {author} {\bibfnamefont
  {N.}~\bibnamefont {Tomita}}, \bibinfo {author} {\bibfnamefont
  {P.}~\bibnamefont {Turin}}, \bibinfo {author} {\bibfnamefont
  {S.}~\bibnamefont {Uozumi}}, \bibinfo {author} {\bibfnamefont
  {S.}~\bibnamefont {Utsunomiya}}, \bibinfo {author} {\bibfnamefont
  {Y.}~\bibnamefont {Uzawa}}, \bibinfo {author} {\bibfnamefont
  {T.}~\bibnamefont {Wada}}, \bibinfo {author} {\bibfnamefont {H.}~\bibnamefont
  {Watanabe}}, \bibinfo {author} {\bibfnamefont {B.}~\bibnamefont {Westbrook}},
  \bibinfo {author} {\bibfnamefont {N.}~\bibnamefont {Whitehorn}}, \bibinfo
  {author} {\bibfnamefont {Y.}~\bibnamefont {Yamada}}, \bibinfo {author}
  {\bibfnamefont {R.}~\bibnamefont {Yamamoto}}, \bibinfo {author}
  {\bibfnamefont {N.}~\bibnamefont {Yamasaki}}, \bibinfo {author}
  {\bibfnamefont {T.}~\bibnamefont {Yamashita}}, \bibinfo {author}
  {\bibfnamefont {T.}~\bibnamefont {Yoshida}}, \bibinfo {author} {\bibfnamefont
  {M.}~\bibnamefont {Yoshida}}, \ and\ \bibinfo {author} {\bibfnamefont
  {K.}~\bibnamefont {Yotsumoto}},\ }\href {\doibase 10.1117/12.2231995}
  {\enquote {\bibinfo {title} {{LiteBIRD: lite satellite for the study of
  B-mode polarization and inflation from cosmic microwave background radiation
  detection}},}\ } (\bibinfo {year} {2016})\BibitemShut {NoStop}%
\bibitem [{\citenamefont {{CMB-S4 Collaboration}}(2016)}]{s42016}%
  \BibitemOpen
  \bibfield  {author} {\bibinfo {author} {\bibnamefont {{CMB-S4
  Collaboration}}},\ }\href
  {http://cmb-s4.org/CMB-S4workshops/index.php/Main_Page#CMB-S4_Science_Book
  http://arxiv.org/abs/1610.02743} {\bibfield  {journal} {\bibinfo  {journal}
  {eprint}\ } (\bibinfo {year} {2016})},\ \Eprint
  {http://arxiv.org/abs/1610.02743} {arXiv:1610.02743} \BibitemShut {NoStop}%
\bibitem [{\citenamefont {Cabass}\ \emph {et~al.}(2016)\citenamefont {Cabass},
  \citenamefont {Pagano}, \citenamefont {Salvati}, \citenamefont {Gerbino},
  \citenamefont {Giusarma},\ and\ \citenamefont {Melchiorri}}]{Cabass2016}%
  \BibitemOpen
  \bibfield  {author} {\bibinfo {author} {\bibfnamefont {G.}~\bibnamefont
  {Cabass}}, \bibinfo {author} {\bibfnamefont {L.}~\bibnamefont {Pagano}},
  \bibinfo {author} {\bibfnamefont {L.}~\bibnamefont {Salvati}}, \bibinfo
  {author} {\bibfnamefont {M.}~\bibnamefont {Gerbino}}, \bibinfo {author}
  {\bibfnamefont {E.}~\bibnamefont {Giusarma}}, \ and\ \bibinfo {author}
  {\bibfnamefont {A.}~\bibnamefont {Melchiorri}},\ }\href {\doibase
  10.1103/PhysRevD.93.063508} {\bibfield  {journal} {\bibinfo  {journal} {Phys.
  Rev. D}\ }\textbf {\bibinfo {volume} {93}},\ \bibinfo {pages} {063508}
  (\bibinfo {year} {2016})},\ \Eprint {http://arxiv.org/abs/1511.05146}
  {arXiv:1511.05146} \BibitemShut {NoStop}%
\bibitem [{\citenamefont {Kamionkowski}\ and\ \citenamefont
  {Kovetz}(2016)}]{Kamionkowski2016}%
  \BibitemOpen
  \bibfield  {author} {\bibinfo {author} {\bibfnamefont {M.}~\bibnamefont
  {Kamionkowski}}\ and\ \bibinfo {author} {\bibfnamefont {E.~D.}\ \bibnamefont
  {Kovetz}},\ }\href {\doibase 10.1146/annurev-astro-081915-023433} {\bibfield
  {journal} {\bibinfo  {journal} {Annu. Rev. Astron.}\ }\textbf {\bibinfo
  {volume} {54}},\ \bibinfo {pages} {227} (\bibinfo {year} {2016})},\ \Eprint
  {http://arxiv.org/abs/1510.06042} {arXiv:1510.06042} \BibitemShut {NoStop}%
\bibitem [{\citenamefont {Delabrouille}\ \emph {et~al.}(2017)\citenamefont
  {Delabrouille}, \citenamefont {de~Bernardis}, \citenamefont {Bouchet},
  \citenamefont {Ach{\'{u}}carro}, \citenamefont {Ade}, \citenamefont
  {Allison}, \citenamefont {Arroja}, \citenamefont {Artal}, \citenamefont
  {Ashdown}, \citenamefont {Baccigalupi}, \citenamefont {Ballardini},
  \citenamefont {Banday}, \citenamefont {Banerji}, \citenamefont {Barbosa},
  \citenamefont {Bartlett}, \citenamefont {Bartolo}, \citenamefont {Basak},
  \citenamefont {Baselmans}, \citenamefont {Basu}, \citenamefont {Battistelli},
  \citenamefont {Battye}, \citenamefont {Baumann}, \citenamefont
  {Beno{\^{i}}t}, \citenamefont {Bersanelli}, \citenamefont {Bideaud},
  \citenamefont {Biesiada}, \citenamefont {Bilicki}, \citenamefont {Bonaldi},
  \citenamefont {Bonato}, \citenamefont {Borrill}, \citenamefont {Boulanger},
  \citenamefont {Brinckmann}, \citenamefont {Brown}, \citenamefont {Bucher},
  \citenamefont {Burigana}, \citenamefont {Buzzelli}, \citenamefont {Cabass},
  \citenamefont {Cai}, \citenamefont {Calvo}, \citenamefont {Caputo},
  \citenamefont {Carvalho}, \citenamefont {Casas}, \citenamefont {Castellano},
  \citenamefont {Catalano}, \citenamefont {Challinor}, \citenamefont {Charles},
  \citenamefont {Chluba}, \citenamefont {Clements}, \citenamefont {Clesse},
  \citenamefont {Colafrancesco}, \citenamefont {Colantoni}, \citenamefont
  {Contreras}, \citenamefont {Coppolecchia}, \citenamefont {Crook},
  \citenamefont {D'Alessandro}, \citenamefont {D'Amico}, \citenamefont
  {da~Silva}, \citenamefont {de~Avillez}, \citenamefont {de~Gasperis},
  \citenamefont {{De Petris}}, \citenamefont {de~Zotti}, \citenamefont
  {Danese}, \citenamefont {D{\'{e}}sert}, \citenamefont {Desjacques},
  \citenamefont {{Di Valentino}}, \citenamefont {Dickinson}, \citenamefont
  {Diego}, \citenamefont {Doyle}, \citenamefont {Durrer}, \citenamefont
  {Dvorkin}, \citenamefont {Eriksen}, \citenamefont {Errard}, \citenamefont
  {Feeney}, \citenamefont {Fern{\'{a}}ndez-Cobos}, \citenamefont {Finelli},
  \citenamefont {Forastieri}, \citenamefont {Franceschet}, \citenamefont
  {Fuskeland}, \citenamefont {Galli}, \citenamefont {G{\'{e}}nova-Santos},
  \citenamefont {Gerbino}, \citenamefont {Giusarma}, \citenamefont {Gomez},
  \citenamefont {Gonz{\'{a}}lez-Nuevo}, \citenamefont {Grandis}, \citenamefont
  {Greenslade}, \citenamefont {Goupy}, \citenamefont {Hagstotz}, \citenamefont
  {Hanany}, \citenamefont {Handley}, \citenamefont {Henrot-Versill{\'{e}}},
  \citenamefont {Hern{\'{a}}ndez-Monteagudo}, \citenamefont {Hervias-Caimapo},
  \citenamefont {Hills}, \citenamefont {Hindmarsh}, \citenamefont {Hivon},
  \citenamefont {Hoang}, \citenamefont {Hooper}, \citenamefont {Hu},
  \citenamefont {Keih{\"{a}}nen}, \citenamefont {Keskitalo}, \citenamefont
  {Kiiveri}, \citenamefont {Kisner}, \citenamefont {Kitching}, \citenamefont
  {Kunz}, \citenamefont {Kurki-Suonio}, \citenamefont {Lagache}, \citenamefont
  {Lamagna}, \citenamefont {Lapi}, \citenamefont {Lasenby}, \citenamefont
  {Lattanzi}, \citenamefont {Brun}, \citenamefont {Lesgourgues}, \citenamefont
  {Liguori}, \citenamefont {Lindholm}, \citenamefont {Lizarraga}, \citenamefont
  {Luzzi}, \citenamefont {Mac{\`{i}}as-P{\'{e}}rez}, \citenamefont {Maffei},
  \citenamefont {Mandolesi}, \citenamefont {Martin}, \citenamefont
  {Martinez-Gonzalez}, \citenamefont {Martins}, \citenamefont {Masi},
  \citenamefont {Massardi}, \citenamefont {Matarrese}, \citenamefont
  {Mazzotta}, \citenamefont {McCarthy}, \citenamefont {Melchiorri},
  \citenamefont {Melin}, \citenamefont {Mennella}, \citenamefont {Mohr},
  \citenamefont {Molinari}, \citenamefont {Monfardini}, \citenamefont
  {Montier}, \citenamefont {Natoli}, \citenamefont {Negrello}, \citenamefont
  {Notari}, \citenamefont {Noviello}, \citenamefont {Oppizzi}, \citenamefont
  {O'Sullivan}, \citenamefont {Pagano}, \citenamefont {Paiella}, \citenamefont
  {Pajer}, \citenamefont {Paoletti}, \citenamefont {Paradiso}, \citenamefont
  {Partridge}, \citenamefont {Patanchon}, \citenamefont {Patil}, \citenamefont
  {Perdereau}, \citenamefont {Piacentini}, \citenamefont {Piat}, \citenamefont
  {Pisano}, \citenamefont {Polastri}, \citenamefont {Polenta}, \citenamefont
  {Pollo}, \citenamefont {Ponthieu}, \citenamefont {Poulin}, \citenamefont
  {Pr{\^{e}}le}, \citenamefont {Quartin}, \citenamefont {Ravenni},
  \citenamefont {Remazeilles}, \citenamefont {Renzi}, \citenamefont {Ringeval},
  \citenamefont {Roest}, \citenamefont {Roman}, \citenamefont {Roukema},
  \citenamefont {Rubino-Martin}, \citenamefont {Salvati}, \citenamefont
  {Scott}, \citenamefont {Serjeant}, \citenamefont {Signorelli}, \citenamefont
  {Starobinsky}, \citenamefont {Sunyaev}, \citenamefont {Tan}, \citenamefont
  {Tartari}, \citenamefont {Tasinato}, \citenamefont {Toffolatti},
  \citenamefont {Tomasi}, \citenamefont {Torrado}, \citenamefont {Tramonte},
  \citenamefont {Trappe}, \citenamefont {Triqueneaux}, \citenamefont
  {Tristram}, \citenamefont {Trombetti}, \citenamefont {Tucci}, \citenamefont
  {Tucker}, \citenamefont {Urrestilla}, \citenamefont {V{\"{a}}liviita},
  \citenamefont {{Van de Weygaert}}, \citenamefont {{Van Tent}}, \citenamefont
  {Vennin}, \citenamefont {Verde}, \citenamefont {Vermeulen}, \citenamefont
  {Vielva}, \citenamefont {Vittorio}, \citenamefont {Voisin}, \citenamefont
  {Wallis}, \citenamefont {Wandelt}, \citenamefont {Wehus}, \citenamefont
  {Weller}, \citenamefont {Young}, \citenamefont {Zannoni},\ and\ \citenamefont
  {Collaboration}}]{Delabrouille2017}%
  \BibitemOpen
  \bibfield  {author} {\bibinfo {author} {\bibfnamefont {J.}~\bibnamefont
  {Delabrouille}}, \bibinfo {author} {\bibfnamefont {P.}~\bibnamefont
  {de~Bernardis}}, \bibinfo {author} {\bibfnamefont {F.~R.}\ \bibnamefont
  {Bouchet}}, \bibinfo {author} {\bibfnamefont {A.}~\bibnamefont
  {Ach{\'{u}}carro}}, \bibinfo {author} {\bibfnamefont {P.~A.~R.}\ \bibnamefont
  {Ade}}, \bibinfo {author} {\bibfnamefont {R.}~\bibnamefont {Allison}},
  \bibinfo {author} {\bibfnamefont {F.}~\bibnamefont {Arroja}}, \bibinfo
  {author} {\bibfnamefont {E.}~\bibnamefont {Artal}}, \bibinfo {author}
  {\bibfnamefont {M.}~\bibnamefont {Ashdown}}, \bibinfo {author} {\bibfnamefont
  {C.}~\bibnamefont {Baccigalupi}}, \bibinfo {author} {\bibfnamefont
  {M.}~\bibnamefont {Ballardini}}, \bibinfo {author} {\bibfnamefont {A.~J.}\
  \bibnamefont {Banday}}, \bibinfo {author} {\bibfnamefont {R.}~\bibnamefont
  {Banerji}}, \bibinfo {author} {\bibfnamefont {D.}~\bibnamefont {Barbosa}},
  \bibinfo {author} {\bibfnamefont {J.}~\bibnamefont {Bartlett}}, \bibinfo
  {author} {\bibfnamefont {N.}~\bibnamefont {Bartolo}}, \bibinfo {author}
  {\bibfnamefont {S.}~\bibnamefont {Basak}}, \bibinfo {author} {\bibfnamefont
  {J.~J.~A.}\ \bibnamefont {Baselmans}}, \bibinfo {author} {\bibfnamefont
  {K.}~\bibnamefont {Basu}}, \bibinfo {author} {\bibfnamefont {E.~S.}\
  \bibnamefont {Battistelli}}, \bibinfo {author} {\bibfnamefont
  {R.}~\bibnamefont {Battye}}, \bibinfo {author} {\bibfnamefont
  {D.}~\bibnamefont {Baumann}}, \bibinfo {author} {\bibfnamefont
  {A.}~\bibnamefont {Beno{\^{i}}t}}, \bibinfo {author} {\bibfnamefont
  {M.}~\bibnamefont {Bersanelli}}, \bibinfo {author} {\bibfnamefont
  {A.}~\bibnamefont {Bideaud}}, \bibinfo {author} {\bibfnamefont
  {M.}~\bibnamefont {Biesiada}}, \bibinfo {author} {\bibfnamefont
  {M.}~\bibnamefont {Bilicki}}, \bibinfo {author} {\bibfnamefont
  {A.}~\bibnamefont {Bonaldi}}, \bibinfo {author} {\bibfnamefont
  {M.}~\bibnamefont {Bonato}}, \bibinfo {author} {\bibfnamefont
  {J.}~\bibnamefont {Borrill}}, \bibinfo {author} {\bibfnamefont
  {F.}~\bibnamefont {Boulanger}}, \bibinfo {author} {\bibfnamefont
  {T.}~\bibnamefont {Brinckmann}}, \bibinfo {author} {\bibfnamefont {M.~L.}\
  \bibnamefont {Brown}}, \bibinfo {author} {\bibfnamefont {M.}~\bibnamefont
  {Bucher}}, \bibinfo {author} {\bibfnamefont {C.}~\bibnamefont {Burigana}},
  \bibinfo {author} {\bibfnamefont {A.}~\bibnamefont {Buzzelli}}, \bibinfo
  {author} {\bibfnamefont {G.}~\bibnamefont {Cabass}}, \bibinfo {author}
  {\bibfnamefont {Z.~Y.}\ \bibnamefont {Cai}}, \bibinfo {author} {\bibfnamefont
  {M.}~\bibnamefont {Calvo}}, \bibinfo {author} {\bibfnamefont
  {A.}~\bibnamefont {Caputo}}, \bibinfo {author} {\bibfnamefont {C.~S.}\
  \bibnamefont {Carvalho}}, \bibinfo {author} {\bibfnamefont {F.~J.}\
  \bibnamefont {Casas}}, \bibinfo {author} {\bibfnamefont {G.}~\bibnamefont
  {Castellano}}, \bibinfo {author} {\bibfnamefont {A.}~\bibnamefont
  {Catalano}}, \bibinfo {author} {\bibfnamefont {A.}~\bibnamefont {Challinor}},
  \bibinfo {author} {\bibfnamefont {I.}~\bibnamefont {Charles}}, \bibinfo
  {author} {\bibfnamefont {J.}~\bibnamefont {Chluba}}, \bibinfo {author}
  {\bibfnamefont {D.~L.}\ \bibnamefont {Clements}}, \bibinfo {author}
  {\bibfnamefont {S.}~\bibnamefont {Clesse}}, \bibinfo {author} {\bibfnamefont
  {S.}~\bibnamefont {Colafrancesco}}, \bibinfo {author} {\bibfnamefont
  {I.}~\bibnamefont {Colantoni}}, \bibinfo {author} {\bibfnamefont
  {D.}~\bibnamefont {Contreras}}, \bibinfo {author} {\bibfnamefont
  {A.}~\bibnamefont {Coppolecchia}}, \bibinfo {author} {\bibfnamefont
  {M.}~\bibnamefont {Crook}}, \bibinfo {author} {\bibfnamefont
  {G.}~\bibnamefont {D'Alessandro}}, \bibinfo {author} {\bibfnamefont
  {G.}~\bibnamefont {D'Amico}}, \bibinfo {author} {\bibfnamefont
  {A.}~\bibnamefont {da~Silva}}, \bibinfo {author} {\bibfnamefont
  {M.}~\bibnamefont {de~Avillez}}, \bibinfo {author} {\bibfnamefont
  {G.}~\bibnamefont {de~Gasperis}}, \bibinfo {author} {\bibfnamefont
  {M.}~\bibnamefont {{De Petris}}}, \bibinfo {author} {\bibfnamefont
  {G.}~\bibnamefont {de~Zotti}}, \bibinfo {author} {\bibfnamefont
  {L.}~\bibnamefont {Danese}}, \bibinfo {author} {\bibfnamefont {F.~X.}\
  \bibnamefont {D{\'{e}}sert}}, \bibinfo {author} {\bibfnamefont
  {V.}~\bibnamefont {Desjacques}}, \bibinfo {author} {\bibfnamefont
  {E.}~\bibnamefont {{Di Valentino}}}, \bibinfo {author} {\bibfnamefont
  {C.}~\bibnamefont {Dickinson}}, \bibinfo {author} {\bibfnamefont {J.~M.}\
  \bibnamefont {Diego}}, \bibinfo {author} {\bibfnamefont {S.}~\bibnamefont
  {Doyle}}, \bibinfo {author} {\bibfnamefont {R.}~\bibnamefont {Durrer}},
  \bibinfo {author} {\bibfnamefont {C.}~\bibnamefont {Dvorkin}}, \bibinfo
  {author} {\bibfnamefont {H.~K.}\ \bibnamefont {Eriksen}}, \bibinfo {author}
  {\bibfnamefont {J.}~\bibnamefont {Errard}}, \bibinfo {author} {\bibfnamefont
  {S.}~\bibnamefont {Feeney}}, \bibinfo {author} {\bibfnamefont
  {R.}~\bibnamefont {Fern{\'{a}}ndez-Cobos}}, \bibinfo {author} {\bibfnamefont
  {F.}~\bibnamefont {Finelli}}, \bibinfo {author} {\bibfnamefont
  {F.}~\bibnamefont {Forastieri}}, \bibinfo {author} {\bibfnamefont
  {C.}~\bibnamefont {Franceschet}}, \bibinfo {author} {\bibfnamefont
  {U.}~\bibnamefont {Fuskeland}}, \bibinfo {author} {\bibfnamefont
  {S.}~\bibnamefont {Galli}}, \bibinfo {author} {\bibfnamefont {R.~T.}\
  \bibnamefont {G{\'{e}}nova-Santos}}, \bibinfo {author} {\bibfnamefont
  {M.}~\bibnamefont {Gerbino}}, \bibinfo {author} {\bibfnamefont
  {E.}~\bibnamefont {Giusarma}}, \bibinfo {author} {\bibfnamefont
  {A.}~\bibnamefont {Gomez}}, \bibinfo {author} {\bibfnamefont
  {J.}~\bibnamefont {Gonz{\'{a}}lez-Nuevo}}, \bibinfo {author} {\bibfnamefont
  {S.}~\bibnamefont {Grandis}}, \bibinfo {author} {\bibfnamefont
  {J.}~\bibnamefont {Greenslade}}, \bibinfo {author} {\bibfnamefont
  {J.}~\bibnamefont {Goupy}}, \bibinfo {author} {\bibfnamefont
  {S.}~\bibnamefont {Hagstotz}}, \bibinfo {author} {\bibfnamefont
  {S.}~\bibnamefont {Hanany}}, \bibinfo {author} {\bibfnamefont
  {W.}~\bibnamefont {Handley}}, \bibinfo {author} {\bibfnamefont
  {S.}~\bibnamefont {Henrot-Versill{\'{e}}}}, \bibinfo {author} {\bibfnamefont
  {C.}~\bibnamefont {Hern{\'{a}}ndez-Monteagudo}}, \bibinfo {author}
  {\bibfnamefont {C.}~\bibnamefont {Hervias-Caimapo}}, \bibinfo {author}
  {\bibfnamefont {M.}~\bibnamefont {Hills}}, \bibinfo {author} {\bibfnamefont
  {M.}~\bibnamefont {Hindmarsh}}, \bibinfo {author} {\bibfnamefont
  {E.}~\bibnamefont {Hivon}}, \bibinfo {author} {\bibfnamefont {D.~T.}\
  \bibnamefont {Hoang}}, \bibinfo {author} {\bibfnamefont {D.~C.}\ \bibnamefont
  {Hooper}}, \bibinfo {author} {\bibfnamefont {B.}~\bibnamefont {Hu}}, \bibinfo
  {author} {\bibfnamefont {E.}~\bibnamefont {Keih{\"{a}}nen}}, \bibinfo
  {author} {\bibfnamefont {R.}~\bibnamefont {Keskitalo}}, \bibinfo {author}
  {\bibfnamefont {K.}~\bibnamefont {Kiiveri}}, \bibinfo {author} {\bibfnamefont
  {T.}~\bibnamefont {Kisner}}, \bibinfo {author} {\bibfnamefont
  {T.}~\bibnamefont {Kitching}}, \bibinfo {author} {\bibfnamefont
  {M.}~\bibnamefont {Kunz}}, \bibinfo {author} {\bibfnamefont {H.}~\bibnamefont
  {Kurki-Suonio}}, \bibinfo {author} {\bibfnamefont {G.}~\bibnamefont
  {Lagache}}, \bibinfo {author} {\bibfnamefont {L.}~\bibnamefont {Lamagna}},
  \bibinfo {author} {\bibfnamefont {A.}~\bibnamefont {Lapi}}, \bibinfo {author}
  {\bibfnamefont {A.}~\bibnamefont {Lasenby}}, \bibinfo {author} {\bibfnamefont
  {M.}~\bibnamefont {Lattanzi}}, \bibinfo {author} {\bibfnamefont {A.~M.
  C.~L.}\ \bibnamefont {Brun}}, \bibinfo {author} {\bibfnamefont
  {J.}~\bibnamefont {Lesgourgues}}, \bibinfo {author} {\bibfnamefont
  {M.}~\bibnamefont {Liguori}}, \bibinfo {author} {\bibfnamefont
  {V.}~\bibnamefont {Lindholm}}, \bibinfo {author} {\bibfnamefont
  {J.}~\bibnamefont {Lizarraga}}, \bibinfo {author} {\bibfnamefont
  {G.}~\bibnamefont {Luzzi}}, \bibinfo {author} {\bibfnamefont {J.~F.}\
  \bibnamefont {Mac{\`{i}}as-P{\'{e}}rez}}, \bibinfo {author} {\bibfnamefont
  {B.}~\bibnamefont {Maffei}}, \bibinfo {author} {\bibfnamefont
  {N.}~\bibnamefont {Mandolesi}}, \bibinfo {author} {\bibfnamefont
  {S.}~\bibnamefont {Martin}}, \bibinfo {author} {\bibfnamefont
  {E.}~\bibnamefont {Martinez-Gonzalez}}, \bibinfo {author} {\bibfnamefont
  {C.~J. A.~P.}\ \bibnamefont {Martins}}, \bibinfo {author} {\bibfnamefont
  {S.}~\bibnamefont {Masi}}, \bibinfo {author} {\bibfnamefont {M.}~\bibnamefont
  {Massardi}}, \bibinfo {author} {\bibfnamefont {S.}~\bibnamefont {Matarrese}},
  \bibinfo {author} {\bibfnamefont {P.}~\bibnamefont {Mazzotta}}, \bibinfo
  {author} {\bibfnamefont {D.}~\bibnamefont {McCarthy}}, \bibinfo {author}
  {\bibfnamefont {A.}~\bibnamefont {Melchiorri}}, \bibinfo {author}
  {\bibfnamefont {J.~B.}\ \bibnamefont {Melin}}, \bibinfo {author}
  {\bibfnamefont {A.}~\bibnamefont {Mennella}}, \bibinfo {author}
  {\bibfnamefont {J.}~\bibnamefont {Mohr}}, \bibinfo {author} {\bibfnamefont
  {D.}~\bibnamefont {Molinari}}, \bibinfo {author} {\bibfnamefont
  {A.}~\bibnamefont {Monfardini}}, \bibinfo {author} {\bibfnamefont
  {L.}~\bibnamefont {Montier}}, \bibinfo {author} {\bibfnamefont
  {P.}~\bibnamefont {Natoli}}, \bibinfo {author} {\bibfnamefont
  {M.}~\bibnamefont {Negrello}}, \bibinfo {author} {\bibfnamefont
  {A.}~\bibnamefont {Notari}}, \bibinfo {author} {\bibfnamefont
  {F.}~\bibnamefont {Noviello}}, \bibinfo {author} {\bibfnamefont
  {F.}~\bibnamefont {Oppizzi}}, \bibinfo {author} {\bibfnamefont
  {C.}~\bibnamefont {O'Sullivan}}, \bibinfo {author} {\bibfnamefont
  {L.}~\bibnamefont {Pagano}}, \bibinfo {author} {\bibfnamefont
  {A.}~\bibnamefont {Paiella}}, \bibinfo {author} {\bibfnamefont
  {E.}~\bibnamefont {Pajer}}, \bibinfo {author} {\bibfnamefont
  {D.}~\bibnamefont {Paoletti}}, \bibinfo {author} {\bibfnamefont
  {S.}~\bibnamefont {Paradiso}}, \bibinfo {author} {\bibfnamefont {R.~B.}\
  \bibnamefont {Partridge}}, \bibinfo {author} {\bibfnamefont {G.}~\bibnamefont
  {Patanchon}}, \bibinfo {author} {\bibfnamefont {S.~P.}\ \bibnamefont
  {Patil}}, \bibinfo {author} {\bibfnamefont {O.}~\bibnamefont {Perdereau}},
  \bibinfo {author} {\bibfnamefont {F.}~\bibnamefont {Piacentini}}, \bibinfo
  {author} {\bibfnamefont {M.}~\bibnamefont {Piat}}, \bibinfo {author}
  {\bibfnamefont {G.}~\bibnamefont {Pisano}}, \bibinfo {author} {\bibfnamefont
  {L.}~\bibnamefont {Polastri}}, \bibinfo {author} {\bibfnamefont
  {G.}~\bibnamefont {Polenta}}, \bibinfo {author} {\bibfnamefont
  {A.}~\bibnamefont {Pollo}}, \bibinfo {author} {\bibfnamefont
  {N.}~\bibnamefont {Ponthieu}}, \bibinfo {author} {\bibfnamefont
  {V.}~\bibnamefont {Poulin}}, \bibinfo {author} {\bibfnamefont
  {D.}~\bibnamefont {Pr{\^{e}}le}}, \bibinfo {author} {\bibfnamefont
  {M.}~\bibnamefont {Quartin}}, \bibinfo {author} {\bibfnamefont
  {A.}~\bibnamefont {Ravenni}}, \bibinfo {author} {\bibfnamefont
  {M.}~\bibnamefont {Remazeilles}}, \bibinfo {author} {\bibfnamefont
  {A.}~\bibnamefont {Renzi}}, \bibinfo {author} {\bibfnamefont
  {C.}~\bibnamefont {Ringeval}}, \bibinfo {author} {\bibfnamefont
  {D.}~\bibnamefont {Roest}}, \bibinfo {author} {\bibfnamefont
  {M.}~\bibnamefont {Roman}}, \bibinfo {author} {\bibfnamefont {B.~F.}\
  \bibnamefont {Roukema}}, \bibinfo {author} {\bibfnamefont {J.~A.}\
  \bibnamefont {Rubino-Martin}}, \bibinfo {author} {\bibfnamefont
  {L.}~\bibnamefont {Salvati}}, \bibinfo {author} {\bibfnamefont
  {D.}~\bibnamefont {Scott}}, \bibinfo {author} {\bibfnamefont
  {S.}~\bibnamefont {Serjeant}}, \bibinfo {author} {\bibfnamefont
  {G.}~\bibnamefont {Signorelli}}, \bibinfo {author} {\bibfnamefont {A.~A.}\
  \bibnamefont {Starobinsky}}, \bibinfo {author} {\bibfnamefont
  {R.}~\bibnamefont {Sunyaev}}, \bibinfo {author} {\bibfnamefont {C.~Y.}\
  \bibnamefont {Tan}}, \bibinfo {author} {\bibfnamefont {A.}~\bibnamefont
  {Tartari}}, \bibinfo {author} {\bibfnamefont {G.}~\bibnamefont {Tasinato}},
  \bibinfo {author} {\bibfnamefont {L.}~\bibnamefont {Toffolatti}}, \bibinfo
  {author} {\bibfnamefont {M.}~\bibnamefont {Tomasi}}, \bibinfo {author}
  {\bibfnamefont {J.}~\bibnamefont {Torrado}}, \bibinfo {author} {\bibfnamefont
  {D.}~\bibnamefont {Tramonte}}, \bibinfo {author} {\bibfnamefont
  {N.}~\bibnamefont {Trappe}}, \bibinfo {author} {\bibfnamefont
  {S.}~\bibnamefont {Triqueneaux}}, \bibinfo {author} {\bibfnamefont
  {M.}~\bibnamefont {Tristram}}, \bibinfo {author} {\bibfnamefont
  {T.}~\bibnamefont {Trombetti}}, \bibinfo {author} {\bibfnamefont
  {M.}~\bibnamefont {Tucci}}, \bibinfo {author} {\bibfnamefont
  {C.}~\bibnamefont {Tucker}}, \bibinfo {author} {\bibfnamefont
  {J.}~\bibnamefont {Urrestilla}}, \bibinfo {author} {\bibfnamefont
  {J.}~\bibnamefont {V{\"{a}}liviita}}, \bibinfo {author} {\bibfnamefont
  {R.}~\bibnamefont {{Van de Weygaert}}}, \bibinfo {author} {\bibfnamefont
  {B.}~\bibnamefont {{Van Tent}}}, \bibinfo {author} {\bibfnamefont
  {V.}~\bibnamefont {Vennin}}, \bibinfo {author} {\bibfnamefont
  {L.}~\bibnamefont {Verde}}, \bibinfo {author} {\bibfnamefont
  {G.}~\bibnamefont {Vermeulen}}, \bibinfo {author} {\bibfnamefont
  {P.}~\bibnamefont {Vielva}}, \bibinfo {author} {\bibfnamefont
  {N.}~\bibnamefont {Vittorio}}, \bibinfo {author} {\bibfnamefont
  {F.}~\bibnamefont {Voisin}}, \bibinfo {author} {\bibfnamefont
  {C.}~\bibnamefont {Wallis}}, \bibinfo {author} {\bibfnamefont
  {B.}~\bibnamefont {Wandelt}}, \bibinfo {author} {\bibfnamefont
  {I.}~\bibnamefont {Wehus}}, \bibinfo {author} {\bibfnamefont
  {J.}~\bibnamefont {Weller}}, \bibinfo {author} {\bibfnamefont
  {K.}~\bibnamefont {Young}}, \bibinfo {author} {\bibfnamefont
  {M.}~\bibnamefont {Zannoni}}, \ and\ \bibinfo {author} {\bibfnamefont
  {f.~t.~C.}\ \bibnamefont {Collaboration}},\ }\href
  {http://arxiv.org/abs/1706.04516} {\bibfield  {journal} {\bibinfo  {journal}
  {eprint}\ } (\bibinfo {year} {2017})},\ \Eprint
  {http://arxiv.org/abs/1706.04516} {arXiv:1706.04516} \BibitemShut {NoStop}%
\bibitem [{\citenamefont {Hildebrand}\ \emph {et~al.}(1999)\citenamefont
  {Hildebrand}, \citenamefont {Dotson}, \citenamefont {Dowell}, \citenamefont
  {Schleuning},\ and\ \citenamefont {Vaillancourt}}]{Hildebrand1999}%
  \BibitemOpen
  \bibfield  {author} {\bibinfo {author} {\bibfnamefont {R.}~\bibnamefont
  {Hildebrand}}, \bibinfo {author} {\bibfnamefont {J.}~\bibnamefont {Dotson}},
  \bibinfo {author} {\bibfnamefont {C.}~\bibnamefont {Dowell}}, \bibinfo
  {author} {\bibfnamefont {D.}~\bibnamefont {Schleuning}}, \ and\ \bibinfo
  {author} {\bibfnamefont {J.}~\bibnamefont {Vaillancourt}},\ }\href {\doibase
  10.1086/307142} {\bibfield  {journal} {\bibinfo  {journal} {Astrophys. J.}\
  }\textbf {\bibinfo {volume} {516}},\ \bibinfo {pages} {834} (\bibinfo {year}
  {1999})}\BibitemShut {NoStop}%
\bibitem [{\citenamefont {Draine}(2004)}]{Draine2004}%
  \BibitemOpen
  \bibfield  {author} {\bibinfo {author} {\bibfnamefont {B.~T.}\ \bibnamefont
  {Draine}},\ }\href {\doibase 10.1002/chin.200335262} {\emph {\bibinfo {title}
  {Cold Universe}}}\ (\bibinfo {year} {2004})\ \Eprint
  {http://arxiv.org/abs/0304488} {arXiv:0304488 [astro-ph]} \BibitemShut
  {NoStop}%
\bibitem [{\citenamefont {Beno{\^{i}}t}\ \emph {et~al.}(2004)\citenamefont
  {Beno{\^{i}}t}, \citenamefont {Ade}, \citenamefont {Amblard}, \citenamefont
  {Ansari}, \citenamefont {Aubourg}, \citenamefont {Bargot}, \citenamefont
  {Bartlett}, \citenamefont {Bernard}, \citenamefont {Bhatia}, \citenamefont
  {Blanchard}, \citenamefont {Bock}, \citenamefont {Boscaleri}, \citenamefont
  {Bouchet}, \citenamefont {Bourrachot}, \citenamefont {Camus}, \citenamefont
  {Couchot}, \citenamefont {{De Bernardis}}, \citenamefont {Delabrouille},
  \citenamefont {D{\'{e}}sert}, \citenamefont {Dor{\'{e}}}, \citenamefont
  {Douspis}, \citenamefont {Dumoulin}, \citenamefont {Dupac}, \citenamefont
  {Filliatre}, \citenamefont {Fosalba}, \citenamefont {Ganga}, \citenamefont
  {Gannaway}, \citenamefont {Gautier}, \citenamefont {Giard}, \citenamefont
  {Giraud-H{\'{e}}raud}, \citenamefont {Gispert}, \citenamefont {Guglielmi},
  \citenamefont {Hamilton}, \citenamefont {Hanany}, \citenamefont
  {Henrot-Versill{\'{e}}}, \citenamefont {Kaplan}, \citenamefont {Lagache},
  \citenamefont {Lamarre}, \citenamefont {Lange}, \citenamefont
  {Mac{\'{i}}as-P{\'{e}}rez}, \citenamefont {Madet}, \citenamefont {Maffei},
  \citenamefont {Magneville}, \citenamefont {Marrone}, \citenamefont {Masi},
  \citenamefont {Mayet}, \citenamefont {Murphy}, \citenamefont {Naraghi},
  \citenamefont {Nati}, \citenamefont {Patanchon}, \citenamefont {Perrin},
  \citenamefont {Piat}, \citenamefont {Ponthieu}, \citenamefont {Prunet},
  \citenamefont {Puget}, \citenamefont {Renault}, \citenamefont {Rosset},
  \citenamefont {Santos}, \citenamefont {Starobinsky}, \citenamefont {Strukov},
  \citenamefont {Sudiwala}, \citenamefont {Teyssier}, \citenamefont {Tristram},
  \citenamefont {Tucker}, \citenamefont {Vanel}, \citenamefont {Vibert},
  \citenamefont {Wakui},\ and\ \citenamefont {Yvon}}]{Benoit2004}%
  \BibitemOpen
  \bibfield  {author} {\bibinfo {author} {\bibfnamefont {A.}~\bibnamefont
  {Beno{\^{i}}t}}, \bibinfo {author} {\bibfnamefont {P.}~\bibnamefont {Ade}},
  \bibinfo {author} {\bibfnamefont {A.}~\bibnamefont {Amblard}}, \bibinfo
  {author} {\bibfnamefont {R.}~\bibnamefont {Ansari}}, \bibinfo {author}
  {\bibfnamefont {{\'{E}}.}~\bibnamefont {Aubourg}}, \bibinfo {author}
  {\bibfnamefont {S.}~\bibnamefont {Bargot}}, \bibinfo {author} {\bibfnamefont
  {J.~G.}\ \bibnamefont {Bartlett}}, \bibinfo {author} {\bibfnamefont {J.-P.}\
  \bibnamefont {Bernard}}, \bibinfo {author} {\bibfnamefont {R.~S.}\
  \bibnamefont {Bhatia}}, \bibinfo {author} {\bibfnamefont {A.}~\bibnamefont
  {Blanchard}}, \bibinfo {author} {\bibfnamefont {J.~J.}\ \bibnamefont {Bock}},
  \bibinfo {author} {\bibfnamefont {A.}~\bibnamefont {Boscaleri}}, \bibinfo
  {author} {\bibfnamefont {F.~R.}\ \bibnamefont {Bouchet}}, \bibinfo {author}
  {\bibfnamefont {A.}~\bibnamefont {Bourrachot}}, \bibinfo {author}
  {\bibfnamefont {P.}~\bibnamefont {Camus}}, \bibinfo {author} {\bibfnamefont
  {F.}~\bibnamefont {Couchot}}, \bibinfo {author} {\bibfnamefont
  {P.}~\bibnamefont {{De Bernardis}}}, \bibinfo {author} {\bibfnamefont
  {J.}~\bibnamefont {Delabrouille}}, \bibinfo {author} {\bibfnamefont {F.-X.}\
  \bibnamefont {D{\'{e}}sert}}, \bibinfo {author} {\bibfnamefont
  {O.}~\bibnamefont {Dor{\'{e}}}}, \bibinfo {author} {\bibfnamefont
  {M.}~\bibnamefont {Douspis}}, \bibinfo {author} {\bibfnamefont
  {L.}~\bibnamefont {Dumoulin}}, \bibinfo {author} {\bibfnamefont
  {X.}~\bibnamefont {Dupac}}, \bibinfo {author} {\bibfnamefont
  {P.}~\bibnamefont {Filliatre}}, \bibinfo {author} {\bibfnamefont
  {P.}~\bibnamefont {Fosalba}}, \bibinfo {author} {\bibfnamefont
  {K.}~\bibnamefont {Ganga}}, \bibinfo {author} {\bibfnamefont
  {F.}~\bibnamefont {Gannaway}}, \bibinfo {author} {\bibfnamefont
  {B.}~\bibnamefont {Gautier}}, \bibinfo {author} {\bibfnamefont
  {M.}~\bibnamefont {Giard}}, \bibinfo {author} {\bibfnamefont
  {Y.}~\bibnamefont {Giraud-H{\'{e}}raud}}, \bibinfo {author} {\bibfnamefont
  {R.}~\bibnamefont {Gispert}}, \bibinfo {author} {\bibfnamefont
  {L.}~\bibnamefont {Guglielmi}}, \bibinfo {author} {\bibfnamefont {J.-C.}\
  \bibnamefont {Hamilton}}, \bibinfo {author} {\bibfnamefont {S.}~\bibnamefont
  {Hanany}}, \bibinfo {author} {\bibfnamefont {S.}~\bibnamefont
  {Henrot-Versill{\'{e}}}}, \bibinfo {author} {\bibfnamefont {J.}~\bibnamefont
  {Kaplan}}, \bibinfo {author} {\bibfnamefont {G.}~\bibnamefont {Lagache}},
  \bibinfo {author} {\bibfnamefont {J.-M.}\ \bibnamefont {Lamarre}}, \bibinfo
  {author} {\bibfnamefont {A.~E.}\ \bibnamefont {Lange}}, \bibinfo {author}
  {\bibfnamefont {J.~F.}\ \bibnamefont {Mac{\'{i}}as-P{\'{e}}rez}}, \bibinfo
  {author} {\bibfnamefont {K.}~\bibnamefont {Madet}}, \bibinfo {author}
  {\bibfnamefont {B.}~\bibnamefont {Maffei}}, \bibinfo {author} {\bibfnamefont
  {C.}~\bibnamefont {Magneville}}, \bibinfo {author} {\bibfnamefont {D.~P.}\
  \bibnamefont {Marrone}}, \bibinfo {author} {\bibfnamefont {S.}~\bibnamefont
  {Masi}}, \bibinfo {author} {\bibfnamefont {F.}~\bibnamefont {Mayet}},
  \bibinfo {author} {\bibfnamefont {A.}~\bibnamefont {Murphy}}, \bibinfo
  {author} {\bibfnamefont {F.}~\bibnamefont {Naraghi}}, \bibinfo {author}
  {\bibfnamefont {F.}~\bibnamefont {Nati}}, \bibinfo {author} {\bibfnamefont
  {G.}~\bibnamefont {Patanchon}}, \bibinfo {author} {\bibfnamefont
  {G.}~\bibnamefont {Perrin}}, \bibinfo {author} {\bibfnamefont
  {M.}~\bibnamefont {Piat}}, \bibinfo {author} {\bibfnamefont {N.}~\bibnamefont
  {Ponthieu}}, \bibinfo {author} {\bibfnamefont {S.}~\bibnamefont {Prunet}},
  \bibinfo {author} {\bibfnamefont {J.-L.}\ \bibnamefont {Puget}}, \bibinfo
  {author} {\bibfnamefont {C.}~\bibnamefont {Renault}}, \bibinfo {author}
  {\bibfnamefont {C.}~\bibnamefont {Rosset}}, \bibinfo {author} {\bibfnamefont
  {D.}~\bibnamefont {Santos}}, \bibinfo {author} {\bibfnamefont
  {A.}~\bibnamefont {Starobinsky}}, \bibinfo {author} {\bibfnamefont
  {I.}~\bibnamefont {Strukov}}, \bibinfo {author} {\bibfnamefont {R.~V.}\
  \bibnamefont {Sudiwala}}, \bibinfo {author} {\bibfnamefont {R.}~\bibnamefont
  {Teyssier}}, \bibinfo {author} {\bibfnamefont {M.}~\bibnamefont {Tristram}},
  \bibinfo {author} {\bibfnamefont {C.}~\bibnamefont {Tucker}}, \bibinfo
  {author} {\bibfnamefont {J.-C.}\ \bibnamefont {Vanel}}, \bibinfo {author}
  {\bibfnamefont {D.}~\bibnamefont {Vibert}}, \bibinfo {author} {\bibfnamefont
  {E.}~\bibnamefont {Wakui}}, \ and\ \bibinfo {author} {\bibfnamefont
  {D.}~\bibnamefont {Yvon}},\ }\href {\doibase 10.1051/0004-6361:20040042}
  {\bibfield  {journal} {\bibinfo  {journal} {Astron. Astrophys.}\ }\textbf
  {\bibinfo {volume} {424}},\ \bibinfo {pages} {571} (\bibinfo {year}
  {2004})}\BibitemShut {NoStop}%
\bibitem [{\citenamefont {Mortonson}\ \emph {et~al.}(2014)\citenamefont
  {Mortonson}, \citenamefont {Seljak}, \citenamefont {et~al.
  Planck~collaboration}, \citenamefont {Seljak}, \citenamefont {Seljak},
  \citenamefont {U.}, \citenamefont {{M. Kamionkowski}}, \citenamefont
  {Stebbins}, \citenamefont {Seljak}, \citenamefont {U.}, \citenamefont {et~al.
  BICEP2~collaboration}, \citenamefont {et~al. Planck~collaboration},
  \citenamefont {et~al. Planck~collaboration}, \citenamefont {Bernard},
  \citenamefont {J.-P.}, \citenamefont {et~al. BICEP2~collaboration},
  \citenamefont {Aumont}, \citenamefont {J.}, \citenamefont {et~al.
  Planck~collaboration}, \citenamefont {{R. Flauger}}, \citenamefont {Spergel},
  \citenamefont {et~al. Planck~collaboration}, \citenamefont {{D. Spergel}},
  \citenamefont {Hlozek}, \citenamefont {{A. Lewis}}, \citenamefont {Lasenby},
  \citenamefont {Bridle}, \citenamefont {S.}, \citenamefont {{C. Cheng}},
  \citenamefont {Wang}, \citenamefont {{B. Audren}}, \citenamefont {Tram},
  \citenamefont {et~al. Planck~collaboration}, \citenamefont {{U. Fuskeland,
  I.K. Wehus}}, \citenamefont {N{\ae}ss},\ and\ \citenamefont
  {et~Al.}}]{Mortonson2014}%
  \BibitemOpen
  \bibfield  {author} {\bibinfo {author} {\bibfnamefont {M.~J.}\ \bibnamefont
  {Mortonson}}, \bibinfo {author} {\bibfnamefont {U.}~\bibnamefont {Seljak}},
  \bibinfo {author} {\bibfnamefont {P.~A.}\ \bibnamefont {et~al.
  Planck~collaboration}}, \bibinfo {author} {\bibfnamefont {U.}~\bibnamefont
  {Seljak}}, \bibinfo {author} {\bibfnamefont {M.~Z.}\ \bibnamefont {Seljak}},
  \bibinfo {author} {\bibnamefont {U.}}, \bibinfo {author} {\bibfnamefont
  {A.~K.}\ \bibnamefont {{M. Kamionkowski}}}, \bibinfo {author} {\bibfnamefont
  {A.}~\bibnamefont {Stebbins}}, \bibinfo {author} {\bibfnamefont {M.~Z.}\
  \bibnamefont {Seljak}}, \bibinfo {author} {\bibnamefont {U.}}, \bibinfo
  {author} {\bibfnamefont {P.~A.}\ \bibnamefont {et~al. BICEP2~collaboration}},
  \bibinfo {author} {\bibfnamefont {P.~A.}\ \bibnamefont {et~al.
  Planck~collaboration}}, \bibinfo {author} {\bibfnamefont {R.~A.}\
  \bibnamefont {et~al. Planck~collaboration}}, \bibinfo {author} {\bibfnamefont
  {P.~c.}\ \bibnamefont {Bernard}}, \bibinfo {author} {\bibnamefont {J.-P.}},
  \bibinfo {author} {\bibfnamefont {P.~A.}\ \bibnamefont {et~al.
  BICEP2~collaboration}}, \bibinfo {author} {\bibfnamefont {P.~c.}\
  \bibnamefont {Aumont}}, \bibinfo {author} {\bibnamefont {J.}}, \bibinfo
  {author} {\bibfnamefont {P.~A.}\ \bibnamefont {et~al. Planck~collaboration}},
  \bibinfo {author} {\bibfnamefont {J.~H.}\ \bibnamefont {{R. Flauger}}},
  \bibinfo {author} {\bibfnamefont {D.}~\bibnamefont {Spergel}}, \bibinfo
  {author} {\bibfnamefont {P.~A.}\ \bibnamefont {et~al. Planck~collaboration}},
  \bibinfo {author} {\bibfnamefont {R.~F.}\ \bibnamefont {{D. Spergel}}},
  \bibinfo {author} {\bibfnamefont {R.}~\bibnamefont {Hlozek}}, \bibinfo
  {author} {\bibfnamefont {A.~C.}\ \bibnamefont {{A. Lewis}}}, \bibinfo
  {author} {\bibfnamefont {A.}~\bibnamefont {Lasenby}}, \bibinfo {author}
  {\bibfnamefont {A.~L.}\ \bibnamefont {Bridle}}, \bibinfo {author}
  {\bibnamefont {S.}}, \bibinfo {author} {\bibfnamefont {Q.-G.~H.}\
  \bibnamefont {{C. Cheng}}}, \bibinfo {author} {\bibfnamefont
  {S.}~\bibnamefont {Wang}}, \bibinfo {author} {\bibfnamefont {D.~F.}\
  \bibnamefont {{B. Audren}}}, \bibinfo {author} {\bibfnamefont
  {T.}~\bibnamefont {Tram}}, \bibinfo {author} {\bibfnamefont {A.~A.}\
  \bibnamefont {et~al. Planck~collaboration}}, \bibinfo {author} {\bibfnamefont
  {H.~E.}\ \bibnamefont {{U. Fuskeland, I.K. Wehus}}}, \bibinfo {author}
  {\bibfnamefont {S.}~\bibnamefont {N{\ae}ss}}, \ and\ \bibinfo {author}
  {\bibfnamefont {D.~B.}\ \bibnamefont {et~Al.}},\ }\href {\doibase
  10.1088/1475-7516/2014/10/035} {\bibfield  {journal} {\bibinfo  {journal} {J.
  Cosmol. Astropart. Phys.}\ ,\ \bibinfo {pages} {035}} (\bibinfo {year}
  {2014})}\BibitemShut {NoStop}%
\bibitem [{\citenamefont {Niemack}\ \emph {et~al.}(2015)\citenamefont
  {Niemack}, \citenamefont {Ade}, \citenamefont {de~Bernardis}, \citenamefont
  {Boulanger}, \citenamefont {Bryan}, \citenamefont {Devlin}, \citenamefont
  {Dunkley}, \citenamefont {Eales}, \citenamefont {Gomez}, \citenamefont
  {Groppi}, \citenamefont {Henderson}, \citenamefont {Hillbrand}, \citenamefont
  {Hubmayr}, \citenamefont {Mauskopf}, \citenamefont {McMahon}, \citenamefont
  {Miville-Desch{\^{e}}nes}, \citenamefont {Pascale}, \citenamefont {Pisano},
  \citenamefont {Novak}, \citenamefont {Scott}, \citenamefont {Soler},\ and\
  \citenamefont {Tucker}}]{Niemack2015}%
  \BibitemOpen
  \bibfield  {author} {\bibinfo {author} {\bibfnamefont {M.~D.}\ \bibnamefont
  {Niemack}}, \bibinfo {author} {\bibfnamefont {P.}~\bibnamefont {Ade}},
  \bibinfo {author} {\bibfnamefont {F.}~\bibnamefont {de~Bernardis}}, \bibinfo
  {author} {\bibfnamefont {F.}~\bibnamefont {Boulanger}}, \bibinfo {author}
  {\bibfnamefont {S.}~\bibnamefont {Bryan}}, \bibinfo {author} {\bibfnamefont
  {M.}~\bibnamefont {Devlin}}, \bibinfo {author} {\bibfnamefont
  {J.}~\bibnamefont {Dunkley}}, \bibinfo {author} {\bibfnamefont
  {S.}~\bibnamefont {Eales}}, \bibinfo {author} {\bibfnamefont
  {H.}~\bibnamefont {Gomez}}, \bibinfo {author} {\bibfnamefont
  {C.}~\bibnamefont {Groppi}}, \bibinfo {author} {\bibfnamefont
  {S.}~\bibnamefont {Henderson}}, \bibinfo {author} {\bibfnamefont
  {S.}~\bibnamefont {Hillbrand}}, \bibinfo {author} {\bibfnamefont
  {J.}~\bibnamefont {Hubmayr}}, \bibinfo {author} {\bibfnamefont
  {P.}~\bibnamefont {Mauskopf}}, \bibinfo {author} {\bibfnamefont
  {J.}~\bibnamefont {McMahon}}, \bibinfo {author} {\bibfnamefont {M.~A.}\
  \bibnamefont {Miville-Desch{\^{e}}nes}}, \bibinfo {author} {\bibfnamefont
  {E.}~\bibnamefont {Pascale}}, \bibinfo {author} {\bibfnamefont
  {G.}~\bibnamefont {Pisano}}, \bibinfo {author} {\bibfnamefont
  {G.}~\bibnamefont {Novak}}, \bibinfo {author} {\bibfnamefont
  {D.}~\bibnamefont {Scott}}, \bibinfo {author} {\bibfnamefont
  {J.}~\bibnamefont {Soler}}, \ and\ \bibinfo {author} {\bibfnamefont
  {C.}~\bibnamefont {Tucker}},\ }\href {\doibase 10.1007/s10909-015-1395-6}
  {\bibfield  {journal} {\bibinfo  {journal} {J. Low Temp. Phys.}\ }\textbf
  {\bibinfo {volume} {184}},\ \bibinfo {pages} {1} (\bibinfo {year} {2015})},\
  \Eprint {http://arxiv.org/abs/1509.05392} {arXiv:1509.05392} \BibitemShut
  {NoStop}%
\bibitem [{\citenamefont {{Planck Collaboration
  XXX}}(2016)}]{PlanckCollaborationXXX2016}%
  \BibitemOpen
  \bibfield  {author} {\bibinfo {author} {\bibnamefont {{Planck Collaboration
  XXX}}},\ }\href {\doibase 10.1051/0004-6361/201425034} {\bibfield  {journal}
  {\bibinfo  {journal} {Astron. Astrophys.}\ }\textbf {\bibinfo {volume}
  {586}},\ \bibinfo {pages} {A133} (\bibinfo {year} {2016})}\BibitemShut
  {NoStop}%
\bibitem [{\citenamefont {{Planck Collaboration
  L}}(2016)}]{PlanckCollaborationL2016}%
  \BibitemOpen
  \bibfield  {author} {\bibinfo {author} {\bibnamefont {{Planck Collaboration
  L}}},\ }\href {http://arxiv.org/abs/1606.07335} {\bibfield  {journal}
  {\bibinfo  {journal} {eprint}\ } (\bibinfo {year} {2016})},\ \Eprint
  {http://arxiv.org/abs/1606.07335} {arXiv:1606.07335} \BibitemShut {NoStop}%
\bibitem [{\citenamefont {Krachmalnicoff}\ \emph {et~al.}(2016)\citenamefont
  {Krachmalnicoff}, \citenamefont {Baccigalupi}, \citenamefont {Aumont},
  \citenamefont {Bersanelli},\ and\ \citenamefont
  {Mennella}}]{Krachmalnicoff2016}%
  \BibitemOpen
  \bibfield  {author} {\bibinfo {author} {\bibfnamefont {N.}~\bibnamefont
  {Krachmalnicoff}}, \bibinfo {author} {\bibfnamefont {C.}~\bibnamefont
  {Baccigalupi}}, \bibinfo {author} {\bibfnamefont {J.}~\bibnamefont {Aumont}},
  \bibinfo {author} {\bibfnamefont {M.}~\bibnamefont {Bersanelli}}, \ and\
  \bibinfo {author} {\bibfnamefont {A.}~\bibnamefont {Mennella}},\ }\href
  {\doibase 10.1051/0004-6361/201527678} {\bibfield  {journal} {\bibinfo
  {journal} {Astron. Astrophys.}\ }\textbf {\bibinfo {volume} {588}},\ \bibinfo
  {pages} {A65} (\bibinfo {year} {2016})}\BibitemShut {NoStop}%
\bibitem [{\citenamefont {Hanson}\ \emph {et~al.}(2013)\citenamefont {Hanson},
  \citenamefont {Hoover}, \citenamefont {Crites}, \citenamefont {Ade},
  \citenamefont {Aird}, \citenamefont {Austermann}, \citenamefont {Beall},
  \citenamefont {Bender}, \citenamefont {Benson}, \citenamefont {Bleem},
  \citenamefont {Bock}, \citenamefont {Carlstrom}, \citenamefont {Chang},
  \citenamefont {Chiang}, \citenamefont {Cho}, \citenamefont {Conley},
  \citenamefont {Crawford}, \citenamefont {de~Haan}, \citenamefont {Dobbs},
  \citenamefont {Everett}, \citenamefont {Gallicchio}, \citenamefont {Gao},
  \citenamefont {George}, \citenamefont {Halverson}, \citenamefont
  {Harrington}, \citenamefont {Henning}, \citenamefont {Hilton}, \citenamefont
  {Holder}, \citenamefont {Holzapfel}, \citenamefont {Hrubes}, \citenamefont
  {Huang}, \citenamefont {Hubmayr}, \citenamefont {Irwin}, \citenamefont
  {Keisler}, \citenamefont {Knox}, \citenamefont {Lee}, \citenamefont {Leitch},
  \citenamefont {Li}, \citenamefont {Liang}, \citenamefont {Luong-Van},
  \citenamefont {Marsden}, \citenamefont {McMahon}, \citenamefont {Mehl},
  \citenamefont {Meyer}, \citenamefont {Mocanu}, \citenamefont {Montroy},
  \citenamefont {Natoli}, \citenamefont {Nibarger}, \citenamefont {Novosad},
  \citenamefont {Padin}, \citenamefont {Pryke}, \citenamefont {Reichardt},
  \citenamefont {Ruhl}, \citenamefont {Saliwanchik}, \citenamefont {Sayre},
  \citenamefont {Schaffer}, \citenamefont {Schulz}, \citenamefont {Smecher},
  \citenamefont {Stark}, \citenamefont {Story}, \citenamefont {Tucker},
  \citenamefont {Vanderlinde}, \citenamefont {Vieira}, \citenamefont {Viero},
  \citenamefont {Wang}, \citenamefont {Yefremenko}, \citenamefont {Zahn},\ and\
  \citenamefont {Zemcov}}]{Hanson2013}%
  \BibitemOpen
  \bibfield  {author} {\bibinfo {author} {\bibfnamefont {D.}~\bibnamefont
  {Hanson}}, \bibinfo {author} {\bibfnamefont {S.}~\bibnamefont {Hoover}},
  \bibinfo {author} {\bibfnamefont {A.}~\bibnamefont {Crites}}, \bibinfo
  {author} {\bibfnamefont {P.~A.~R.}\ \bibnamefont {Ade}}, \bibinfo {author}
  {\bibfnamefont {K.~A.}\ \bibnamefont {Aird}}, \bibinfo {author}
  {\bibfnamefont {J.~E.}\ \bibnamefont {Austermann}}, \bibinfo {author}
  {\bibfnamefont {J.~A.}\ \bibnamefont {Beall}}, \bibinfo {author}
  {\bibfnamefont {A.~N.}\ \bibnamefont {Bender}}, \bibinfo {author}
  {\bibfnamefont {B.~A.}\ \bibnamefont {Benson}}, \bibinfo {author}
  {\bibfnamefont {L.~E.}\ \bibnamefont {Bleem}}, \bibinfo {author}
  {\bibfnamefont {J.~J.}\ \bibnamefont {Bock}}, \bibinfo {author}
  {\bibfnamefont {J.~E.}\ \bibnamefont {Carlstrom}}, \bibinfo {author}
  {\bibfnamefont {C.~L.}\ \bibnamefont {Chang}}, \bibinfo {author}
  {\bibfnamefont {H.~C.}\ \bibnamefont {Chiang}}, \bibinfo {author}
  {\bibfnamefont {H.-M.}\ \bibnamefont {Cho}}, \bibinfo {author} {\bibfnamefont
  {A.}~\bibnamefont {Conley}}, \bibinfo {author} {\bibfnamefont {T.~M.}\
  \bibnamefont {Crawford}}, \bibinfo {author} {\bibfnamefont {T.}~\bibnamefont
  {de~Haan}}, \bibinfo {author} {\bibfnamefont {M.~A.}\ \bibnamefont {Dobbs}},
  \bibinfo {author} {\bibfnamefont {W.}~\bibnamefont {Everett}}, \bibinfo
  {author} {\bibfnamefont {J.}~\bibnamefont {Gallicchio}}, \bibinfo {author}
  {\bibfnamefont {J.}~\bibnamefont {Gao}}, \bibinfo {author} {\bibfnamefont
  {E.~M.}\ \bibnamefont {George}}, \bibinfo {author} {\bibfnamefont {N.~W.}\
  \bibnamefont {Halverson}}, \bibinfo {author} {\bibfnamefont {N.}~\bibnamefont
  {Harrington}}, \bibinfo {author} {\bibfnamefont {J.~W.}\ \bibnamefont
  {Henning}}, \bibinfo {author} {\bibfnamefont {G.~C.}\ \bibnamefont {Hilton}},
  \bibinfo {author} {\bibfnamefont {G.~P.}\ \bibnamefont {Holder}}, \bibinfo
  {author} {\bibfnamefont {W.~L.}\ \bibnamefont {Holzapfel}}, \bibinfo {author}
  {\bibfnamefont {J.~D.}\ \bibnamefont {Hrubes}}, \bibinfo {author}
  {\bibfnamefont {N.}~\bibnamefont {Huang}}, \bibinfo {author} {\bibfnamefont
  {J.}~\bibnamefont {Hubmayr}}, \bibinfo {author} {\bibfnamefont {K.~D.}\
  \bibnamefont {Irwin}}, \bibinfo {author} {\bibfnamefont {R.}~\bibnamefont
  {Keisler}}, \bibinfo {author} {\bibfnamefont {L.}~\bibnamefont {Knox}},
  \bibinfo {author} {\bibfnamefont {A.~T.}\ \bibnamefont {Lee}}, \bibinfo
  {author} {\bibfnamefont {E.}~\bibnamefont {Leitch}}, \bibinfo {author}
  {\bibfnamefont {D.}~\bibnamefont {Li}}, \bibinfo {author} {\bibfnamefont
  {C.}~\bibnamefont {Liang}}, \bibinfo {author} {\bibfnamefont
  {D.}~\bibnamefont {Luong-Van}}, \bibinfo {author} {\bibfnamefont
  {G.}~\bibnamefont {Marsden}}, \bibinfo {author} {\bibfnamefont {J.~J.}\
  \bibnamefont {McMahon}}, \bibinfo {author} {\bibfnamefont {J.}~\bibnamefont
  {Mehl}}, \bibinfo {author} {\bibfnamefont {S.~S.}\ \bibnamefont {Meyer}},
  \bibinfo {author} {\bibfnamefont {L.}~\bibnamefont {Mocanu}}, \bibinfo
  {author} {\bibfnamefont {T.~E.}\ \bibnamefont {Montroy}}, \bibinfo {author}
  {\bibfnamefont {T.}~\bibnamefont {Natoli}}, \bibinfo {author} {\bibfnamefont
  {J.~P.}\ \bibnamefont {Nibarger}}, \bibinfo {author} {\bibfnamefont
  {V.}~\bibnamefont {Novosad}}, \bibinfo {author} {\bibfnamefont
  {S.}~\bibnamefont {Padin}}, \bibinfo {author} {\bibfnamefont
  {C.}~\bibnamefont {Pryke}}, \bibinfo {author} {\bibfnamefont {C.~L.}\
  \bibnamefont {Reichardt}}, \bibinfo {author} {\bibfnamefont {J.~E.}\
  \bibnamefont {Ruhl}}, \bibinfo {author} {\bibfnamefont {B.~R.}\ \bibnamefont
  {Saliwanchik}}, \bibinfo {author} {\bibfnamefont {J.~T.}\ \bibnamefont
  {Sayre}}, \bibinfo {author} {\bibfnamefont {K.~K.}\ \bibnamefont {Schaffer}},
  \bibinfo {author} {\bibfnamefont {B.}~\bibnamefont {Schulz}}, \bibinfo
  {author} {\bibfnamefont {G.}~\bibnamefont {Smecher}}, \bibinfo {author}
  {\bibfnamefont {A.~A.}\ \bibnamefont {Stark}}, \bibinfo {author}
  {\bibfnamefont {K.~T.}\ \bibnamefont {Story}}, \bibinfo {author}
  {\bibfnamefont {C.}~\bibnamefont {Tucker}}, \bibinfo {author} {\bibfnamefont
  {K.}~\bibnamefont {Vanderlinde}}, \bibinfo {author} {\bibfnamefont {J.~D.}\
  \bibnamefont {Vieira}}, \bibinfo {author} {\bibfnamefont {M.~P.}\
  \bibnamefont {Viero}}, \bibinfo {author} {\bibfnamefont {G.}~\bibnamefont
  {Wang}}, \bibinfo {author} {\bibfnamefont {V.}~\bibnamefont {Yefremenko}},
  \bibinfo {author} {\bibfnamefont {O.}~\bibnamefont {Zahn}}, \ and\ \bibinfo
  {author} {\bibfnamefont {M.}~\bibnamefont {Zemcov}},\ }\href {\doibase
  10.1103/PhysRevLett.111.141301} {\bibfield  {journal} {\bibinfo  {journal}
  {Phys. Rev. Lett.}\ }\textbf {\bibinfo {volume} {111}},\ \bibinfo {pages}
  {141301} (\bibinfo {year} {2013})}\BibitemShut {NoStop}%
\bibitem [{\citenamefont {{POLARBEAR Collaboration}}(2014)}]{Ade2014}%
  \BibitemOpen
  \bibfield  {author} {\bibinfo {author} {\bibnamefont {{POLARBEAR
  Collaboration}}},\ }\href {\doibase 10.1103/PhysRevLett.112.131302}
  {\bibfield  {journal} {\bibinfo  {journal} {Phys. Rev. Lett.}\ }\textbf
  {\bibinfo {volume} {112}},\ \bibinfo {pages} {131302} (\bibinfo {year}
  {2014})},\ \Eprint {http://arxiv.org/abs/1312.6645} {arXiv:1312.6645}
  \BibitemShut {NoStop}%
\bibitem [{\citenamefont {van Engelen}\ \emph {et~al.}(2015)\citenamefont {van
  Engelen}, \citenamefont {Sherwin}, \citenamefont {Sehgal}, \citenamefont
  {Addison}, \citenamefont {Allison}, \citenamefont {Battaglia}, \citenamefont
  {de~Bernardis}, \citenamefont {Bond}, \citenamefont {Calabrese},
  \citenamefont {Coughlin}, \citenamefont {Crichton}, \citenamefont {Datta},
  \citenamefont {Devlin}, \citenamefont {Dunkley}, \citenamefont
  {D{\"{u}}nner}, \citenamefont {Gallardo}, \citenamefont {Grace},
  \citenamefont {Gralla}, \citenamefont {Hajian}, \citenamefont {Hasselfield},
  \citenamefont {Henderson}, \citenamefont {Hill}, \citenamefont {Hilton},
  \citenamefont {Hincks}, \citenamefont {Hlozek}, \citenamefont {Huffenberger},
  \citenamefont {Hughes}, \citenamefont {Koopman}, \citenamefont {Kosowsky},
  \citenamefont {Louis}, \citenamefont {Lungu}, \citenamefont {Madhavacheril},
  \citenamefont {Maurin}, \citenamefont {McMahon}, \citenamefont {Moodley},
  \citenamefont {Munson}, \citenamefont {Naess}, \citenamefont {Nati},
  \citenamefont {Newburgh}, \citenamefont {Niemack}, \citenamefont {Nolta},
  \citenamefont {Page}, \citenamefont {Pappas}, \citenamefont {Partridge},
  \citenamefont {Schmitt}, \citenamefont {Sievers}, \citenamefont {Simon},
  \citenamefont {Spergel}, \citenamefont {Staggs}, \citenamefont {Switzer},
  \citenamefont {Ward},\ and\ \citenamefont {Wollack}}]{VanEngelen2014}%
  \BibitemOpen
  \bibfield  {author} {\bibinfo {author} {\bibfnamefont {A.}~\bibnamefont {van
  Engelen}}, \bibinfo {author} {\bibfnamefont {B.~D.}\ \bibnamefont {Sherwin}},
  \bibinfo {author} {\bibfnamefont {N.}~\bibnamefont {Sehgal}}, \bibinfo
  {author} {\bibfnamefont {G.~E.}\ \bibnamefont {Addison}}, \bibinfo {author}
  {\bibfnamefont {R.}~\bibnamefont {Allison}}, \bibinfo {author} {\bibfnamefont
  {N.}~\bibnamefont {Battaglia}}, \bibinfo {author} {\bibfnamefont
  {F.}~\bibnamefont {de~Bernardis}}, \bibinfo {author} {\bibfnamefont {J.~R.}\
  \bibnamefont {Bond}}, \bibinfo {author} {\bibfnamefont {E.}~\bibnamefont
  {Calabrese}}, \bibinfo {author} {\bibfnamefont {K.}~\bibnamefont {Coughlin}},
  \bibinfo {author} {\bibfnamefont {D.}~\bibnamefont {Crichton}}, \bibinfo
  {author} {\bibfnamefont {R.}~\bibnamefont {Datta}}, \bibinfo {author}
  {\bibfnamefont {M.~J.}\ \bibnamefont {Devlin}}, \bibinfo {author}
  {\bibfnamefont {J.}~\bibnamefont {Dunkley}}, \bibinfo {author} {\bibfnamefont
  {R.}~\bibnamefont {D{\"{u}}nner}}, \bibinfo {author} {\bibfnamefont
  {P.}~\bibnamefont {Gallardo}}, \bibinfo {author} {\bibfnamefont
  {E.}~\bibnamefont {Grace}}, \bibinfo {author} {\bibfnamefont
  {M.}~\bibnamefont {Gralla}}, \bibinfo {author} {\bibfnamefont
  {A.}~\bibnamefont {Hajian}}, \bibinfo {author} {\bibfnamefont
  {M.}~\bibnamefont {Hasselfield}}, \bibinfo {author} {\bibfnamefont
  {S.}~\bibnamefont {Henderson}}, \bibinfo {author} {\bibfnamefont {J.~C.}\
  \bibnamefont {Hill}}, \bibinfo {author} {\bibfnamefont {M.}~\bibnamefont
  {Hilton}}, \bibinfo {author} {\bibfnamefont {A.~D.}\ \bibnamefont {Hincks}},
  \bibinfo {author} {\bibfnamefont {R.}~\bibnamefont {Hlozek}}, \bibinfo
  {author} {\bibfnamefont {K.~M.}\ \bibnamefont {Huffenberger}}, \bibinfo
  {author} {\bibfnamefont {J.~P.}\ \bibnamefont {Hughes}}, \bibinfo {author}
  {\bibfnamefont {B.}~\bibnamefont {Koopman}}, \bibinfo {author} {\bibfnamefont
  {A.}~\bibnamefont {Kosowsky}}, \bibinfo {author} {\bibfnamefont
  {T.}~\bibnamefont {Louis}}, \bibinfo {author} {\bibfnamefont
  {M.}~\bibnamefont {Lungu}}, \bibinfo {author} {\bibfnamefont
  {M.}~\bibnamefont {Madhavacheril}}, \bibinfo {author} {\bibfnamefont
  {L.}~\bibnamefont {Maurin}}, \bibinfo {author} {\bibfnamefont
  {J.}~\bibnamefont {McMahon}}, \bibinfo {author} {\bibfnamefont
  {K.}~\bibnamefont {Moodley}}, \bibinfo {author} {\bibfnamefont
  {C.}~\bibnamefont {Munson}}, \bibinfo {author} {\bibfnamefont
  {S.}~\bibnamefont {Naess}}, \bibinfo {author} {\bibfnamefont
  {F.}~\bibnamefont {Nati}}, \bibinfo {author} {\bibfnamefont {L.}~\bibnamefont
  {Newburgh}}, \bibinfo {author} {\bibfnamefont {M.~D.}\ \bibnamefont
  {Niemack}}, \bibinfo {author} {\bibfnamefont {M.~R.}\ \bibnamefont {Nolta}},
  \bibinfo {author} {\bibfnamefont {L.~A.}\ \bibnamefont {Page}}, \bibinfo
  {author} {\bibfnamefont {C.}~\bibnamefont {Pappas}}, \bibinfo {author}
  {\bibfnamefont {B.}~\bibnamefont {Partridge}}, \bibinfo {author}
  {\bibfnamefont {B.~L.}\ \bibnamefont {Schmitt}}, \bibinfo {author}
  {\bibfnamefont {J.~L.}\ \bibnamefont {Sievers}}, \bibinfo {author}
  {\bibfnamefont {S.}~\bibnamefont {Simon}}, \bibinfo {author} {\bibfnamefont
  {D.~N.}\ \bibnamefont {Spergel}}, \bibinfo {author} {\bibfnamefont {S.~T.}\
  \bibnamefont {Staggs}}, \bibinfo {author} {\bibfnamefont {E.~R.}\
  \bibnamefont {Switzer}}, \bibinfo {author} {\bibfnamefont {J.~T.}\
  \bibnamefont {Ward}}, \ and\ \bibinfo {author} {\bibfnamefont {E.~J.}\
  \bibnamefont {Wollack}},\ }\href {\doibase 10.1088/0004-637X/808/1/7}
  {\bibfield  {journal} {\bibinfo  {journal} {Astrophys. J.}\ }\textbf
  {\bibinfo {volume} {808}},\ \bibinfo {pages} {7} (\bibinfo {year} {2015})},\
  \Eprint {http://arxiv.org/abs/1412.0626} {arXiv:1412.0626} \BibitemShut
  {NoStop}%
\bibitem [{\citenamefont {Story}\ \emph {et~al.}(2015)\citenamefont {Story},
  \citenamefont {Hanson}, \citenamefont {Ade}, \citenamefont {Aird},
  \citenamefont {Austermann}, \citenamefont {{J. A. Beall}}, \citenamefont
  {Bender}, \citenamefont {Benson}, \citenamefont {Bleem}, \citenamefont
  {Carlstrom}, \citenamefont {Chang}, \citenamefont {Chiang}, \citenamefont
  {Cho}, \citenamefont {Citron}, \citenamefont {Crawford}, \citenamefont
  {Crites}, \citenamefont {de~Haan}, \citenamefont {Dobbs}, \citenamefont
  {Everett}, \citenamefont {Gallicchio}, \citenamefont {Gao}, \citenamefont
  {George}, \citenamefont {Gilbert}, \citenamefont {Halverson}, \citenamefont
  {Harrington}, \citenamefont {Henning}, \citenamefont {Hilton}, \citenamefont
  {Holder}, \citenamefont {Holzapfel}, \citenamefont {Hoover}, \citenamefont
  {Hou}, \citenamefont {Hrubes}, \citenamefont {Huang}, \citenamefont
  {Hubmayr}, \citenamefont {Irwin}, \citenamefont {Keisler}, \citenamefont
  {Knox}, \citenamefont {Lee}, \citenamefont {Leitch}, \citenamefont {Li},
  \citenamefont {Liang}, \citenamefont {Luong-Van}, \citenamefont {McMahon},
  \citenamefont {Mehl}, \citenamefont {Meyer}, \citenamefont {Mocanu},
  \citenamefont {Montroy}, \citenamefont {Natoli}, \citenamefont {Nibarger},
  \citenamefont {Novosad}, \citenamefont {Padin}, \citenamefont {Pryke},
  \citenamefont {Reichardt}, \citenamefont {Ruhl}, \citenamefont {Saliwanchik},
  \citenamefont {Sayre}, \citenamefont {Schaffer}, \citenamefont {Smecher},
  \citenamefont {Stark}, \citenamefont {Tucker}, \citenamefont {Vanderlinde},
  \citenamefont {Vieira}, \citenamefont {Wang}, \citenamefont {Whitehorn},
  \citenamefont {Yefremenko},\ and\ \citenamefont {Zahn}}]{Story2014}%
  \BibitemOpen
  \bibfield  {author} {\bibinfo {author} {\bibfnamefont {K.~T.}\ \bibnamefont
  {Story}}, \bibinfo {author} {\bibfnamefont {D.}~\bibnamefont {Hanson}},
  \bibinfo {author} {\bibfnamefont {P.~A.~R.}\ \bibnamefont {Ade}}, \bibinfo
  {author} {\bibfnamefont {K.~A.}\ \bibnamefont {Aird}}, \bibinfo {author}
  {\bibfnamefont {J.~E.}\ \bibnamefont {Austermann}}, \bibinfo {author}
  {\bibnamefont {{J. A. Beall}}}, \bibinfo {author} {\bibfnamefont {A.~N.}\
  \bibnamefont {Bender}}, \bibinfo {author} {\bibfnamefont {B.~A.}\
  \bibnamefont {Benson}}, \bibinfo {author} {\bibfnamefont {L.~E.}\
  \bibnamefont {Bleem}}, \bibinfo {author} {\bibfnamefont {J.~E.}\ \bibnamefont
  {Carlstrom}}, \bibinfo {author} {\bibfnamefont {C.~L.}\ \bibnamefont
  {Chang}}, \bibinfo {author} {\bibfnamefont {H.~C.}\ \bibnamefont {Chiang}},
  \bibinfo {author} {\bibfnamefont {H.-M.}\ \bibnamefont {Cho}}, \bibinfo
  {author} {\bibfnamefont {R.}~\bibnamefont {Citron}}, \bibinfo {author}
  {\bibfnamefont {T.~M.}\ \bibnamefont {Crawford}}, \bibinfo {author}
  {\bibfnamefont {A.~T.}\ \bibnamefont {Crites}}, \bibinfo {author}
  {\bibfnamefont {T.}~\bibnamefont {de~Haan}}, \bibinfo {author} {\bibfnamefont
  {M.~A.}\ \bibnamefont {Dobbs}}, \bibinfo {author} {\bibfnamefont
  {W.}~\bibnamefont {Everett}}, \bibinfo {author} {\bibfnamefont
  {J.}~\bibnamefont {Gallicchio}}, \bibinfo {author} {\bibfnamefont
  {J.}~\bibnamefont {Gao}}, \bibinfo {author} {\bibfnamefont {E.~M.}\
  \bibnamefont {George}}, \bibinfo {author} {\bibfnamefont {A.}~\bibnamefont
  {Gilbert}}, \bibinfo {author} {\bibfnamefont {N.~W.}\ \bibnamefont
  {Halverson}}, \bibinfo {author} {\bibfnamefont {N.}~\bibnamefont
  {Harrington}}, \bibinfo {author} {\bibfnamefont {J.~W.}\ \bibnamefont
  {Henning}}, \bibinfo {author} {\bibfnamefont {G.~C.}\ \bibnamefont {Hilton}},
  \bibinfo {author} {\bibfnamefont {G.~P.}\ \bibnamefont {Holder}}, \bibinfo
  {author} {\bibfnamefont {W.~L.}\ \bibnamefont {Holzapfel}}, \bibinfo {author}
  {\bibfnamefont {S.}~\bibnamefont {Hoover}}, \bibinfo {author} {\bibfnamefont
  {Z.}~\bibnamefont {Hou}}, \bibinfo {author} {\bibfnamefont {J.~D.}\
  \bibnamefont {Hrubes}}, \bibinfo {author} {\bibfnamefont {N.}~\bibnamefont
  {Huang}}, \bibinfo {author} {\bibfnamefont {J.}~\bibnamefont {Hubmayr}},
  \bibinfo {author} {\bibfnamefont {K.~D.}\ \bibnamefont {Irwin}}, \bibinfo
  {author} {\bibfnamefont {R.}~\bibnamefont {Keisler}}, \bibinfo {author}
  {\bibfnamefont {L.}~\bibnamefont {Knox}}, \bibinfo {author} {\bibfnamefont
  {A.~T.}\ \bibnamefont {Lee}}, \bibinfo {author} {\bibfnamefont {E.~M.}\
  \bibnamefont {Leitch}}, \bibinfo {author} {\bibfnamefont {D.}~\bibnamefont
  {Li}}, \bibinfo {author} {\bibfnamefont {C.}~\bibnamefont {Liang}}, \bibinfo
  {author} {\bibfnamefont {D.}~\bibnamefont {Luong-Van}}, \bibinfo {author}
  {\bibfnamefont {J.~J.}\ \bibnamefont {McMahon}}, \bibinfo {author}
  {\bibfnamefont {J.}~\bibnamefont {Mehl}}, \bibinfo {author} {\bibfnamefont
  {S.~S.}\ \bibnamefont {Meyer}}, \bibinfo {author} {\bibfnamefont
  {L.}~\bibnamefont {Mocanu}}, \bibinfo {author} {\bibfnamefont {T.~E.}\
  \bibnamefont {Montroy}}, \bibinfo {author} {\bibfnamefont {T.}~\bibnamefont
  {Natoli}}, \bibinfo {author} {\bibfnamefont {J.~P.}\ \bibnamefont
  {Nibarger}}, \bibinfo {author} {\bibfnamefont {V.}~\bibnamefont {Novosad}},
  \bibinfo {author} {\bibfnamefont {S.}~\bibnamefont {Padin}}, \bibinfo
  {author} {\bibfnamefont {C.}~\bibnamefont {Pryke}}, \bibinfo {author}
  {\bibfnamefont {C.~L.}\ \bibnamefont {Reichardt}}, \bibinfo {author}
  {\bibfnamefont {J.~E.}\ \bibnamefont {Ruhl}}, \bibinfo {author}
  {\bibfnamefont {B.~R.}\ \bibnamefont {Saliwanchik}}, \bibinfo {author}
  {\bibfnamefont {J.~T.}\ \bibnamefont {Sayre}}, \bibinfo {author}
  {\bibfnamefont {K.~K.}\ \bibnamefont {Schaffer}}, \bibinfo {author}
  {\bibfnamefont {G.}~\bibnamefont {Smecher}}, \bibinfo {author} {\bibfnamefont
  {A.~A.}\ \bibnamefont {Stark}}, \bibinfo {author} {\bibfnamefont
  {C.}~\bibnamefont {Tucker}}, \bibinfo {author} {\bibfnamefont
  {K.}~\bibnamefont {Vanderlinde}}, \bibinfo {author} {\bibfnamefont {J.~D.}\
  \bibnamefont {Vieira}}, \bibinfo {author} {\bibfnamefont {G.}~\bibnamefont
  {Wang}}, \bibinfo {author} {\bibfnamefont {N.}~\bibnamefont {Whitehorn}},
  \bibinfo {author} {\bibfnamefont {V.}~\bibnamefont {Yefremenko}}, \ and\
  \bibinfo {author} {\bibfnamefont {O.}~\bibnamefont {Zahn}},\ }\href {\doibase
  10.1088/0004-637X/810/1/50} {\bibfield  {journal} {\bibinfo  {journal}
  {Astrophys. J.}\ }\textbf {\bibinfo {volume} {810}},\ \bibinfo {pages} {50}
  (\bibinfo {year} {2015})},\ \Eprint {http://arxiv.org/abs/1412.4760}
  {arXiv:1412.4760} \BibitemShut {NoStop}%
\bibitem [{\citenamefont {{Planck Collaboration
  XV}}(2016)}]{PlanckCollaboration2015a}%
  \BibitemOpen
  \bibfield  {author} {\bibinfo {author} {\bibnamefont {{Planck Collaboration
  XV}}},\ }\href {\doibase 10.1051/0004-6361/201525941} {\bibfield  {journal}
  {\bibinfo  {journal} {Astron. Astrophys.}\ }\textbf {\bibinfo {volume}
  {594}},\ \bibinfo {pages} {A15} (\bibinfo {year} {2016})},\ \Eprint
  {http://arxiv.org/abs/1502.01591} {arXiv:1502.01591} \BibitemShut {NoStop}%
\bibitem [{\citenamefont {Lewis}\ and\ \citenamefont
  {Challinor}(2006)}]{Lewis2006}%
  \BibitemOpen
  \bibfield  {author} {\bibinfo {author} {\bibfnamefont {A.}~\bibnamefont
  {Lewis}}\ and\ \bibinfo {author} {\bibfnamefont {A.}~\bibnamefont
  {Challinor}},\ }\href {\doibase 10.1016/j.physrep.2006.03.002} {\bibfield
  {journal} {\bibinfo  {journal} {Phys. Rep.}\ }\textbf {\bibinfo {volume}
  {429}},\ \bibinfo {pages} {1} (\bibinfo {year} {2006})},\ \Eprint
  {http://arxiv.org/abs/0601594} {arXiv:0601594 [astro-ph]} \BibitemShut
  {NoStop}%
\bibitem [{\citenamefont {Sherwin}\ and\ \citenamefont
  {Schmittfull}(2015)}]{Sherwin2015}%
  \BibitemOpen
  \bibfield  {author} {\bibinfo {author} {\bibfnamefont {B.~D.}\ \bibnamefont
  {Sherwin}}\ and\ \bibinfo {author} {\bibfnamefont {M.}~\bibnamefont
  {Schmittfull}},\ }\href {\doibase 10.1103/PhysRevD.92.043005} {\bibfield
  {journal} {\bibinfo  {journal} {Phys. Rev. D}\ }\textbf {\bibinfo {volume}
  {92}},\ \bibinfo {pages} {43005} (\bibinfo {year} {2015})},\ \Eprint
  {http://arxiv.org/abs/arXiv:1502.05356v1} {arXiv:arXiv:1502.05356v1}
  \BibitemShut {NoStop}%
\bibitem [{\citenamefont {Knox}\ and\ \citenamefont {Song}(2002)}]{Knox2002}%
  \BibitemOpen
  \bibfield  {author} {\bibinfo {author} {\bibfnamefont {L.}~\bibnamefont
  {Knox}}\ and\ \bibinfo {author} {\bibfnamefont {Y.-S.}\ \bibnamefont
  {Song}},\ }\href {\doibase 10.1103/PhysRevLett.89.011303} {\bibfield
  {journal} {\bibinfo  {journal} {Phys. Rev. Lett.}\ }\textbf {\bibinfo
  {volume} {89}},\ \bibinfo {pages} {11303} (\bibinfo {year} {2002})},\ \Eprint
  {http://arxiv.org/abs/0202286} {arXiv:0202286 [astro-ph]} \BibitemShut
  {NoStop}%
\bibitem [{\citenamefont {Kesden}\ \emph {et~al.}(2002)\citenamefont {Kesden},
  \citenamefont {Cooray},\ and\ \citenamefont {Kamionkowski}}]{Kesden2002}%
  \BibitemOpen
  \bibfield  {author} {\bibinfo {author} {\bibfnamefont {M.}~\bibnamefont
  {Kesden}}, \bibinfo {author} {\bibfnamefont {A.}~\bibnamefont {Cooray}}, \
  and\ \bibinfo {author} {\bibfnamefont {M.}~\bibnamefont {Kamionkowski}},\
  }\href {\doibase 10.1103/PhysRevLett.89.011304} {\bibfield  {journal}
  {\bibinfo  {journal} {Phys. Rev. Lett.}\ }\textbf {\bibinfo {volume} {89}},\
  \bibinfo {pages} {011304} (\bibinfo {year} {2002})}\BibitemShut {NoStop}%
\bibitem [{\citenamefont {Seljak}\ and\ \citenamefont
  {Hirata}(2004)}]{Seljak2003a}%
  \BibitemOpen
  \bibfield  {author} {\bibinfo {author} {\bibfnamefont {U.}~\bibnamefont
  {Seljak}}\ and\ \bibinfo {author} {\bibfnamefont {C.~M.}\ \bibnamefont
  {Hirata}},\ }\href {\doibase 10.1103/PhysRevD.69.043005} {\bibfield
  {journal} {\bibinfo  {journal} {Phys. Rev. D}\ }\textbf {\bibinfo {volume}
  {69}},\ \bibinfo {pages} {043005} (\bibinfo {year} {2004})},\ \Eprint
  {http://arxiv.org/abs/0310163} {arXiv:0310163 [astro-ph]} \BibitemShut
  {NoStop}%
\bibitem [{\citenamefont {Simard}\ \emph {et~al.}(2015)\citenamefont {Simard},
  \citenamefont {Hanson},\ and\ \citenamefont {Holder}}]{Simard2015a}%
  \BibitemOpen
  \bibfield  {author} {\bibinfo {author} {\bibfnamefont {G.}~\bibnamefont
  {Simard}}, \bibinfo {author} {\bibfnamefont {D.}~\bibnamefont {Hanson}}, \
  and\ \bibinfo {author} {\bibfnamefont {G.}~\bibnamefont {Holder}},\ }\href
  {\doibase 10.1088/0004-637X/807/2/166} {\bibfield  {journal} {\bibinfo
  {journal} {Astrophys. J.}\ }\textbf {\bibinfo {volume} {807}},\ \bibinfo
  {pages} {166} (\bibinfo {year} {2015})}\BibitemShut {NoStop}%
\bibitem [{\citenamefont {{Planck Collaboration
  XI}}(2016)}]{PlanckCollaborationXI2015}%
  \BibitemOpen
  \bibfield  {author} {\bibinfo {author} {\bibnamefont {{Planck Collaboration
  XI}}},\ }\href {\doibase 10.1051/0004-6361/201526926} {\bibfield  {journal}
  {\bibinfo  {journal} {Astron. Astrophys.}\ }\textbf {\bibinfo {volume}
  {594}},\ \bibinfo {pages} {A11} (\bibinfo {year} {2016})},\ \Eprint
  {http://arxiv.org/abs/1507.02704} {arXiv:1507.02704} \BibitemShut {NoStop}%
\bibitem [{\citenamefont {{Planck Collaboration
  XIII}}(2016)}]{PlanckCollaborationXIII2015}%
  \BibitemOpen
  \bibfield  {author} {\bibinfo {author} {\bibnamefont {{Planck Collaboration
  XIII}}},\ }\href {\doibase 10.1051/0004-6361/201525830} {\bibfield  {journal}
  {\bibinfo  {journal} {Astron. Astrophys.}\ }\textbf {\bibinfo {volume}
  {594}},\ \bibinfo {pages} {A13} (\bibinfo {year} {2016})},\ \Eprint
  {http://arxiv.org/abs/1502.01589} {arXiv:1502.01589} \BibitemShut {NoStop}%
\bibitem [{\citenamefont {Song}\ \emph {et~al.}(2003)\citenamefont {Song},
  \citenamefont {Cooray}, \citenamefont {Knox},\ and\ \citenamefont
  {Zaldarriaga}}]{Song2003}%
  \BibitemOpen
  \bibfield  {author} {\bibinfo {author} {\bibfnamefont {Y.-S.}\ \bibnamefont
  {Song}}, \bibinfo {author} {\bibfnamefont {A.}~\bibnamefont {Cooray}},
  \bibinfo {author} {\bibfnamefont {L.}~\bibnamefont {Knox}}, \ and\ \bibinfo
  {author} {\bibfnamefont {M.}~\bibnamefont {Zaldarriaga}},\ }\href {\doibase
  10.1086/375188} {\bibfield  {journal} {\bibinfo  {journal} {Astrophys. J.}\
  }\textbf {\bibinfo {volume} {590}},\ \bibinfo {pages} {664} (\bibinfo {year}
  {2003})},\ \Eprint {http://arxiv.org/abs/0209001v1} {arXiv:0209001v1
  [arXiv:astro-ph]} \BibitemShut {NoStop}%
\bibitem [{\citenamefont {Dole}\ \emph {et~al.}(2006)\citenamefont {Dole},
  \citenamefont {Lagache}, \citenamefont {Puget}, \citenamefont {Caputi},
  \citenamefont {Fern{\'{a}}ndez-Conde}, \citenamefont {{Le Floc'h}},
  \citenamefont {Papovich}, \citenamefont {P{\'{e}}rez-Gonz{\'{a}}lez},
  \citenamefont {Rieke},\ and\ \citenamefont {Blaylock}}]{Dole2006}%
  \BibitemOpen
  \bibfield  {author} {\bibinfo {author} {\bibfnamefont {H.}~\bibnamefont
  {Dole}}, \bibinfo {author} {\bibfnamefont {G.}~\bibnamefont {Lagache}},
  \bibinfo {author} {\bibfnamefont {J.-L.}\ \bibnamefont {Puget}}, \bibinfo
  {author} {\bibfnamefont {K.~I.}\ \bibnamefont {Caputi}}, \bibinfo {author}
  {\bibfnamefont {N.}~\bibnamefont {Fern{\'{a}}ndez-Conde}}, \bibinfo {author}
  {\bibfnamefont {E.}~\bibnamefont {{Le Floc'h}}}, \bibinfo {author}
  {\bibfnamefont {C.}~\bibnamefont {Papovich}}, \bibinfo {author}
  {\bibfnamefont {P.~G.}\ \bibnamefont {P{\'{e}}rez-Gonz{\'{a}}lez}}, \bibinfo
  {author} {\bibfnamefont {G.~H.}\ \bibnamefont {Rieke}}, \ and\ \bibinfo
  {author} {\bibfnamefont {M.}~\bibnamefont {Blaylock}},\ }\href {\doibase
  10.1051/0004-6361:20054446} {\bibfield  {journal} {\bibinfo  {journal}
  {Astron. Astrophys.}\ }\textbf {\bibinfo {volume} {451}},\ \bibinfo {pages}
  {417} (\bibinfo {year} {2006})}\BibitemShut {NoStop}%
\bibitem [{\citenamefont {{Planck Collaboration
  XVII}}(2014)}]{PlanckCollaborationXVII2014}%
  \BibitemOpen
  \bibfield  {author} {\bibinfo {author} {\bibnamefont {{Planck Collaboration
  XVII}}},\ }\href {\doibase 10.1051/0004-6361/201321540} {\bibfield  {journal}
  {\bibinfo  {journal} {Astron. Astrophys.}\ }\textbf {\bibinfo {volume}
  {571}},\ \bibinfo {pages} {A18} (\bibinfo {year} {2014})}\BibitemShut
  {NoStop}%
\bibitem [{\citenamefont {Larsen}\ \emph {et~al.}(2016)\citenamefont {Larsen},
  \citenamefont {Challinor}, \citenamefont {Sherwin},\ and\ \citenamefont
  {Mak}}]{Larsen2016}%
  \BibitemOpen
  \bibfield  {author} {\bibinfo {author} {\bibfnamefont {P.}~\bibnamefont
  {Larsen}}, \bibinfo {author} {\bibfnamefont {A.}~\bibnamefont {Challinor}},
  \bibinfo {author} {\bibfnamefont {B.~D.}\ \bibnamefont {Sherwin}}, \ and\
  \bibinfo {author} {\bibfnamefont {D.}~\bibnamefont {Mak}},\ }\href {\doibase
  10.1103/PhysRevLett.117.151102} {\bibfield  {journal} {\bibinfo  {journal}
  {Phys. Rev. Lett.}\ }\textbf {\bibinfo {volume} {117}},\ \bibinfo {pages}
  {151102} (\bibinfo {year} {2016})},\ \Eprint
  {http://arxiv.org/abs/1607.05733} {arXiv:1607.05733} \BibitemShut {NoStop}%
\bibitem [{\citenamefont {Manzotti}\ \emph {et~al.}(2017)\citenamefont
  {Manzotti}, \citenamefont {Story}, \citenamefont {Wu}, \citenamefont
  {Austermann}, \citenamefont {Beall}, \citenamefont {Bender}, \citenamefont
  {Benson}, \citenamefont {Bleem}, \citenamefont {Bock}, \citenamefont
  {Carlstrom}, \citenamefont {Chang}, \citenamefont {Chiang}, \citenamefont
  {Cho}, \citenamefont {Citron}, \citenamefont {Conley}, \citenamefont
  {Crawford}, \citenamefont {Crites}, \citenamefont {de~Haan}, \citenamefont
  {Dobbs}, \citenamefont {Dodelson}, \citenamefont {Everett}, \citenamefont
  {Gallicchio}, \citenamefont {George}, \citenamefont {Gilbert}, \citenamefont
  {Halverson}, \citenamefont {Harrington}, \citenamefont {Henning},
  \citenamefont {Hilton}, \citenamefont {Holder}, \citenamefont {Holzapfel},
  \citenamefont {Hoover}, \citenamefont {Hou}, \citenamefont {Hrubes},
  \citenamefont {Huang}, \citenamefont {Hubmayr}, \citenamefont {Irwin},
  \citenamefont {Keisler}, \citenamefont {Knox}, \citenamefont {Lee},
  \citenamefont {Leitch}, \citenamefont {Li}, \citenamefont {McMahon},
  \citenamefont {Meyer}, \citenamefont {Mocanu}, \citenamefont {Natoli},
  \citenamefont {Nibarger}, \citenamefont {Novosad}, \citenamefont {Padin},
  \citenamefont {Pryke}, \citenamefont {Reichardt}, \citenamefont {Ruhl},
  \citenamefont {Saliwanchik}, \citenamefont {Sayre}, \citenamefont {Schaffer},
  \citenamefont {Smecher}, \citenamefont {Stark}, \citenamefont {Vanderlinde},
  \citenamefont {Vieira}, \citenamefont {Viero}, \citenamefont {Wang},
  \citenamefont {Whitehorn}, \citenamefont {Yefremenko},\ and\ \citenamefont
  {Zemcov}}]{Manzotti2017}%
  \BibitemOpen
  \bibfield  {author} {\bibinfo {author} {\bibfnamefont {A.}~\bibnamefont
  {Manzotti}}, \bibinfo {author} {\bibfnamefont {K.~T.}\ \bibnamefont {Story}},
  \bibinfo {author} {\bibfnamefont {W.~L.~K.}\ \bibnamefont {Wu}}, \bibinfo
  {author} {\bibfnamefont {J.~E.}\ \bibnamefont {Austermann}}, \bibinfo
  {author} {\bibfnamefont {J.~A.}\ \bibnamefont {Beall}}, \bibinfo {author}
  {\bibfnamefont {A.~N.}\ \bibnamefont {Bender}}, \bibinfo {author}
  {\bibfnamefont {B.~A.}\ \bibnamefont {Benson}}, \bibinfo {author}
  {\bibfnamefont {L.~E.}\ \bibnamefont {Bleem}}, \bibinfo {author}
  {\bibfnamefont {J.~J.}\ \bibnamefont {Bock}}, \bibinfo {author}
  {\bibfnamefont {J.~E.}\ \bibnamefont {Carlstrom}}, \bibinfo {author}
  {\bibfnamefont {C.~L.}\ \bibnamefont {Chang}}, \bibinfo {author}
  {\bibfnamefont {H.~C.}\ \bibnamefont {Chiang}}, \bibinfo {author}
  {\bibfnamefont {H.-M.}\ \bibnamefont {Cho}}, \bibinfo {author} {\bibfnamefont
  {R.}~\bibnamefont {Citron}}, \bibinfo {author} {\bibfnamefont
  {A.}~\bibnamefont {Conley}}, \bibinfo {author} {\bibfnamefont {T.~M.}\
  \bibnamefont {Crawford}}, \bibinfo {author} {\bibfnamefont {A.~T.}\
  \bibnamefont {Crites}}, \bibinfo {author} {\bibfnamefont {T.}~\bibnamefont
  {de~Haan}}, \bibinfo {author} {\bibfnamefont {M.~A.}\ \bibnamefont {Dobbs}},
  \bibinfo {author} {\bibfnamefont {S.}~\bibnamefont {Dodelson}}, \bibinfo
  {author} {\bibfnamefont {W.}~\bibnamefont {Everett}}, \bibinfo {author}
  {\bibfnamefont {J.}~\bibnamefont {Gallicchio}}, \bibinfo {author}
  {\bibfnamefont {E.~M.}\ \bibnamefont {George}}, \bibinfo {author}
  {\bibfnamefont {A.}~\bibnamefont {Gilbert}}, \bibinfo {author} {\bibfnamefont
  {N.~W.}\ \bibnamefont {Halverson}}, \bibinfo {author} {\bibfnamefont
  {N.}~\bibnamefont {Harrington}}, \bibinfo {author} {\bibfnamefont {J.~W.}\
  \bibnamefont {Henning}}, \bibinfo {author} {\bibfnamefont {G.~C.}\
  \bibnamefont {Hilton}}, \bibinfo {author} {\bibfnamefont {G.~P.}\
  \bibnamefont {Holder}}, \bibinfo {author} {\bibfnamefont {W.~L.}\
  \bibnamefont {Holzapfel}}, \bibinfo {author} {\bibfnamefont {S.}~\bibnamefont
  {Hoover}}, \bibinfo {author} {\bibfnamefont {Z.}~\bibnamefont {Hou}},
  \bibinfo {author} {\bibfnamefont {J.~D.}\ \bibnamefont {Hrubes}}, \bibinfo
  {author} {\bibfnamefont {N.}~\bibnamefont {Huang}}, \bibinfo {author}
  {\bibfnamefont {J.}~\bibnamefont {Hubmayr}}, \bibinfo {author} {\bibfnamefont
  {K.~D.}\ \bibnamefont {Irwin}}, \bibinfo {author} {\bibfnamefont
  {R.}~\bibnamefont {Keisler}}, \bibinfo {author} {\bibfnamefont
  {L.}~\bibnamefont {Knox}}, \bibinfo {author} {\bibfnamefont {A.~T.}\
  \bibnamefont {Lee}}, \bibinfo {author} {\bibfnamefont {E.~M.}\ \bibnamefont
  {Leitch}}, \bibinfo {author} {\bibfnamefont {D.}~\bibnamefont {Li}}, \bibinfo
  {author} {\bibfnamefont {J.~J.}\ \bibnamefont {McMahon}}, \bibinfo {author}
  {\bibfnamefont {S.~S.}\ \bibnamefont {Meyer}}, \bibinfo {author}
  {\bibfnamefont {L.~M.}\ \bibnamefont {Mocanu}}, \bibinfo {author}
  {\bibfnamefont {T.}~\bibnamefont {Natoli}}, \bibinfo {author} {\bibfnamefont
  {J.~P.}\ \bibnamefont {Nibarger}}, \bibinfo {author} {\bibfnamefont
  {V.}~\bibnamefont {Novosad}}, \bibinfo {author} {\bibfnamefont
  {S.}~\bibnamefont {Padin}}, \bibinfo {author} {\bibfnamefont
  {C.}~\bibnamefont {Pryke}}, \bibinfo {author} {\bibfnamefont {C.~L.}\
  \bibnamefont {Reichardt}}, \bibinfo {author} {\bibfnamefont {J.~E.}\
  \bibnamefont {Ruhl}}, \bibinfo {author} {\bibfnamefont {B.~R.}\ \bibnamefont
  {Saliwanchik}}, \bibinfo {author} {\bibfnamefont {J.~T.}\ \bibnamefont
  {Sayre}}, \bibinfo {author} {\bibfnamefont {K.~K.}\ \bibnamefont {Schaffer}},
  \bibinfo {author} {\bibfnamefont {G.}~\bibnamefont {Smecher}}, \bibinfo
  {author} {\bibfnamefont {A.~A.}\ \bibnamefont {Stark}}, \bibinfo {author}
  {\bibfnamefont {K.}~\bibnamefont {Vanderlinde}}, \bibinfo {author}
  {\bibfnamefont {J.~D.}\ \bibnamefont {Vieira}}, \bibinfo {author}
  {\bibfnamefont {M.~P.}\ \bibnamefont {Viero}}, \bibinfo {author}
  {\bibfnamefont {G.}~\bibnamefont {Wang}}, \bibinfo {author} {\bibfnamefont
  {N.}~\bibnamefont {Whitehorn}}, \bibinfo {author} {\bibfnamefont
  {V.}~\bibnamefont {Yefremenko}}, \ and\ \bibinfo {author} {\bibfnamefont
  {M.}~\bibnamefont {Zemcov}},\ }\href {http://arxiv.org/abs/1701.04396} {\
  (\bibinfo {year} {2017})},\ \Eprint {http://arxiv.org/abs/1701.04396}
  {arXiv:1701.04396} \BibitemShut {NoStop}%
\bibitem [{\citenamefont {Hu}(2001)}]{Hu2001}%
  \BibitemOpen
  \bibfield  {author} {\bibinfo {author} {\bibfnamefont {W.}~\bibnamefont
  {Hu}},\ }\href {\doibase 10.1086/323253} {\bibfield  {journal} {\bibinfo
  {journal} {Astrophys. J.}\ }\textbf {\bibinfo {volume} {557}},\ \bibinfo
  {pages} {L79} (\bibinfo {year} {2001})}\BibitemShut {NoStop}%
\bibitem [{\citenamefont {Hu}\ and\ \citenamefont {Okamoto}(2002)}]{Hu2002b}%
  \BibitemOpen
  \bibfield  {author} {\bibinfo {author} {\bibfnamefont {W.}~\bibnamefont
  {Hu}}\ and\ \bibinfo {author} {\bibfnamefont {T.}~\bibnamefont {Okamoto}},\
  }\href {\doibase 10.1086/341110} {\bibfield  {journal} {\bibinfo  {journal}
  {Astrophys. J.}\ }\textbf {\bibinfo {volume} {574}},\ \bibinfo {pages} {566}
  (\bibinfo {year} {2002})},\ \Eprint {http://arxiv.org/abs/0111606}
  {arXiv:0111606 [astro-ph]} \BibitemShut {NoStop}%
\bibitem [{\citenamefont {Hirata}\ and\ \citenamefont
  {Seljak}(2003{\natexlab{a}})}]{Hirata2003a}%
  \BibitemOpen
  \bibfield  {author} {\bibinfo {author} {\bibfnamefont {C.~M.}\ \bibnamefont
  {Hirata}}\ and\ \bibinfo {author} {\bibfnamefont {U.}~\bibnamefont
  {Seljak}},\ }\href {\doibase 10.1103/PhysRevD.67.043001} {\bibfield
  {journal} {\bibinfo  {journal} {Phys. Rev. D}\ }\textbf {\bibinfo {volume}
  {67}},\ \bibinfo {pages} {043001} (\bibinfo {year}
  {2003}{\natexlab{a}})}\BibitemShut {NoStop}%
\bibitem [{\citenamefont {Hirata}\ and\ \citenamefont
  {Seljak}(2003{\natexlab{b}})}]{Hirata2003}%
  \BibitemOpen
  \bibfield  {author} {\bibinfo {author} {\bibfnamefont {C.~M.}\ \bibnamefont
  {Hirata}}\ and\ \bibinfo {author} {\bibfnamefont {U.}~\bibnamefont
  {Seljak}},\ }\href {\doibase 10.1103/PhysRevD.68.083002} {\bibfield
  {journal} {\bibinfo  {journal} {Phys. Rev. D}\ }\textbf {\bibinfo {volume}
  {68}},\ \bibinfo {pages} {083002} (\bibinfo {year}
  {2003}{\natexlab{b}})}\BibitemShut {NoStop}%
\bibitem [{\citenamefont {Anderes}\ \emph {et~al.}(2011)\citenamefont
  {Anderes}, \citenamefont {Knox},\ and\ \citenamefont {{Van
  Engelen}}}]{Anderes2011}%
  \BibitemOpen
  \bibfield  {author} {\bibinfo {author} {\bibfnamefont {E.}~\bibnamefont
  {Anderes}}, \bibinfo {author} {\bibfnamefont {L.}~\bibnamefont {Knox}}, \
  and\ \bibinfo {author} {\bibfnamefont {A.}~\bibnamefont {{Van Engelen}}},\
  }\href {\doibase 10.1103/PhysRevD.83.043523} {\bibfield  {journal} {\bibinfo
  {journal} {Phys. Rev. D}\ }\textbf {\bibinfo {volume} {83}},\ \bibinfo
  {pages} {043523} (\bibinfo {year} {2011})},\ \Eprint
  {http://arxiv.org/abs/1012.1833} {arXiv:1012.1833 [astro-ph.CO]} \BibitemShut
  {NoStop}%
\bibitem [{\citenamefont {Anderes}\ \emph {et~al.}(2015)\citenamefont
  {Anderes}, \citenamefont {Wandelt},\ and\ \citenamefont
  {Lavaux}}]{Anderes2015}%
  \BibitemOpen
  \bibfield  {author} {\bibinfo {author} {\bibfnamefont {E.}~\bibnamefont
  {Anderes}}, \bibinfo {author} {\bibfnamefont {B.~D.}\ \bibnamefont
  {Wandelt}}, \ and\ \bibinfo {author} {\bibfnamefont {G.}~\bibnamefont
  {Lavaux}},\ }\href {\doibase 10.1088/0004-637X/808/2/152} {\bibfield
  {journal} {\bibinfo  {journal} {Astrophys. J.}\ }\textbf {\bibinfo {volume}
  {808}},\ \bibinfo {pages} {152} (\bibinfo {year} {2015})}\BibitemShut
  {NoStop}%
\bibitem [{\citenamefont {Millea}\ \emph {et~al.}(2017)\citenamefont {Millea},
  \citenamefont {Anderes},\ and\ \citenamefont {Wandelt}}]{Millea2017}%
  \BibitemOpen
  \bibfield  {author} {\bibinfo {author} {\bibfnamefont {M.}~\bibnamefont
  {Millea}}, \bibinfo {author} {\bibfnamefont {E.}~\bibnamefont {Anderes}}, \
  and\ \bibinfo {author} {\bibfnamefont {B.~D.}\ \bibnamefont {Wandelt}},\
  }\href {https://arxiv.org/abs/1708.06753
  https://arxiv.org/pdf/1708.06753.pdf} {\bibfield  {journal} {\bibinfo
  {journal} {arXiv}\ } (\bibinfo {year} {2017})},\ \Eprint
  {http://arxiv.org/abs/1708.06753} {arXiv:1708.06753} \BibitemShut {NoStop}%
\bibitem [{\citenamefont {Lesgourgues}(2011)}]{Lesg2011}%
  \BibitemOpen
  \bibfield  {author} {\bibinfo {author} {\bibfnamefont {J.}~\bibnamefont
  {Lesgourgues}},\ }\href@noop {} {\bibfield  {journal} {\bibinfo  {journal}
  {eprint}\ } (\bibinfo {year} {2011})},\ \Eprint
  {http://arxiv.org/abs/1104.2932} {arXiv:1104.2932 [astro-ph.IM]} \BibitemShut
  {NoStop}%
\bibitem [{\citenamefont {Anderes}(2013)}]{Anderes2013}%
  \BibitemOpen
  \bibfield  {author} {\bibinfo {author} {\bibfnamefont {E.}~\bibnamefont
  {Anderes}},\ }\href {\doibase 10.1103/PhysRevD.88.083517} {\bibfield
  {journal} {\bibinfo  {journal} {Phys. Rev. D}\ }\textbf {\bibinfo {volume}
  {88}},\ \bibinfo {pages} {083517} (\bibinfo {year} {2013})},\ \Eprint
  {http://arxiv.org/abs/arXiv:1301.2576v1} {arXiv:arXiv:1301.2576v1}
  \BibitemShut {NoStop}%
\bibitem [{\citenamefont {Teng}\ \emph {et~al.}(2011)\citenamefont {Teng},
  \citenamefont {Kuo},\ and\ \citenamefont {Wu}}]{Teng2011}%
  \BibitemOpen
  \bibfield  {author} {\bibinfo {author} {\bibfnamefont {W.-H.}\ \bibnamefont
  {Teng}}, \bibinfo {author} {\bibfnamefont {C.-L.}\ \bibnamefont {Kuo}}, \
  and\ \bibinfo {author} {\bibfnamefont {J.-H.~P.}\ \bibnamefont {Wu}},\ }\href
  {http://arxiv.org/abs/1102.5729} {\bibfield  {journal} {\bibinfo  {journal}
  {eprint}\ } (\bibinfo {year} {2011})},\ \Eprint
  {http://arxiv.org/abs/1102.5729} {arXiv:1102.5729} \BibitemShut {NoStop}%
\bibitem [{\citenamefont {Namikawa}\ and\ \citenamefont
  {Nagata}(2015)}]{Namikawa2015}%
  \BibitemOpen
  \bibfield  {author} {\bibinfo {author} {\bibfnamefont {T.}~\bibnamefont
  {Namikawa}}\ and\ \bibinfo {author} {\bibfnamefont {R.}~\bibnamefont
  {Nagata}},\ }\href {\doibase 10.1088/1475-7516/2015/10/004} {\bibfield
  {journal} {\bibinfo  {journal} {J. Cosmol. Astropart. Phys.}\ }\textbf
  {\bibinfo {volume} {2015}},\ \bibinfo {pages} {004} (\bibinfo {year}
  {2015})},\ \Eprint {http://arxiv.org/abs/1506.09209} {arXiv:1506.09209}
  \BibitemShut {NoStop}%
\bibitem [{\citenamefont {Sehgal}\ \emph {et~al.}(2017)\citenamefont {Sehgal},
  \citenamefont {Madhavacheril}, \citenamefont {Sherwin},\ and\ \citenamefont
  {{Van Engelen}}}]{Sehgal2016}%
  \BibitemOpen
  \bibfield  {author} {\bibinfo {author} {\bibfnamefont {N.}~\bibnamefont
  {Sehgal}}, \bibinfo {author} {\bibfnamefont {M.~S.}\ \bibnamefont
  {Madhavacheril}}, \bibinfo {author} {\bibfnamefont {B.}~\bibnamefont
  {Sherwin}}, \ and\ \bibinfo {author} {\bibfnamefont {A.}~\bibnamefont {{Van
  Engelen}}},\ }\href {\doibase 10.1103/PhysRevD.95.103512} {\bibfield
  {journal} {\bibinfo  {journal} {Phys. Rev. D}\ }\textbf {\bibinfo {volume}
  {95}},\ \bibinfo {pages} {103512} (\bibinfo {year} {2017})},\ \Eprint
  {http://arxiv.org/abs/1612.03898} {arXiv:1612.03898} \BibitemShut {NoStop}%
\bibitem [{\citenamefont {Carron}\ \emph {et~al.}(2017)\citenamefont {Carron},
  \citenamefont {Lewis},\ and\ \citenamefont {Challinor}}]{Carron2017}%
  \BibitemOpen
  \bibfield  {author} {\bibinfo {author} {\bibfnamefont {J.}~\bibnamefont
  {Carron}}, \bibinfo {author} {\bibfnamefont {A.}~\bibnamefont {Lewis}}, \
  and\ \bibinfo {author} {\bibfnamefont {A.}~\bibnamefont {Challinor}},\ }\href
  {http://arxiv.org/abs/1701.01712} {\bibfield  {journal} {\bibinfo  {journal}
  {eprint}\ } (\bibinfo {year} {2017})},\ \Eprint
  {http://arxiv.org/abs/1701.01712} {arXiv:1701.01712} \BibitemShut {NoStop}%
\end{thebibliography}%

\appendix
\section{Signal Covariance Matrix}
\label{sec:app1}

There are eight different terms in the map covariance matrix $\Sigma_r$:
$\{ \tilde\Sigma^{SE,SE},\tilde\Sigma^{CE,SE},\tilde\Sigma^{CE,CE} \}$,
$\{ \tilde\Sigma^{SB,SB},\tilde\Sigma^{CB,SB},\tilde\Sigma^{CB,CB} \}$,
and $\{ \Sigma^{NQ,NQ},\Sigma^{NU,NU} \}$ (see Equations (\ref{eq:elements})).
In this subsection, we show how the marginalization over uncertainty in the $\phi$ estimate
is done for the six lensed signal terms,
and leave the two noise terms to the next subsection.
Take $\tilde\Sigma^{XX} (X = SE)$ as an example,
\begin{widetext}
\be
\begin{aligned}
\left<\tilde\Sigma^{XX}_{r,\phi} \right>_{n\phi}
& = \left<\widetilde X(\mathbf x) \widetilde X(\mathbf y)  \right>_{n\phi} \\
& = \left<X(x + \nabla \mu\phi(\mathbf x) + \nabla n\phi(\mathbf x))
 X(\mathbf y + \nabla \mu\phi(\mathbf y) + \nabla n\phi(\mathbf y)) \right>_{n\phi} \\
& = \int \frac{d^2\Bell}{(2\pi)^2} e^{i \Bell\cdot (\mathbf x- \mathbf y+\nabla\mu\phi(\mathbf x) - \nabla\mu\phi(\mathbf y))}
C_\ell^{XX} \left< e^{i\Bell\cdot(\nabla n\phi(\mathbf x) -\nabla n\phi(\mathbf y))}\right>_{n\phi} \\
& = \int \frac{d^2\Bell}{(2\pi)^2} e^{i \Bell\cdot (\mathbf x-\mathbf y+\nabla\mu\phi(\mathbf x) - \nabla\mu\phi(\mathbf y))}
C_\ell^{XX} \exp\{-\Bell\cdot \left[\Sigma^{n\phi}(0)-\Sigma^{n\phi}(\mathbf x-\mathbf y)\right] \cdot \Bell\}  \\
&\simeq \int \frac{d^2\Bell}{(2\pi)^2} e^{i \Bell\cdot (\mathbf x-\mathbf y+\nabla\mu\phi(\mathbf x) - \nabla\mu\phi(\mathbf y))}
C_\ell^{XX} \left(1- \Bell\cdot \left[\Sigma^{n\phi}(0)-\Sigma^{n\phi}(\mathbf x-\mathbf y)\right] \cdot \Bell\right)  \\
& = \int \frac{d^2\Bell}{(2\pi)^2} e^{i \Bell\cdot (\mathbf x-\mathbf y+\nabla\mu\phi(\mathbf x) - \nabla\mu\phi(\mathbf y))} C_\ell^{XX}  \\
& - \sum_{p,q=1}^2 \left[\Sigma^{n\phi}(0)-\Sigma^{n\phi}(\mathbf x-\mathbf y)\right]_{p,q}
\int \frac{d^2\Bell}{(2\pi)^2} \ell_p \ell_q e^{i \Bell\cdot (\mathbf x-\mathbf y+\nabla\mu\phi(\mathbf x) - \nabla\mu\phi(\mathbf y))} C_\ell^{XX} \\
& = {\rm Cov}(X(\mathbf w), X(0))
+ \sum_{p,q=1}^2 \left[\Sigma^{n\phi}(0)-\Sigma^{n\phi}(\mathbf x-\mathbf y)\right]_{p,q}
\partial_{p,q} {\rm Cov}(X(\mathbf w), X(0))
\end{aligned}
\ee
where we have used cumulant expansion at the 4th equal sign,  $\left[\Sigma^{n\phi}(\mathbf x-\mathbf y)\right]_{p,q}$
is the covariance of $\nabla(n\phi)$, i.e.,
\be
\begin{aligned}
\label{eq:a2}
    \left[\Sigma^{n\phi}(\mathbf x-\mathbf y)\right]_{p,q}
  & = \left< \nabla_p n\phi(\mathbf x) \nabla_q n\phi(\mathbf y)\right>_{n\phi}
    = \int \frac{d^2\Bell}{(2\pi)^2} \ell_p \ell_q e^{i\Bell\cdot (\mathbf x-\mathbf y)} N_\ell^{\phi\phi},
\end{aligned}
\ee

\end{widetext}
and ${\rm Cov}(X(\mathbf w), X(0))$ is the covariance of $X$
at separation $\mathbf w =\mathbf x-\mathbf y + \nabla\mu\phi(\mathbf x)-\nabla\mu\phi(\mathbf y)$,
\be
\label{eq:a3}
    {\rm Cov}(X(\mathbf w), X(0)) = \int \frac{d^2\Bell}{(2\pi)^2} e^{i \Bell\cdot \mathbf w} C_\ell^{XX}.
\ee

The above two dimensional integrals (Equations (\ref{eq:a2}-\ref{eq:a3})) can be simplified as  one dimensional integrals
as follows. Take Equation (\ref{eq:a3}) as an example,
\be
\label{eq:a4}
\begin{aligned}
{\rm Cov}\left( X(\mathbf w), X(0) \right)
& = 4\partial_1^2\partial_2^2 \int \frac{d^2\Bell}{(2\pi)^2} e^{i\Bell\cdot \mathbf w} C_\ell^{\mE\mE}\\
&\equiv 4\partial_1^2\partial_2^2 K^{\mathcal E}(w),
\end{aligned}
\ee
where we have used $C_\ell^{XX} = 4\ell_1^2 \ell_2^2 (C_\ell^{EE}/\ell^4)$
and defined $C_\ell^{\mE\mE} \equiv C_\ell^{EE}/\ell^4 $.
Exploiting the integral representation of Bessel functions, we rewrite $K^{\mathcal E}(\mathbf w)$ as a one dimensional
integral
\be
\begin{aligned}
K^{\mathcal E}(\mathbf w)
&=\int \frac{d^2\Bell}{(2\pi)^2} e^{i\Bell\cdot \mathbf w} C_\ell^{\mE\mE} \\
&=\frac{1}{2\pi}\int  J_0\left(\ell w \right)  C_\ell^{\mE\mE} \ell d\ell,
\end{aligned}
\ee
which has no angular dependence.
For derivative calculation, we define $\hat K(w^2) \equiv K^{\mathcal E}(\mathbf w)$,
then
\be
\label{eq:a6}
\begin{aligned}
&\partial_1^2 \partial_2^2 K^{\mathcal E}(\mathbf w)
=\partial_1^2 \partial_2^2 \hat K (w^2) \\
&= 16w_1^2w_2^2 \ \hat K^{(4)}(w^2)  \\
& + 8(w_1^2 + w_2^2) \ \hat K^{(3)}(w^2)  + 4\hat K^{(2)}(w^2).
\end{aligned}
\ee
Using the property
\be
\frac{d}{dz} z^{-s} J_s(z) = - z^{-s} J_{s+1}(z),
\ee
the $n$-th order derivative $\hat K^{(n)}$ is explicitly expressed as
\be
\label{eq:a8}
\hat K^{(n)}(w^2)
= \frac{1}{2\pi} \int \left(-\frac{\ell}{2w} \right)^n  J_n(\ell w) \ell d\ell.
\ee
Collecting Equations (\ref{eq:a4}, \ref{eq:a6}, \ref{eq:a8}),  ${\rm Cov}\left( X(\mathbf w), X(0) \right)$
is decomposed into a few one dimensional integrals.
The calculation of $\partial_{p,q}{\rm Cov}\left( X(\mathbf w), X(0) \right)$ and
$\left[\Sigma^{n\phi}(\mathbf x-\mathbf y)\right]_{p,q}$ is conducted in the same way.
For other lensed terms, the above formulas apply similarly.

\section{Noise Covariance Matrix}
\label{sec:app2}

In Section \ref{sec:app1}, we completely ignore the consequence of the finite beam size in
the signal covariance evaluation, since the signal suppression by the beam convolution
can be interpreted as the noise enhancement by the beam deconvolution.
For noise field $n(\mathbf x)$,  we denote the deconvolved noise field as
$X(\mathbf x) = \varphi^{-1}_x[ n(\mathbf x) ]$, with
\be
\begin{aligned}
    \varphi^{-1}_x[ n(\mathbf x) ]
    &= \int \frac{d^2\Bell}{2\pi} e^{i\Bell\cdot \mathbf x}  \frac{n_{\Bell}}{\varphi_{\Bell} }\\
    &=\int \frac{d^2\Bell}{2\pi} \frac{d^2\mathbf x'}{2\pi}
    e^{i\Bell\cdot (\mathbf x- \mathbf x')}  \frac{n(\mathbf x')}{\varphi_{\Bell} },
\end{aligned}
\ee
\begin{widetext}
where for Gaussian beam profile
$\varphi(\mathbf x) = \frac{1}{2\pi\sigma_b^2} \exp\left(-\frac{\mathbf x^2}{2\sigma_b^2}\right)$,
$\varphi_{\Bell} = \exp \left(-\frac{l^2 \sigma_b^2}{2} \right)$, and $\sigma_b^2 = \theta^2_{\rm FWHM}/(8 \ln 2)$.
Then
\begin{align}
    \Sigma^{XX} = \left< X(\mathbf x) X(\mathbf y)\right>
    &= \int \frac{d^2\Bell}{2\pi}
    \frac{d^2\mathbf x'}{2\pi} \frac{d^2\mathbf k}{2\pi} \frac{d^2\mathbf y'}{2\pi}
    e^{i\Bell\cdot (\mathbf x- \mathbf x')}
     e^{i\mathbf k\cdot (\mathbf y- \mathbf y')}  \frac{1}{ \varphi_{\Bell}\varphi_{\mathbf k} }
    \left<n(\mathbf x') n(\mathbf y') \right>,
\end{align}
For simple white noise
$\left< n(\mathbf x) n(\mathbf y)\right> =\Delta_{\rm P}^2 \delta_D(\mathbf x-\mathbf y)$, we have
\begin{align}
    \left< X(\mathbf x) X(\mathbf y)\right>
    &= \int \frac{d^2\Bell}{(2\pi)^2}
    e^{i\Bell\cdot (\mathbf x -\mathbf y)}
    \frac{\Delta_{\rm P}^2 }{ \varphi_{\Bell} \varphi_{-\Bell} }
    = \Delta_{\rm P}^2 \int \frac{d^2\Bell}{(2\pi)^2}
    e^{i\Bell\cdot (\mathbf x -\mathbf y)}\ e^{\ell^2\sigma_b^2},
\end{align}

where
$\Delta_{\rm P}$ is polarization noise and we usually take $\Delta_{\rm P} = \sqrt{2} \Delta_{\rm T}$.
For more realistic non-stationary noise $\left< n(\mathbf x) n(\mathbf y)\right>
= \sigma^2(\mathbf x)\Delta_{\rm P}^2 \delta_D(\mathbf x-\mathbf y)$,
the covariance matrix of the deconvolved noise field $X(\mathbf x)$ is written as
\be
\begin{aligned}
    \left< X(\mathbf x) X(\mathbf y)\right>
    &= \left<\varphi_x^{-1}[n(\mathbf x)]  \ \varphi_y^{-1}[n(\mathbf y)]\right> \\
    &= \left<\varphi_x^{-1}[n(\mathbf x)]  \ \int \frac{d^2\Bell}{2\pi} \frac{d^2\mathbf y'}{2\pi}
    e^{i\Bell\cdot (\mathbf y- \mathbf y')}  \frac{n(\mathbf y')}{\varphi_{\Bell}} \right>\\
    &= \varphi_x^{-1} \left[  \int \frac{d^2\Bell}{2\pi} \frac{d^2\mathbf y'}{2\pi}
    e^{i\Bell\cdot (\mathbf y- \mathbf y')}  \frac{1}{\varphi_{\Bell}} \left<n(\mathbf x)n(\mathbf y') \right> \right] \\
    &=\varphi_x^{-1} \left[ \sigma^2(\mathbf x) \Delta_{\rm P}^2  \int \frac{d^2\Bell}{(2\pi)^2}
    e^{i\Bell\cdot (\mathbf y- \mathbf x)}  e^{\frac{\ell^2\sigma_b^2}{2}} \right]
\end{aligned}
\ee
where we have exchanged the order of deconvolution and ensemble average at the 3rd equal sign,
since deconvolution is a linear operator.

\be
\begin{aligned}
    \left< X(\mathbf x) X(\mathbf y)\right>
    &= \Delta_{\rm P}^2 \int \frac{d^2\Bell}{(2\pi)^2}
    \frac{d^2\mathbf k}{2\pi}     e^{i\Bell\cdot \mathbf x}
         e^{i\mathbf k\cdot \mathbf y}   \varphi_{\Bell}\varphi_{\mathbf k}
        \left[ \int \frac{d^2\mathbf x'}{2\pi}  e^{-i(\Bell + \mathbf k)\cdot \mathbf x'} \delta^2(\mathbf x') \right]\\
    &= \Delta_{\rm P}^2 \int \frac{d^2\Bell}{(2\pi)^2}
    \frac{d^2\mathbf k}{2\pi}     e^{i\Bell\cdot \mathbf x}
         e^{i\mathbf k\cdot \mathbf y}   \varphi_{\Bell}\varphi_{\mathbf k}\
          (\delta^2)_{\Bell + \mathbf k} \\
    &= \Delta_{\rm P}^2 \int \frac{d^2\Bell}{(2\pi)^2}
    e^{i\Bell\cdot \mathbf x} \varphi_{\Bell}
          \int \frac{d^2\mathbf k}{2\pi}
               e^{i\mathbf k\cdot \mathbf y}   \varphi_{\mathbf k}\
                (\delta^2)_{\Bell + \mathbf k}
\end{aligned}
\ee

\end{widetext}

\section{Inverse of Covariance Matrix}
\label{sec:app3}
The inverse covariance matrix $\Sigma_r^{-1}$ evaluation  is the key to
the $r$ likelihood in Equation~(\ref{eq:likeli}).
To avoid repeating the similar computation for every different $r$, we can single out the $r$ dependence
rewriting the covariance matrix in the form
$\Sigma_r = \Sigma^{\rm en} + r \Sigma^{\rm b}$,
where

\[
\begin{aligned}
\Sigma^{\rm en}
&=
\left(
\begin{tabular}{cc}
    $\tilde \Sigma^{CE,CE} + \Sigma^{NQ,NQ}$ & $\tilde\Sigma^{CE,SE}$ \\
    $\tilde \Sigma^{CE,SE}$ & $\tilde\Sigma^{SE,SE}+\Sigma^{NU, NU}$
\end{tabular}
\right), \\
\Sigma^{\rm b}
&=
\left(
\begin{tabular}{cc}
    $\tilde \Sigma^{SB^0,SB^0}$ & $\tilde \Sigma^{CB^0,SB^0}$ \\
    $\tilde \Sigma^{CB^0,SB^0}$ & $\tilde \Sigma^{CB^0,CB^0}$
\end{tabular}
\right).
\end{aligned}
\]

Both $\Sigma^{\rm en}$ and $\Sigma^{\rm b}$ are symmetric and positive definite.
We first decompose $\Sigma^{\rm b}$ as $\Sigma^{\rm b} = V\Lambda V^\intercal$,
with $\Lambda$ being a diagonal matrix composed of its eigenvalues,
and $V$ being a matrix composed of its eigenvectors.
Now we do a little manipulation to the covariance matrix
\be
\begin{aligned}
\Sigma_r
&= \Sigma^{\rm en}  + r V\Lambda V^\intercal \\
&= V\sqrt{\Lambda} \left(\sqrt{\Lambda^{-1}}V^\intercal \Sigma^{\rm en}V\sqrt{\Lambda^{-1}}  + r I\right) \sqrt{\Lambda} V^\intercal,
\end{aligned}
\ee
where we have used the orthogonality $V^\intercal = V^{-1}$. One more eigendecomposition,
$\sqrt{\Lambda^{-1}}V^\intercal \Sigma^{\rm en} V\sqrt{\Lambda^{-1}} = \hat V \hat \Lambda \hat V^\intercal$,
enables us further transform $\Sigma_r$ as
\be
\begin{aligned}
\Sigma_r
&= V\sqrt{\Lambda}\hat V \left( \hat \Lambda + r I\right)  \hat V^\intercal \sqrt{\Lambda} V^\intercal\\
&= V\sqrt{\Lambda}\hat V \left( \hat \Lambda + r I\right)  (V\sqrt{\Lambda}\hat V )^\intercal.
\end{aligned}
\ee
Here we can obtain the inverse matrix $\Sigma_r^{-1}$ at little cost, using the orthogonality of $V$ and $\hat V$.
And more beautifully,  all the matrices $V, \Lambda$ and $\hat V, \hat\Lambda$ have no $r$ dependence,
hence we obtain the inverse covariance matrix as a function of $r$ at the same computation cost
of a single inverse matrix computation.

\end{document}